\documentclass[fleqn,usenatbib]{mnras}

\usepackage[T1]{fontenc}
\usepackage{ae,aecompl}

\usepackage{graphicx}	
\usepackage{amsmath}	
\usepackage{amssymb}	
\usepackage{siunitx}

\usepackage[export]{adjustbox}

\usepackage{setspace}
\renewcommand*{\href}[3][]{#3}

\usepackage{subcaption}
\usepackage{etoolbox}
\makeatletter
\patchcmd\@combinedblfloats{\box\@outputbox}{\unvbox\@outputbox}{}{%
   \errmessage{\noexpand\@combinedblfloats could not be patched}%
}%
 \makeatother

\newcommand{\be}{\begin{equation}}
\newcommand{\ee}{\end{equation}}
\newcommand{\bea}{\begin{eqnarray}}
\newcommand{\eea}{\end{eqnarray}}

\def\Eqn#1{Equation~(\ref{eqn:#1})}
\def\Sec#1{Section~\ref{sec:#1}}
\def\App#1{Appendix~\ref{sec:#1}}
\def\Fig#1{Fig.~\ref{fig:#1}}

\def\ifm#1{\relax\ifmmode#1\else$\mathsurround=0pt #1$\fi}
\def\kms{\ifmmode\,{\rm km}\,{\rm s}^{-1}\else km$\,$s$^{-1}$\fi}

\def\Msun{\,{\rm M_{\odot}}}
\def\kcm3{\ifmmode\,{\rm K}\,{\rm cm}^{-3}\else K$\,$cm$^{-3}$\fi}

\def\Mstar{M_{\star}}

\def\MstarChab{M_{\rm \star,Chab}}

\def\SFRChab{{\rm SFR}_{\rm Chab}}

\def\MBH{M_{\rm BH}}

\def\MDM{M_{\rm DM}}

\newcommand{\lom}{LoM-50}
\newcommand{\him}{HiM-50}

\newcommand{\CapColour}{green}
\newcommand{\HimColour}{red}

\def\ltsima{$\; \buildrel < \over \sim \;$}
\def\simlt{\lower.5ex\hbox{\ltsima}}
\def\gtsima{$\; \buildrel > \over \sim \;$}
\def\simgt{\lower.5ex\hbox{\gtsima}}

\def\r200{r_{200}}
\def\m200{m_{200}}
\def\V200{V_{200}}
\def\M200{M_{200}}
\def\R200{R_{200}}

\title[Variable IMFs with EAGLE -- I. Simulations]{Calibrated, cosmological hydrodynamical simulations with variable IMFs I: Method and effect on global galaxy scaling relations}

\author[C. Barber et al.]{
Christopher Barber,$^{1}$\thanks{Email: \href{mailto:cbar@strw.leidenuniv.nl}{cbar@strw.leidenuniv.nl}}
Robert A. Crain,$^{2}$
and Joop Schaye$^{1}$
\\
$^{1}$Leiden Observatory, Leiden University, PO Box 9513, NL-2300 RA Leiden, The Netherlands\\
$^{2}$Astrophysics Research Institute, Liverpool John Moores University, 146 Brownlow Hill, Liverpool L3 5RF, UK
}

\date{Accepted XXX. Received YYY; in original form ZZZ}

\pubyear{2016}

\graphicspath{{Figures/}}

\begin{document}
\label{firstpage}
\pagerange{\pageref{firstpage}--\pageref{lastpage}}
\maketitle

\begin{abstract}

The recently inferred variations in the stellar initial mass function (IMF) among local high-mass early-type galaxies may require a reinterpretation of observations of galaxy populations and may have important consequences for the predictions of models of galaxy formation and evolution.  We present a new pair of cosmological, hydrodynamical simulations based on the EAGLE model that self-consistently adopt an IMF that respectively becomes bottom- or top-heavy in high-pressure environments for individual star-forming gas particles. In such models, the excess stellar mass-to-light ($M/L$) ratio with respect to a reference IMF is increased due to an overabundance of low-mass dwarf stars or stellar remnants, respectively. Crucially, both pressure-dependent IMFs have been calibrated to reproduce the observed trends of increasing excess $M/L$ with central stellar velocity dispersion ($\sigma_e$) in early-type galaxies, while maintaining agreement with the observables used to calibrate the EAGLE model, namely the galaxy luminosity function, half-light radii of late-type galaxies, and black hole masses. We find that while the $M/L$ excess is a good measure of the IMF for low-mass slope variations, it depends strongly on the age of the stellar population for high-mass slope variations. The normalization of the [Mg/Fe]$-\sigma_e$ relation is decreased (increased) for bottom- (top-)heavy IMF variations, while the slope is not strongly affected. Bottom-heavy variations have little impact on galaxy metallicities, half-light radii of early-type galaxies, or star formation rates, while top-heavy variations significantly increase these quantities for high-mass galaxies, leading to tension with observations. 

\end{abstract}


\section{Introduction}

The stellar initial mass function (IMF) is a crucial ingredient in the interpretation of galaxy observations as well as for predictions of models of galaxy formation. It defines the translation between physical quantities and observables, and is one of the largest sources of uncertainty in model predictions. In the Milky Way (MW), the IMF seems to be insensitive to environment, with a steep high-mass slope that flattens below $\sim 1 \Msun$ \citep{Kroupa2001, Chabrier2003a, Bastian2010}. Observational and theoretical studies alike nearly always adopt such a universal IMF in stellar evolution models, applying it to all galaxies, regardless of the conditions under which their stars were formed. 

In the past decade, evidence for variations in the IMF has been steadily mounting, leading to a near-consensus that the IMF becomes ``heavier'' in the regions of high global stellar velocity dispersion, $\sigma$, found in the centres of high-mass early-type galaxies (ETGs). In unresolved systems, the IMF is often parametrized by the excess stellar mass-to-light ratio ($M/L$) of the stars relative to the $M/L$ one would derive spectroscopically assuming a standard IMF. The $M/L$-excess (hereafter MLE; also known as the ``IMF mismatch parameter'')\footnote{We introduce the notation ``MLE'' rather than the more-popular ``$\alpha$'' for the IMF mismatch parameter to avoid confusion in discussions involving abundances of $\alpha$-elements.} is constrained observationally via several independent methods, including gravitational lensing \citep[e.g.][]{Auger2010, Treu2010, Spiniello2011, Barnabe2013, Sonnenfeld2015, Posacki2015, Smith2015b, Collier2018}, stellar population synthesis (SPS) modelling of IMF-sensitive spectral absorption features \citep[e.g.][]{Cenarro2003, vanDokkum2010, Conroy2012b, Spiniello2012, Ferreras2013, LaBarbera2013, LaBarbera2015, Spiniello2014, Rosani2018}, or dynamical modelling of the stellar kinematics \citep[e.g.][]{Thomas2011, Dutton2012, Tortora2013, Cappellari2013b, Li2017}, with many of these studies employing a combination thereof. All three methods point to a strong trend of increasing MLE, and thus a ``heavier'' IMF, with $\sigma$. Some studies find additional (and sometimes stronger) trends between the IMF and metallicity and/or alpha enhancement, but there is still much debate on this issue \citep{Conroy2012b, LaBarbera2013, McDermid2014, LaBarbera2015, Martin-Navarro2015c}.

Puzzlingly, constraints on the IMF seem to be inconsistent on a case-by-case basis. \citet{Smith2014} has shown that for a sample of 34 ETGs, while both SPS and dynamical modelling imply heavier IMFs in high-mass ETGs, there seems to be no correlation between the MLE values derived using the two methods. \citet{Newman2017} compared the MLE derived using lensing, stellar dynamics, and SPS modelling for 3 SNELLS lenses \citep{Smith2015b}, also finding inconsistent results between the methods. Conversely, \citet{Lyubenova2016} finds consistent MLE values for SPS and dynamical modelling for a sample of 27 ETGs, arguing that inconsistencies found in other studies may be due to differences in aperture sizes, SPS models employed, or non-optimal dark matter halo corrections. These findings imply that the systematic errors involved in some of these analyses may not be well understood. Indeed, \citet{Clauwens2015} have shown that IMF trends inferred from stellar kinematics arise also in models assuming a universal IMF if the measurements and/or modelling errors have been underestimated. Furthermore, they found that the data shows an IMF dependence on distance from the Galaxy, suggesting the presence of systematic errors. These results imply that further study is required.

Although the majority of the evidence points toward a non-universal IMF, it is not clear {\it how} it varies. Dynamical modelling and gravitational lensing constrain only the dynamical $M/L$, and indicate that it is higher than expected assuming a stellar population with a fixed IMF. This generally implies that either the IMF is more bottom-heavy, leading to more low-mass dwarf stars that contribute significantly to the mass but not the total luminosity, or that the IMF is top-heavy, implying the extra mass comes from stellar remnants: black holes (BHs), neutron stars, and white dwarfs. Some spectroscopic IMF studies are thought to be able to constrain the shape of the low-mass end of the IMF, as a number of absorption features are sensitive to the surface gravity of stars and thus measure the ratio of dwarf-to-giant stars. The majority of these studies find that this ratio is higher in high-$\sigma$ galaxies, but the means by which this is achieved is similarly unclear, since the increased ratio of dwarf-to-giant stars can be achieved either through a steepening of the IMF low-mass slope \citep[e.g.][]{Conroy2012b, Conroy2017}, or steepening of the high-mass slope \citep[e.g.][]{LaBarbera2013}. On the other hand, H$\alpha$ and $g-r$ colours of local star-forming galaxies from the GAMA survey imply that the high-mass end of the IMF becomes shallower in strongly star-bursting environments \citep{Gunawardhana2011}. The large variety of parametrizations of IMF variations makes comparison between different methods difficult, and has dramatic consequences for the uncertainty in the physical properties of galaxies inferred from observational surveys \citep{Clauwens2016}.

The consequences of a variable IMF on the predictions from galaxy formation models are unclear. While the IMF determines the present-day stellar $M/L$ ratios of galaxies, it also governs the strength of stellar feedback and metal yields. For example, a more top-heavy (bottom-heavy) IMF produces more (fewer) high-mass stars that end their lives as supernovae and return mass and energy to the interstellar medium (ISM), affecting the production and distribution of metals throughout the ISM. Metallicity affects the rate at which gas cools and forms future generations of stars, while stellar feedback governs the balance between the flow of gas into, and out of, galaxies, thus regulating star formation. The situation becomes even more complex with the inclusion of supermassive BHs, whose growth depends on the ability of supernova feedback to remove gas from the central regions of galaxies where such BHs reside \citep{Bower2017}. BH gas accretion generates AGN feedback, which is important for quenching star formation in high-mass galaxies. These processes are non-linear and deeply intertwined, rendering the question of how variations in the IMF would impact galaxies in such models non-trivial.

To address this question, recent studies have begun investigating the effect of IMF variations by post-processing cosmological simulations and semi-analytic models, and by conducting self-consistent, small-scale, numerical simulations. In a post-processing analysis of the Illustris simulations, \citet{Blancato2017} study how variations in the IMF of individual star particles manifests as global IMF trends between galaxies, finding that the IMF of individual particles must vary much more strongly than the global trends imply in order to obtain the observed MLE-$\sigma$ trends. \citet{Sonnenfeld2017} use an evolutionary model based on dark matter-only numerical simulations to predict the evolution of the IMF in early-type galaxies due to dry mergers from $z=2$ to 0, finding that dry mergers tend to decrease the MLE of individual galaxies over time, while the correlation between the IMF and $\sigma$ should remain time-invariant. Much can be learned from post-processing of such large-scale simulations, but such studies by construction neglect the effect that a variable IMF may have on galaxy properties during their formation and evolution due to the change in stellar feedback and metal yields.

IMF variations have also been investigated in semi-analytic models (SAMs) of galaxy formation. \citet{Fontanot2014} find that the variations at the high-mass end of the IMF have a much stronger effect on galaxy properties than variations at the low-mass end. By implementing the ``integrated galactic IMF theory'' \citep{Kroupa2003}, which predicts that the IMF should become top-heavy in galaxies with high SFRs, into SAMs, \citet{Fontanot2017} and \citet{Gargiulo2015b} both find that models with a variable IMF are better able to reproduce observed abundance scaling relations than those with a universal IMF. While such SAMs are useful as a computationally inexpensive method of exploring many types of IMF variations, they lack the ability to resolve the internal properties of galaxies, which may be important for IMF studies in light of recent evidence for significant radial gradients in the IMF in individual high-mass ETGs (\citealt{Martin-Navarro2015b, LaBarbera2016, vanDokkum2017, Oldham2018, Sarzi2018}; but see \citealt{Davis2017, Alton2017}). 

Hydrodynamical simulations with the ability to resolve the internal structure of galaxies have been run with self-consistent IMF variations for a limited number of idealized galaxies. \citet{Bekki2013b} implement a density and metallicity-dependent IMF prescription to idealized chemodynamical simulations of dwarf-to-MW mass galaxies. They find that a top-heavy IMF in the high-density environments of actively star-forming galaxies suppresses the formation of dense clumps and thus suppresses star formation, as well as increasing the overall metallicities and $\alpha$-enhancement of such galaxies. \citet{Gutcke2017} apply the local metallicity-dependent IMF of \citet{Martin-Navarro2015c} to numerical simulations of 6 MW-analogues using AREPO, finding a strong effect on the metallicity evolution in such systems. \citet{Guszejnov2017} apply various prescriptions of IMF variations from giant molecular cloud (GMC) theory in a simulation of an individual MW analogue galaxy, finding that most prescriptions produce variations within the MW that are much stronger than observed. Such simulations are an excellent starting point to study the effect of IMF variations on galaxy formation and evolution, but are currently limited in statistics, especially for high-mass ETGs where the IMF is observed to vary the strongest.

In this paper, we present a pair of fully cosmological, hydrodynamical simulations, based on the EAGLE project \citep[][hereafter referred to as S15 and C15, respectively]{Schaye2015, Crain2015}, each of which includes a prescription for varying the IMF on a per-particle basis to become either bottom-heavy or top-heavy in high-pressure environments, while self-consistently modelling its consequences for feedback and heavy element synthesis. While a pressure-dependent IMF has been studied before using self-consistent, cosmological, hydrodynamical simulations as part of the OWLS project \citep{Schaye2010, Haas2013}, the adopted IMF was in that case not motivated by the recent observations discussed above, and the OWLS models did not agree well with basic observables such as the galaxy luminosity function. In contrast, our prescription has been calibrated to broadly reproduce the observed relationship between the MLE and the central stellar velocity dispersion, and we verify that the simulations maintain good agreement with the observed luminosity function. It is the goal of this paper to investigate the effect that a variable IMF has on the properties of the galaxy population in the EAGLE model of galaxy formation, such as the galaxy stellar mass function, luminosity function, star formation rates, metallicities, alpha-enhancement, and sizes. In doing so, we may inform how the IMF should correlate with many galaxy observables, both across the galaxy population as well as within individual galaxies.

This paper is organized as follows. In \Sec{methods} we describe the EAGLE simulations and the calibration of IMF variation prescriptions to match the local empirical MLE-$\sigma$ correlations, and discuss how these prescriptions are self-consistently incorporated into the EAGLE model. \Sec{IMF_vs_sigma} introduces the variable IMF simulations and details the resulting correlations between the galaxy-averaged IMF and central stellar velocity dispersion. In \Sec{calibration_diagnostics} we show that IMF variations have little effect on galaxy observables used to calibrate the reference EAGLE model, while \Sec{IMF_effect} investigates the impact on predicted galaxy properties such as metallicity, alpha-enhancement, SFR, and sizes. Our conclusions are summarized in \Sec{conclusions}. \App{Appendix_calibration} gives extra details regarding aperture effects and the IMF calibration, while \App{Appendix_self_consistency} shows the effect of incrementally making individual physical processes in the simulations consistent with a variable IMF. 

Paper II in this series will discuss trends between the MLE and global galaxy observables and determine which correlate most strongly with the MLE. In Paper III we will discuss the spatially-resolved IMF trends within individual high-mass galaxies and the redshift-dependence of the MLE-$\sigma$ relation. The simulation data is publicly available at \url{http://icc.dur.ac.uk/Eagle/database.php}.

\section{Methods}
\label{sec:methods}
In this section we describe the EAGLE model (\Sec{eagle}) and our procedure of calibrating IMF variations in post-processing to match observed trends with galaxy velocity dispersion (\Sec{calibration}), followed by a description of the modifications to the EAGLE model necessary to produce simulations that are self-consistent when including a variable IMF (\Sec{mods}).

\subsection{The EAGLE simulations}
\label{sec:eagle}

In this study we use the EAGLE model (S15, C15) to study the effect of a variable IMF on predictions of galaxy properties. Here we briefly summarize the simulation model, but refer the reader to S15 for a full description.

EAGLE, short for ``Evolution and Assembly of GaLaxies and their Environments'', is a suite of hydrodynamical, cosmological simulations aimed at studying the formation and evolution of galaxies from the early Universe to $z=0$. It was run with a modified version of the Tree-PM Smoothed Particle Hydrodynamics (SPH) code Gadget-3, last described by \citet{Springel2005}. The modifications to the SPH implementation, collectively known as Anarchy \citep[][appendix A of S15]{Schaller2015b}, improve the treatment of artificial viscosity, time-stepping, and alleviate issues stemming from unphysical surface tension at contact discontinuities. Cubic volumes of up to (100 comoving Mpc)$^3$  were simulated at various resolutions -- in this paper we focus only on the ``intermediate'' resolution simulations, with $m_{\rm gas} = 1.6 \times 10^6 \Msun$ and $m_{\rm DM} = 9.7 \times 10^6 \Msun$ for gas and dark matter particles, respectively. The gravitational softening is kept fixed at 2.66 co-moving kpc for $z > 2.8$ and at 0.70 proper kpc at lower redshifts. A Lambda cold dark matter cosmogony is assumed, with cosmological parameters chosen for consistency with Planck 2013: \citep[$\Omega_{\rm b} = 0.04825$, $\Omega_{\rm m} = 0.307$, $\Omega_{\Lambda}=0.693$, $h = 0.6777$;][]{Planck2014}.\\

Physical processes acting on scales below the resolution limit of the simulation (termed ``subgrid physics'') are modelled using analytic prescriptions whose inputs are quantities resolved by the simulation. The efficiency of feedback associated with the formation of stars and the growth of BHs was calibrated to match the observed $z=0.1$ galaxy stellar mass function (GSMF), galaxy sizes, and the $\MBH$-$\Mstar$ relation.

Radiative cooling and photo-heating of gas are implemented element-by-element for the 11 elements most important for these processes, computing their heating and cooling rates via Cloudy assuming a \citet{Haardt2001} UV and X-ray background \citep{Wiersma2009a}.

Star formation is implemented by converting gas particles into star particles stochastically with a probability proportional to their pressure, such that the simulations reproduce by construction the empirical Kennicutt-Schmidt relation \citep{Schaye2008}, renormalized for a \citet[][hereafter ``Chabrier'']{Chabrier2003a} IMF. For a self-gravitating gaseous disk this star formation law is equivalent to the observed \citet{KennicuttJr.1998} surface density law. The density threshold for star formation increases with decreasing metallicity according to the model of \citet{Schaye2004} to account for the metallicity dependence of the transition from the warm (i.e. $T \sim 10^4$ K) atomic to the cold ($T \ll 10^4$ K), molecular interstellar gas phase. Once stars are formed, their subsequent mass loss is computed assuming a Chabrier IMF and the metallicity-dependent stellar lifetimes of \citet{Portinari1998}. Heavy element synthesis and mass loss in winds from asymptotic giant branch stars, high-mass stars, and ejecta from core-collapse and type Ia supernovae are accounted for \citep{Wiersma2009b}. Stellar feedback is implemented by stochastically injecting a fixed amount of thermal energy into some number of the surrounding gas particles \citep{DallaVecchia2012}, where the probability of heating depends on the local density and metallicity (S15).

Supermassive black holes are seeded in haloes that reach a Friends of Friends (FoF) mass of $10^{10} \Msun/h$ by injecting a subgrid seed BH of mass $10^5\Msun/h$ into the most bound gas particle \citep{Springel2005a}. BHs grow at the minimum of the Eddington rate and the \citet{Bondi1944} rate for spherically symmetric accretion, taking into account angular momentum of in-falling gas \citep{Rosas-Guevara2015}. BHs provide AGN feedback by building up an energy reservoir until they can heat at least one of their nearest neighbours by a minimum temperature, at which point they may stochastically heat their SPH neighbours \citep{Booth2009}. This procedure prevents gas from cooling too quickly after being heated, preventing over-cooling.

We classify DM haloes using a FoF algorithm with a linking length of 0.2 times the mean inter-particle spacing \citep{Davis1985}. Baryons are assigned to the halo (if any) of their nearest DM particle. Self-bound substructures within haloes, termed ``subhaloes'', are then identified using the SUBFIND algorithm \citep{Springel2001,Dolag2009}.  The ``central'' subhalo within a halo is defined as the one containing the gas particle most tightly bound to the group, while all others are classified as ``satellites''. We only consider subhaloes containing at least 100 star particles as resolved ``galaxies''. For consistency with S15, we define stellar mass, $\Mstar$, as the mass of stars within a spherical aperture of radius 30 proper kpc around each galaxy. To compare with observations, we measured all other quantities, such as stellar velocity dispersion ($\sigma_e$), $M/L$, metallicity ($Z$), and alpha enhancement ([Mg/Fe]), within a 2D circular aperture with the SDSS $r$-band projected half-light radius, $r_e$, of each galaxy, observed along the $z$-axis of the simulation.

In the next section we discuss how we can use the Reference EAGLE simulations to calibrate a prescription that varies the IMF to match the observed trend between MLE and $\sigma$.

\subsection{IMF calibration}
\label{sec:calibration}

\begin{figure}
  \centering
\includegraphics[width=0.45\textwidth]{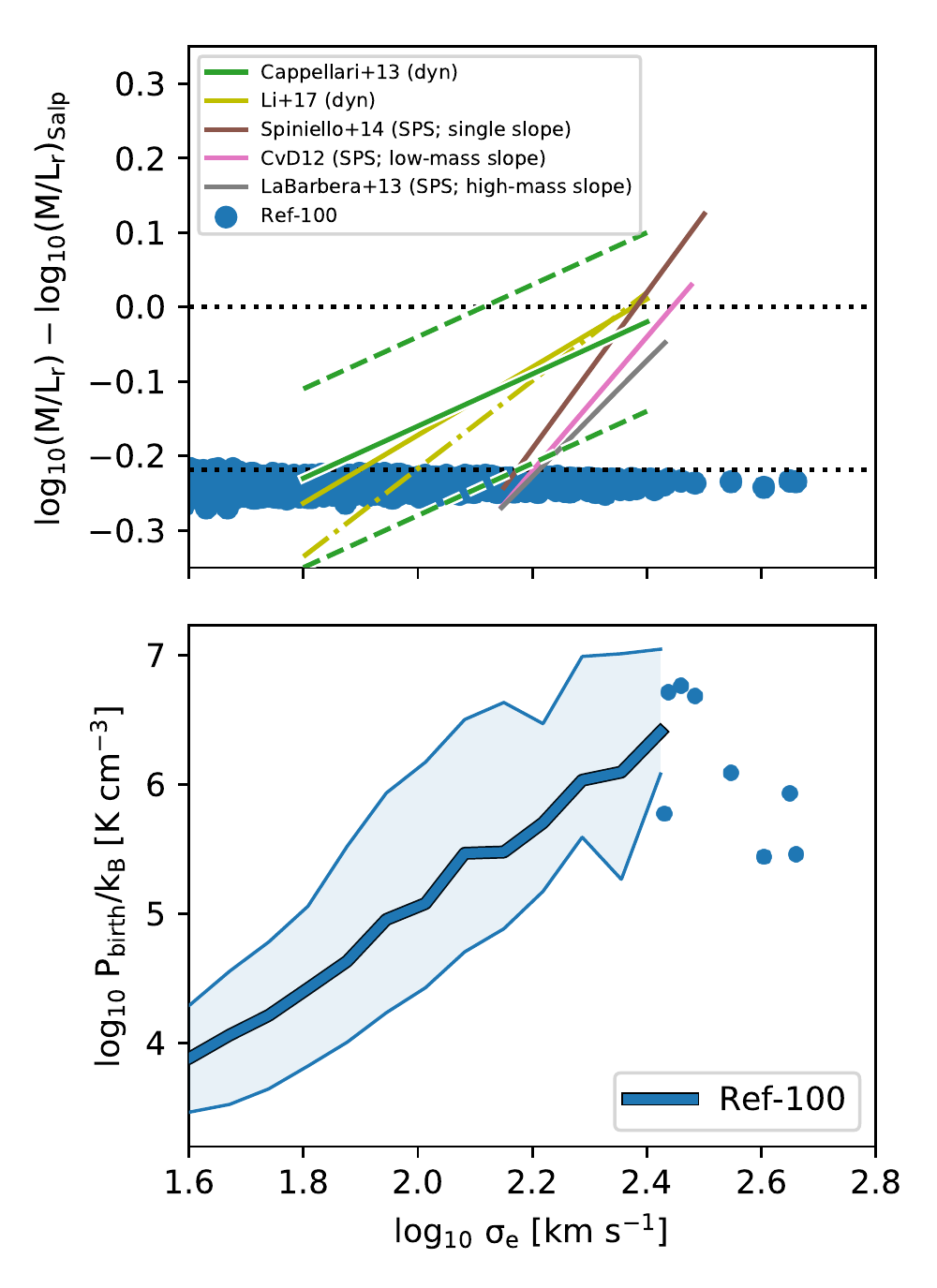}
\caption{IMF-dependent properties of galaxies in the (100 Mpc)$^3$ EAGLE reference simulation (Ref-100) at $z=0.1$. All quantities are measured within the half-light radius, $r_e$. Top panel: Stellar mass-to-light ratio excess (MLE) as a function of stellar velocity dispersion, $\sigma_e$. Horizontal dotted lines at MLE $=0$ and $-0.22$ show the expected MLE for a Salpeter and Chabrier IMF, respectively. The observed trend from \citet{Cappellari2013b} is shown as a \CapColour{} solid line, with the intrinsic scatter shown as dashed lines. Also shown are the observed trends from \citet{Li2017} (yellow solid and dash-dotted for two different SPS models, respectively), \citet{Spiniello2014} (brown solid), \citet{Conroy2012b} (pink solid) and \citet{LaBarbera2013} (grey solid). In brackets we indicate the method of IMF determination, either dynamical or spectroscopic, where for the latter case we also indicate the region of the IMF that is varied in the study. The reference model clearly does not reproduce the observed variation.  
Bottom panel: Stellar birth ISM pressure as a function of $\sigma_e$. The thick and thin lines show the median and 10-90th percentiles in $\sigma_e$ bins. Where a bin has fewer than 10 galaxies, individual galaxies are shown. It is due to this correlation that we are able to vary the IMF for each individual star-forming gas particle as a function of its pressure in order to achieve a trend in integrated galactic IMF with $\sigma_e$, as observed.}  
\label{fig:problem}
\end{figure}

The first goal of this paper is to implement a variable IMF into the EAGLE simulations that yields the observed trends of IMF with galaxy properties. While it is debated how the IMF varies as a function of metallicity or alpha-abundances, there is mounting evidence that the MLE in the centres of massive elliptical galaxies increases with stellar velocity dispersion \citep[e.g.][]{Treu2010, LaBarbera2013, Cappellari2013b, Spiniello2014, Li2017}. This increase could be either due to a higher number of low-luminosity dwarf stars (``bottom-heavy'' IMF) or stellar remnants (``top-heavy'' IMF). 

We follow \citet[][ hereafter C13]{Cappellari2013b} and define the MLE relative to the $(M/L)$ one would obtain assuming a Salpeter IMF:
\be
{\rm MLE}_i = \log_{10}(M/L_i) - \log_{10}(M/L_i)_{\rm Salp},
\label{eqn:MLE}
\ee
where $i$ denotes the observational filter in which the luminosity is measured. In the upper panel of \Fig{problem}, in \CapColour{} we plot the observed relation between SDSS r-band MLE$_r$ and stellar velocity dispersion, $\sigma_e$, both measured within $r_e$, obtained by C13 for high-mass elliptical galaxies in the ATLAS$^{\rm 3D}$ survey. Also included are observed trends from \citet{Conroy2012b}, \citet{LaBarbera2013}, \citet{Spiniello2014}, and \citet{Li2017}. Note that for \citet{Li2017} we show the fits for elliptical and lenticular galaxies using two different SPS models to derive $(M/L)_{\rm Salp}$. For comparison, we also show the same relation for galaxies in the (100 Mpc)$^3$ reference EAGLE simulation (hereafter Ref-100). As expected, the EAGLE galaxies lie along a line of constant MLE$_r \approx -0.22$, corresponding to the asymptotic value reached by a stellar population with constant star formation rate and a Chabrier IMF, and are clearly inconsistent with the observational trends. The goal of this paper is to implement a prescription for an IMF variation that reproduces the observed C13 relation, and to investigate the effect it has on galaxy properties and observables. 

In principle, in order to achieve this correlation, one could simply vary the IMF with the velocity dispersion of the galaxy in which it is born. However, this prescription lacks a physical basis, as there should not be any reason why a star born in a low-mass halo at high redshift should have direct ``knowledge'' of the stellar velocity dispersion of its host galaxy. Indeed, it would have to know the future velocity dispersion of its host galaxy at $z\approx 0$ at the time it was born, which is infeasible to simulate. A more physical approach is to vary the IMF with respect to some gas property local to a star-forming gas particle at the time it is formed. This affords us the ability to seek connections between physical conditions and $z=0$ observables, and to perform controlled experiments whereby the various consequences of a variable IMF are selectively enabled/disabled. Moreover, it is a philosophical choice of the EAGLE project to only allow subgrid routines to be `driven' by physically meaningful properties, such as gas density, metallicity, or temperature.

Many physical models of the formation of the IMF on the scales of giant molecular clouds (GMCs) predict the IMF to depend on the temperature, density, and/or pressure of the GMC from which the stars form \citep[e.g.][]{Bate2005, Jappsen2005, Bate2009, Krumholz2011, Hopkins2012, Hennebelle2013}.  One could in principle simply apply these models to star-forming gas particles in the EAGLE simulation using their individual densities and temperatures \citep[as is done in ][]{Guszejnov2017}, but it is not clear that such an approach is appropriate here given the much coarser resolution of EAGLE compared to current GMC-scale IMF simulations. Indeed, EAGLE does not resolve the cold phase of the ISM. An alternate approach is to vary the IMF with some parameter of the star-forming gas that is found to vary with stellar velocity dispersion, and attempt to calibrate this local dependence to obtain the observed global IMF-velocity dispersion relationship. One enticing possibility is to vary the IMF with the pressure at which gas particles are converted to star particles in the simulation \citep{Schaye2010, Haas2013}. Although the cold interstellar gas phase, which EAGLE does not attempt to model, will have very different densities and temperatures than the gas in EAGLE, pressure equilibrium implies that its pressure may be much more similar. However, note that the pressure in the simulation is smoothed on scales of $\sim 10^2 - 10^3$ pc, corresponding to $L_{\rm Jeans}$ of the warm ISM. Note as well that since the local star formation rate (SFR) in EAGLE galaxies depends only on pressure, varying the IMF with pressure is equivalent to varying it with local SFR density. 

In the lower panel of \Fig{problem} we plot the mean $r$-band light-weighted ISM pressure at which stellar particles within (2D projected) $r_e$ were formed, as a function of $\sigma_e$ for galaxies in Ref-100 at $z=0.1$. We see a strong correlation, where stars in galaxies with larger $\sigma_e$ formed at higher pressures. Thus, by invoking an IMF that varies with birth ISM pressure, we can potentially match the observed MLE$_r$--$\sigma_e$ correlation.\\

\begin{figure*}
\includegraphics[width=0.45\textwidth]{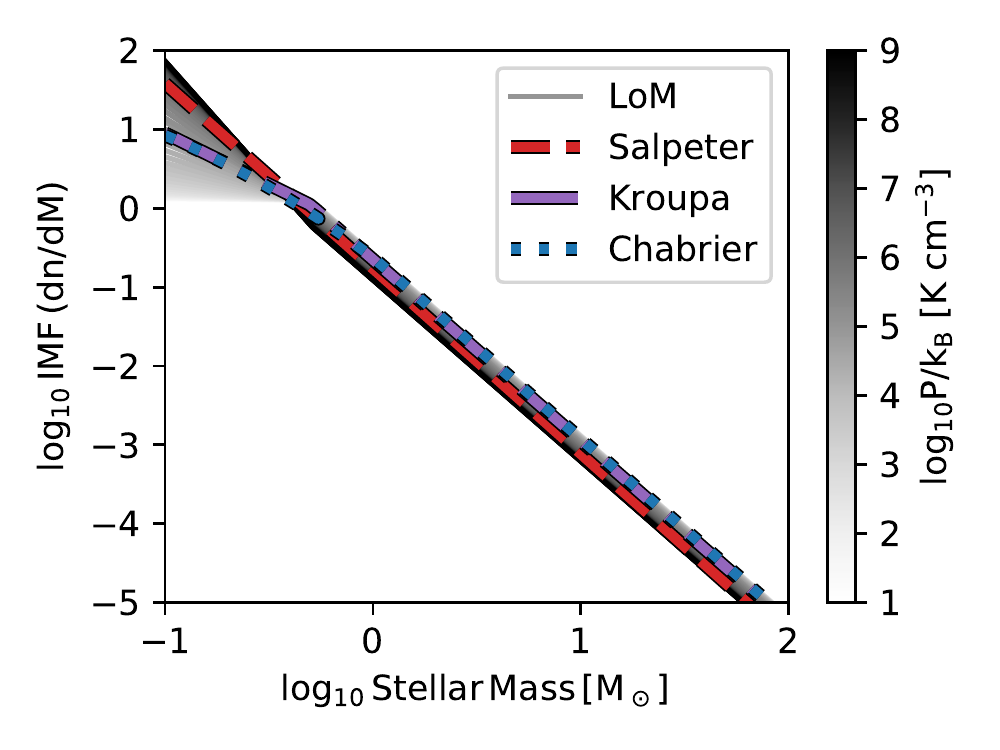}
\includegraphics[width=0.45\textwidth]{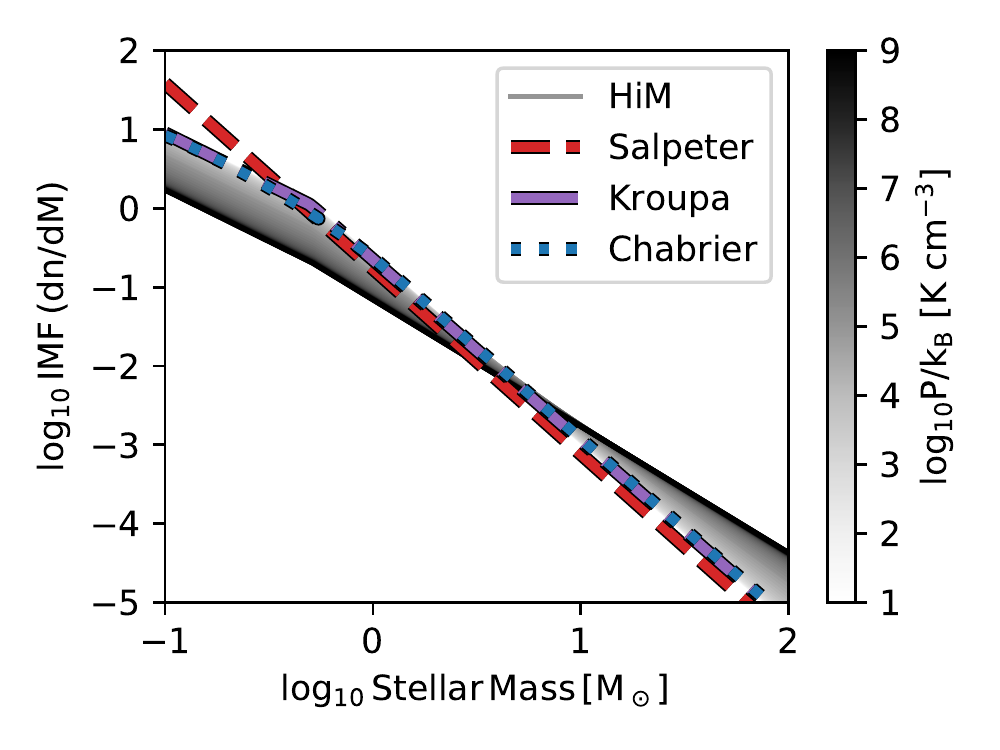}
\includegraphics[width=0.45\textwidth]{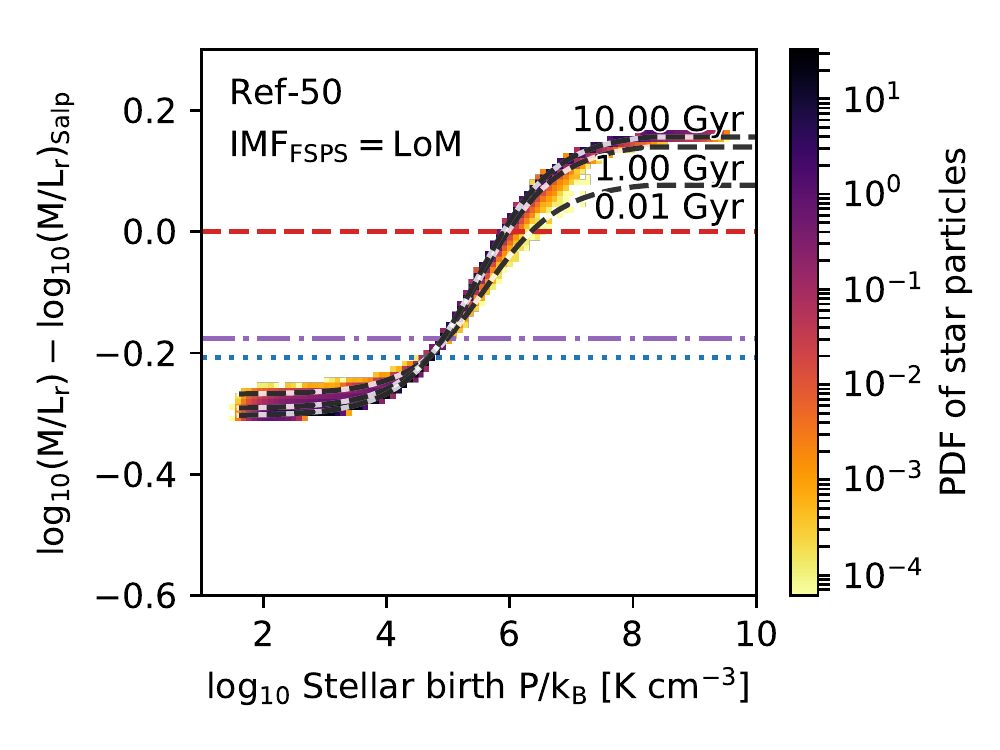}
\includegraphics[width=0.45\textwidth]{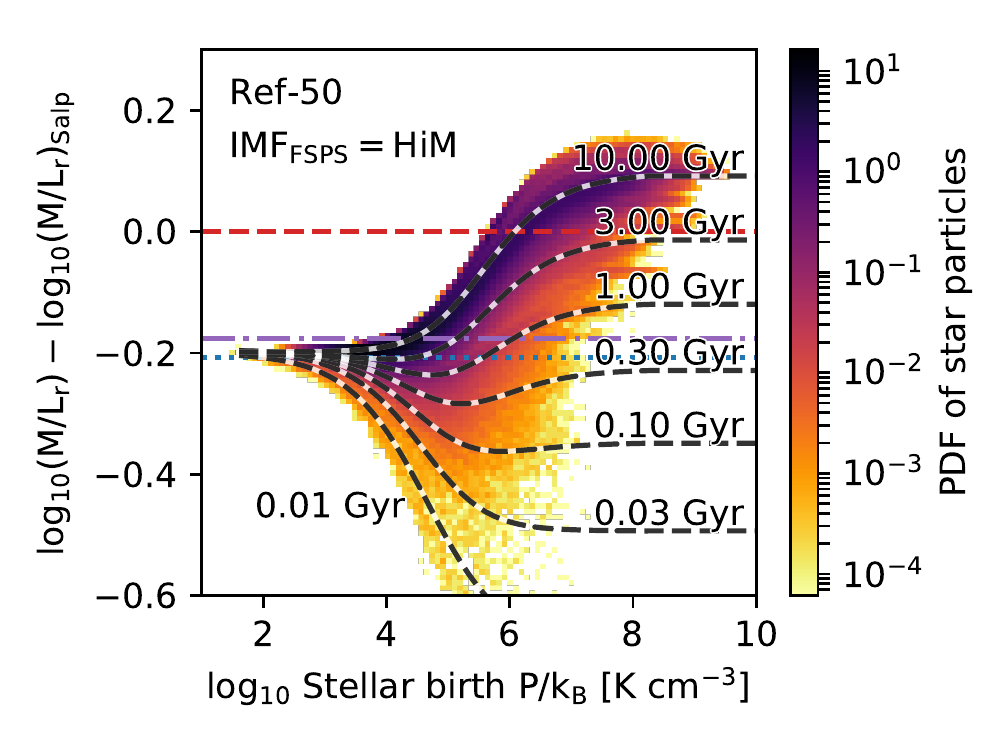}
\caption{Top row: The two variable IMF prescriptions used in this study shown for a range in stellar birth ISM pressures (see greyscale bar). Top Left: Variable IMF in which the slope below 0.5 $\Msun$ is varied (hereafter called LoM) such that the IMF transitions from bottom-light at low $P$ to bottom-heavy at high $P$. Top Right: As in top-left but instead varying the IMF slope {\it above} 0.5 $\Msun$ (hereafter HiM) such that it becomes top-heavy at high $P$. For all IMFs the integrated mass is normalized to $1 \Msun$, causing the low-mass end of the HiM IMF  to greatly decrease in normalization at high pressures.
Bottom panels: 2D probability distribution functions of the $r$-band mass-to-light ratio excess (MLE$_r$) of individual star particles as a function of the pressure of the ISM out of which the star particles formed, for the Ref-50 simulation post-processed assuming LoM and HiM in the bottom-left and -right panels, respectively. Black dashed lines show the MLE-$P$ relation for SSPs at the indicated fixed ages. For reference, in all panels Salpeter, Kroupa, and Chabrier IMFs are shown in red dashed, purple dash-dotted, and blue dotted lines, respectively. Note the small scatter at fixed birth $P$ for LoM, despite the wide range in the ages and metallicities of the stars. This shows that the MLE is a good proxy for the IMF when the high-mass slope is close to Salpeter. However, HiM yields a larger scatter in the MLE because in this case the MLE increases strongly with age at fixed $P$.}
\label{fig:IMF}
\end{figure*}

\begin{figure*}
  \centering
\includegraphics[width=0.32\textwidth]{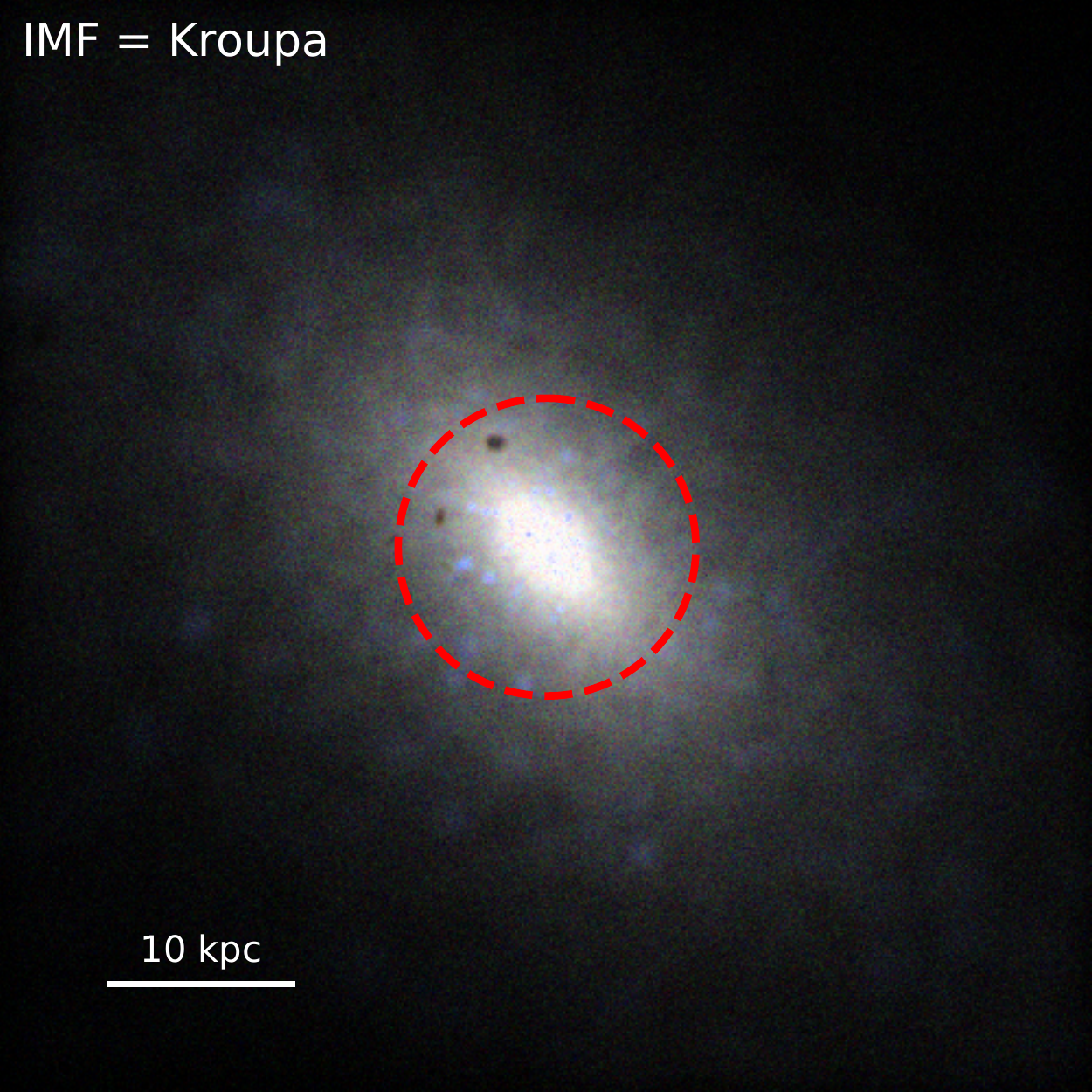}
\includegraphics[width=0.32\textwidth]{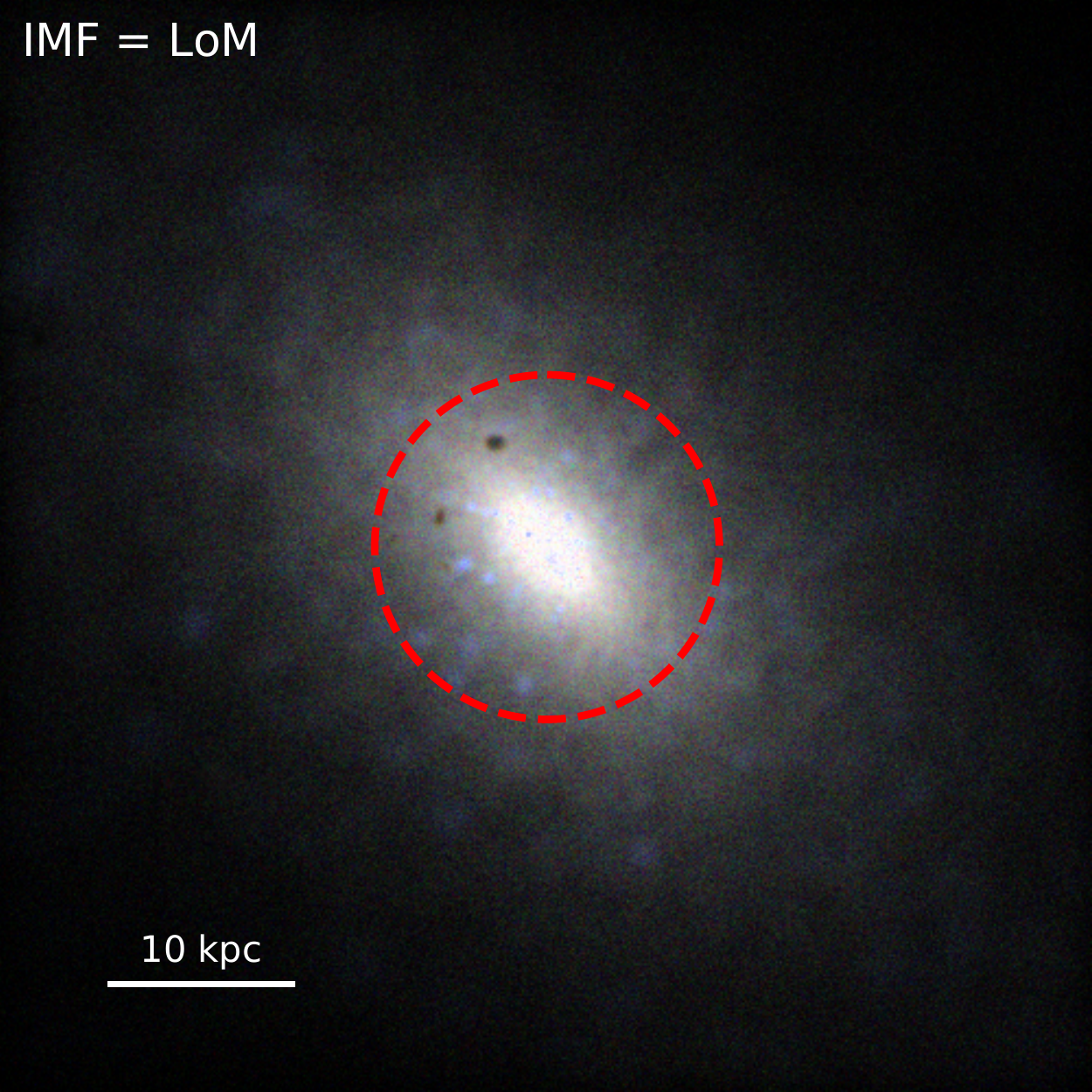}
\includegraphics[width=0.32\textwidth]{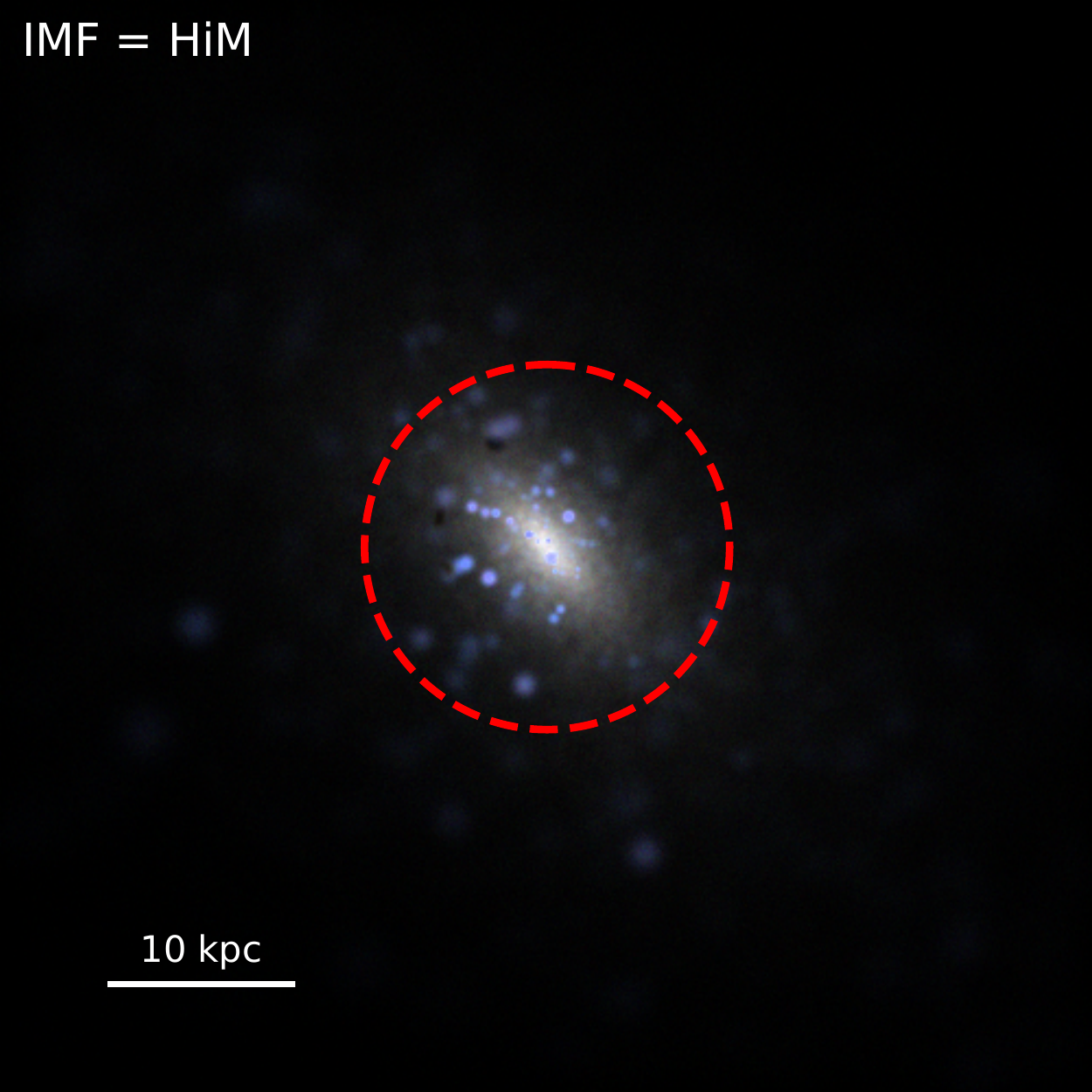}
\caption{Images of a massive elliptical galaxy in the Ref-50 simulation, post-processed using SKIRT assuming 3 different variable IMF prescriptions. The images are 60 proper kpc per side and 60 proper kpc deep, centred on the galaxy. From left to right: Kroupa (universal), LoM, and HiM IMF prescriptions are implemented. The 2D projected $r$-band half-light radius is indicated in each panel as a dashed red circle. RGB colour channels correspond to SDSS $g$, $r$, and $i$ peak wavelengths, respectively, normalized using the \citet{Lupton2004} scaling procedure. Assuming the LoM prescription, we produce a nearly identical image to that assuming a Chabrier IMF, while assuming the HiM prescription significantly reduces the luminosity of the diffuse stellar light due to a reduced fraction of low- and intermediate-mass stars, while increasing the fraction of very young stars.  }  
\label{fig:pictures}
\end{figure*}

To calibrate the IMF pressure-dependence to match the C13 trend, we post-processed the Ref-100 simulation using the Flexible Stellar Population Synthesis (FSPS) software package \citep{Conroy2009, Conroy2010}. With FSPS, it is possible to generate tables of masses and luminosities in many common observational filters for simple stellar populations (SSPs) as a function of their age, metallicity, and IMF. Here we used the Basel spectral library \citep{Lejeune1997, Lejeune1998, Westera2002} with Padova isochrones \citep{Marigo2007, Marigo2008}, but note that using the different available libraries would not affect our conclusions. Using FSPS in post-processing on Ref-100, star particles were reassigned masses and luminosities via interpolation of these tables, given their age, metallicity, initial mass, and birth ISM pressure. As a check, we verified that, for a Chabrier IMF, the SSP masses derived in post-processing using FSPS match the output masses of EAGLE stellar particles computed using the \citet{Wiersma2009b} models built into the simulation to within 2 per cent. However, the agreement between the models is not as good for IMFs with shallow high-mass slopes. Differences in how BH remnants from high-mass stars are treated between the two models result in small differences in mass for a Chabrier-like IMF, but when the high-mass IMF slope is shallow, BH masses begin to become important and these differences are amplified, resulting in $\approx 0.1$ dex lower $\Mstar$ from the \citet{Wiersma2009b} models than with FSPS for high-mass ($\Mstar>10^{11}\Msun$) galaxies with shallow high-mass IMF slopes (applicable to the HiM prescription, below). For consistency, we use stellar masses computed via FSPS for stellar $M/L$ ratios as well as $\Mstar$ throughout this paper. Note as well that we do not perform radiative transfer to estimate dust extinction. We do not expect dust to be very important here since we investigate mostly old, gas-poor galaxies and measure luminosities in the $K$ or $r$-band, which are not as strongly affected by dust extinction as bluer wavelengths. However, we do neglect the luminosities of stellar particles with ages younger than 10 Myr, as such stars should still be embedded in their birth clouds, and thus are not expected to be observable \citep{Charlot2000}. 

We define the IMF piecewise as $dn/dM \propto M^x$, such that a \citet{Salpeter1955} IMF has $x = -2.35$ for all $M$, and a \citet[][hereafter ``Kroupa'']{Kroupa2001} IMF has a slope of $x=-1.3$ and $-2.3$ for stellar masses below and above 0.5 $\Msun$, respectively. Consistent with the EAGLE reference model, we integrate the IMF from 0.1 to 100 $\Msun$.

We began the calibration with a Kroupa IMF which is practically indistinguishable from the Chabrier IMF over this mass range, but is easier to work with due to its simpler double power-law shape. We tried different methods of varying the IMF, including varying the low-mass slope, high-mass slope, and the stellar mass at which the IMF transitions between these slopes.  Varying only the transition mass to make the IMF more bottom-heavy in high-pressure environments (without changing the low-mass cut-off of $0.1 \Msun$) did not yield a strong enough variation in the IMF to reproduce the observed trends. We briefly experimented with instead increasing the transition mass to make the IMF top-heavy in high-pressure environments  \citep[e.g.][]{Fontanot2018}, and found similar results to our ``HiM'' prescription, outlined below. 

We chose to vary the IMF with pressure according to two different prescriptions: one in which the low-mass slope is varied while the high-mass slope is kept fixed (hereafter referred to as LoM) and another where the high-mass slope is varied, keeping the low-mass slope fixed (HiM). These IMF prescriptions are depicted in the top row of \Fig{IMF}. In both prescriptions, we vary the IMF slope between two fixed values $x_{\rm lowP}$ and $x_{\rm highP}$ that are asymptotically reached at low and high pressure, respectively, transitioning between them smoothly via a sigmoid function,
\be
x = \frac{x_{\rm low P} - x_{\rm high P}}{1 + \exp(2[\log_{10}(P/P_{\rm trans})])} + x_{\rm high P}.
\label{eqn:sigmoid}
\ee
Here $P_{\rm trans}$ defines the pressure (and thus the typical $\sigma$) at which the IMF transitions from light to heavy.  We find that in both cases a value $\log_{10}(P_{\rm trans}/k_B / [\kcm3]) = 5$ (corresponding to $\sigma_e \approx 80\kms$ works well for reproducing the C13 trend.

In the LoM case (top left panel of \Fig{IMF}), the slope from 0.5 to 100 $\Msun$ is kept fixed at $x=-2.3$ (as for a Kroupa IMF) but the low-mass slope (0.1 to 0.5 $\Msun$) is varied from $x_{\rm LoM, low P}=0$ at low pressure to $x_{\rm LoM, high P}=-3$ at high pressure. Note that this is by no means the only IMF variation prescription that reproduces the C13 trend, especially given the degeneracies between the slopes and the parameters of the sigmoid function, but we find that it is simple, intuitive, and works quite well at producing a clean trend between MLE$_r$ and $\sigma_e$.

In the lower left panel of \Fig{IMF} we plot the resulting MLE$_r$ as a function of birth ISM pressure for individual star particles in Ref-50, post-processed with the LoM IMF prescription. With this IMF, stars born with $P/k_B \lesssim 10^4\kcm3$ are bottom-light, while those with $P/k_B \gtrsim 10^6\kcm3$ are bottom-heavy, with a smooth transition between these values. Such a prescription increases the fraction of dwarf stars in the stellar population at high pressure. This increases the mass and decreases the luminosity of ageing star particles, both leading to an increased MLE. Note the small amount of scatter at fixed birth $P$, despite the fact that stars of all ages and metallicities are plotted here. Thus, for low-mass slope variations, the MLE-parameter seems to be a good proxy for the IMF.

For our second variable IMF prescription, HiM (shown in the top right panel of \Fig{IMF}), we instead keep the IMF slope below 0.5 $\Msun$ fixed at $x = -1.3$ (the Kroupa value), while making the slope above this mass shallower at high pressures, again varying according to the sigmoid function of \Eqn{sigmoid}. Specifically, we have $x_{\rm HiM, low P} = -2.3$ and $x_{\rm HiM, high P} = -1.6$, again with $\log_{10}(P_{\rm trans}/k_B / [\kcm3]) = 5$. Similar ``top-heavy'' forms of IMF variations have been proposed in the literature to explain the observed properties of strongly star-forming galaxies at both high and low redshifts \citep[e.g.][]{Baugh2005, Meurer2009, Habergham2010, Gunawardhana2011, Narayanan2012, Narayanan2013, Zhang2018}.

The lower right panel of \Fig{IMF} shows MLE$_r$ as a function of birth ISM pressure for individual star particles in Ref-50, this time post-processed assuming the HiM variable IMF. This prescription allows us to increase the MLE at high pressure by adding more stellar remnants such as black holes and neutron stars, while at the same time reducing the total luminosity of old stellar populations. Note that here the mass of ageing star particles is overall lower due to the increased stellar mass loss associated with the increased fraction of high-mass stars, but the stronger decrease in luminosity results in a net increase in the $M/L$ ratio. Here we see much larger scatter than for LoM due to the fact that the MLE$_r$ for a given star particle is no longer independent of age. The age-independence of the MLE for LoM was solely due to the fact that the high-mass slope is approximately the same as the reference (Salpeter) IMF. In that case, ageing the population removes roughly an equal fraction of mass and luminosity from the LoM IMF as it does from an SSP with a Salpeter IMF. For stars with a shallower high-mass IMF slope, the MLE$_r$ is initially small, owing to the high luminosity, but increases over time as the luminosity decreases faster with age than for a Salpeter IMF.  The resulting global correlations between the MLE$_r$ and birth ISM pressure for individual galaxies in self-consistent simulations that include these IMF variations are shown in \App{Appendix_calibration}.

These two IMF variation prescriptions were carefully calibrated by post-processing the Ref-100 simulation to reproduce the C13 trend between MLE$_r$ and $\sigma_e$. Further details of this calibration procedure can be found in \App{Appendix_calibration}. In the next sections we will confirm that this trend is still reproduced when the variable IMF is implemented self-consistently into a full cosmological hydrodynamical simulation.

As an aside, we also experimented with making the IMF become ``top-light'', meaning that the high-mass slope becomes steeper, rather than shallower, at high pressure. This prescription was inspired by observational studies that infer IMF variations spectroscopically using the MILES SPS models \citep{Vazdekis2010, Vazdekis2012}, which allow users to vary only the high-mass slope of the IMF, using the (perhaps confusingly nicknamed) ``bimodal'' IMF of \citet{Vazdekis1996}. Such studies find that the fraction of dwarf to giant stars increases with increasing $\sigma$ in high-mass ETGs, which for this parameterization results in a steeper high-mass slope (or a top-light IMF) \citep[e.g.][]{LaBarbera2013, LaBarbera2015}. While we were able to obtain a match to the C13 trend with this bimodal parameterization in post-processing of the reference EAGLE simulations, we opted to use the LoM prescription instead due to the fact that the latter already increases the fraction of dwarf stars with less of an effect on feedback or metal production, making it more likely that the variable IMF model would match the galaxy observables used to originally calibrate the EAGLE reference model. Indeed, it has been shown by \citet{Martin-Navarro2016} that the bimodal IMF prescription can have significant effects on the [Mg/Fe] abundances in massive ETGs. Confirmation of the validity of such a top-light IMF prescription would require a fully self-consistent simulation, which we have not performed for this prescription. It would be interesting for future work to test how well these SPS models can fit IMF-sensitive absorption features using a ``LoM'' IMF variation parameterization instead.

We also attempted to implement the local metallicity-dependent IMF prescription from \citet{Martin-Navarro2015c}, where the high-mass slope of this bimodal IMF is shallower (steeper) than a Kroupa IMF at low (high) metallicities. This prescription was recently used by \citet{Clauwens2016} to reinterpret observational galaxy surveys and was implemented into hydrodynamical simulations of MW-analogues by \citet{Gutcke2017}. In post-processing of Ref-100, we found no clear trend between the MLE and $\sigma_e$ when implementing this IMF variation prescription. We suspect that this may be partially due to the relatively flat mass-metallicity relation in high-mass galaxies in intermediate-resolution EAGLE (S15).

To provide an idea of the effect of these variable IMF prescriptions on the light output of galaxies, we generate images of galaxies using a modified version of the SKIRT radiative transfer code \citep{Camps2015}. These modifications allow the user to generate images using SED templates from FSPS, for different variable IMF prescriptions. This new functionality in SKIRT is publicly available in a very general form at \url{http://www.skirt.ugent.be/}. In particular, it allows the user to specify for each star particle either the low-mass or high-mass slope of the IMF while keeping the other end fixed at the Kroupa value. In this way, one may vary the IMF according to many desired prescriptions, not only those presented in this paper.

We show in \Fig{pictures} RGB images of the SDSS gri central wavelengths of a massive elliptical galaxy from Ref-50, assuming a Kroupa (left panel), LoM (middle) and HiM (right) IMF. For a Kroupa IMF we see clumps of blue, young star particles embedded in a white, diffuse, intragalactic stellar background. For LoM, the image looks almost identical to the Kroupa image since this IMF mostly adds very dim, low-mass stars to the stellar population, which do not strongly affect the light. On the other hand, in the HiM case the diffuse starlight is much dimmer than the young stars. This is to be expected because for the HiM IMF, older populations should be overall dimmer because a much higher proportion of their mass is invested into the high-mass stars that have since died off. Note, however, that since a top-heavy IMF produces in general more metals per stellar mass formed, the impact of dust on the HiM image is likely underestimated.

\subsection{Preparations for self-consistent simulations with a variable IMF}
\label{sec:mods}

\begin{figure}
  \centering
\includegraphics[width=0.45\textwidth]{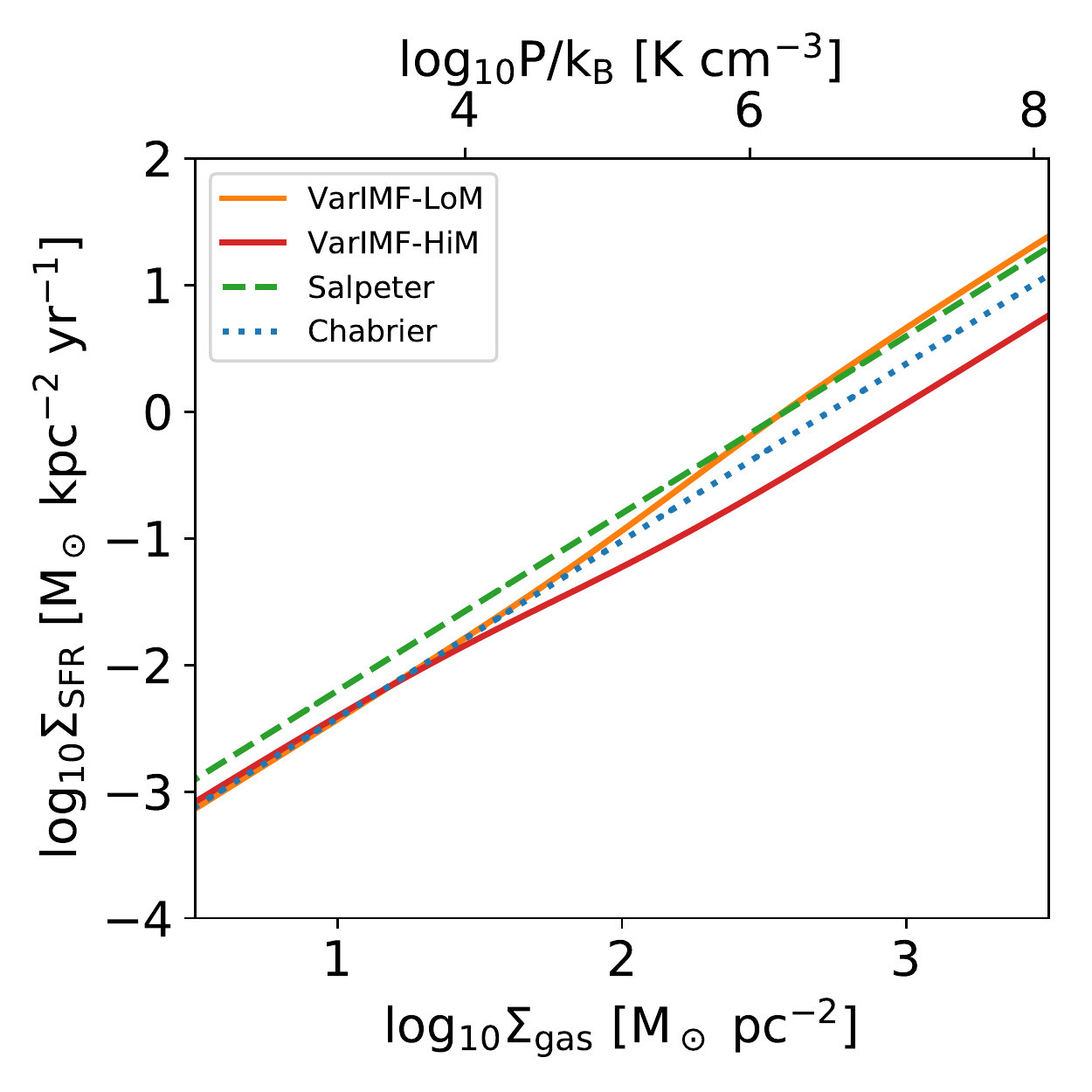}
\caption{Star-formation law recalibration applied to gas particles in our simulations with a variable IMF. For reference, the laws calibrated for a Chabrier (used in Ref-50) and Salpeter IMF are shown in blue-dotted and green-dashed lines, respectively. The recalibrations that are used in our simulations with LoM and HiM are shown as orange and \HimColour{} solid lines, respectively. The pressure corresponding to the gas surface densities on the lower axis is shown along the top axis. To remain consistent with the observed KS law, at high pressure, SFRs are increased (decreased) by $\approx 0.3$ dex relative to the reference simulations for simulations using the LoM (HiM) variable IMF prescription.}
\label{fig:KSlaw}
\end{figure}

The best way to test the full effect of a variable IMF on simulated galaxies is to run a new simulation that explicitly includes this IMF. This is because, for example, the IMF affects the metals released into the ISM by stars, which then affect cooling rates, which further affect future star formation, and so on. Additionally, the IMF affects the available energy from supernovae to provide feedback and regulate star formation. Such effects cannot be accounted for in post-processing. In this section we describe modifications to the EAGLE code that were implemented to maintain self-consistency when adopting a variable IMF. 

In EAGLE, the star formation law reproduces the empirical Kennicutt-Schmidt (KS) law \citep{KennicuttJr.1998}. This relation was originally derived by converting H$\alpha$ fluxes to SFRs assuming a Salpeter IMF. In the reference EAGLE model, S15 accounted for the lower $(M/L)$ obtained from the assumed Chabrier IMF by dividing the normalization of the KS law by a factor 1.65. This factor is the asymptotic ratio between the number of ionizing photons per solar mass formed after 100 Myr of evolution with a constant SFR as predicted by the \citet{Bruzual2003} model for a Chabrier and a Salpeter IMF. Because our IMF is not fixed, but varies with pressure, if we wish to maintain the same relationship between H$\alpha$ surface brightness and $\Sigma_{\rm gas}$, we need to instead divide by a factor that is not constant but varies with pressure. 

We recalibrate the star formation law by using the FSPS software to compute, for a given pressure, the ratio of the luminosity in the GALEX FUV-band for a stellar population with a constant star formation rate, between the variable IMF and a population with a Salpeter IMF, i.e.,
\be
f_{\rm KS,mod}(P) = \frac{L_{\rm FUV}({\rm VarIMF}(P)) }{L_{\rm FUV}({\rm Salpeter})}.
\ee

In \Fig{KSlaw} we plot the recalibrated star formation law as orange and \HimColour{} solid lines for LoM and HiM, respectively, and compare them to the original (Salpeter-derived) relation and EAGLE's Chabrier IMF-corrected version. We used Equation 8 of \citet{Schaye2008} to convert gas pressure to gas surface density, assuming a gas mass fraction of unity and ratio of specific heats of 5/3. In the low-$P$ regime, the normalization remains close to the reference EAGLE value, but at high pressures we multiply (divide) the normalization relative to reference EAGLE by a factor of $\approx 2$ for LoM (HiM). Note that for a Chabrier IMF, by the above method we obtain $f_{\rm KS,mod}(P) \simeq 1.57$, not far from the factor 1.65 assumed in EAGLE. The difference here comes from the differences in the FSPS and BC03 models, and has no noticeable effect on our results.

We also make self-consistent the mass evolution of the stellar populations as well as the heavy element synthesis and mass ejected into the ISM from stellar winds and supernovae. This modification is straightforward since these processes already include an integration over the IMF in the EAGLE code.

Another consideration is that the IMF has a direct impact on the number of massive stars and thus the amount of stellar feedback energy that is returned to the ISM per unit stellar mass formed. We also make this self-consistent, which effectively results in a factor $\approx 2$ less (more) feedback energy produced per stellar mass formed at high pressures for LoM (HiM). In the reference model, such a large change in the feedback efficiency can have significant effects on many galaxy properties (C15). However, in the case of our variable IMF simulations, the modified star formation law counteracts this effect, making the time-averaged feedback energy consistent with the reference model at fixed gas surface density. We refer the reader to \App{Appendix_self_consistency} for further details regarding the individual impact of each of these effects on galaxy properties. As we will show in \Sec{calibration_diagnostics}, performing variable-IMF simulations with these modifications yields excellent agreement with the observational diagnostics that were originally used to calibrate the subgrid feedback physics in the reference EAGLE model. 

The SNIa rate per star particle in EAGLE depends only on the particle's initial mass and an empirical delay time distribution function, calibrated to match the observed (IMF-independent) evolution of the SNIa rate density (S15).
Because of the strong dependency of alpha-enhancement on SNIa rates, having these rates match observations directly is important. While an IMF-dependent SNIa rate model would be ideal from a theoretical point of view, it is precluded by the large uncertainties in parameters that would factor into such a model, such as white dwarf binary fractions, binary separations, and merger rates. While the SNIa rates therefore do not depend directly on the IMF, they do depend on the star formation history of the simulation which can be affected by the IMF. We will show in \Sec{calibration_diagnostics_physical} that the SNIa rates are not strongly affected in our variable IMF simulations. 

In the next section we will present our simulations and discuss the resulting trend between the galaxy-averaged IMF and central stellar velocity dispersion. We discuss the impact of these variable IMF prescriptions on galaxy properties such as metal abundances and SFR in \Sec{IMF_effect}.

\section{Self-consistent simulations with a variable IMF}
\label{sec:sims}
We ran two new (50 Mpc)$^3$ simulations with the same physics and resolution as the reference EAGLE model, except that we imposed two different IMF variation prescriptions. The IMF becomes either bottom-heavy (LoM) or top-heavy (HiM) when the pressure of the ISM out of which star particles are born is high. In \Sec{calibration} we described how in post-processing we calibrated the pressure dependencies to match the observed trend of excess $M/L$-ratio with stellar velocity dispersion of \citet{Cappellari2013b}. Including the IMF variation prescriptions explicitly in these simulations allows the IMF variations to affect self-consistently the mass evolution, metal yields, and the stellar energetic feedback during the simulations. The simulations also include a recalibrated KS law normalization to account for the change in UV luminosity per stellar mass formed due to the variable IMF prescriptions (see \Sec{mods} for details). Throughout we will refer to these simulations with bottom-heavy and top-heavy IMF prescriptions as \lom{} and \him{}, respectively, and the reference (50 Mpc)$^3$ box (with a universal Chabrier IMF) as Ref-50. In \Sec{IMF_vs_sigma} we present the resulting trends between the IMF, MLE$_r$ and $\sigma_e$ in high-mass galaxies, while in \Sec{calibration_diagnostics} we show that both simulations agree well with the observational diagnostics used to calibrate the subgrid physics in the reference EAGLE model. Unless otherwise specified, all quantities are measured within a 2D aperture of radius $r_e$, the projected half-light radius in the $r$-band. This choice is motivated by the fact that many IMF studies \citep[e.g.][]{Cappellari2013b} measure the IMF within such an aperture. 

\subsection{IMF vs stellar velocity dispersion}
\label{sec:IMF_vs_sigma}

\begin{figure*}
\includegraphics[width=\textwidth]{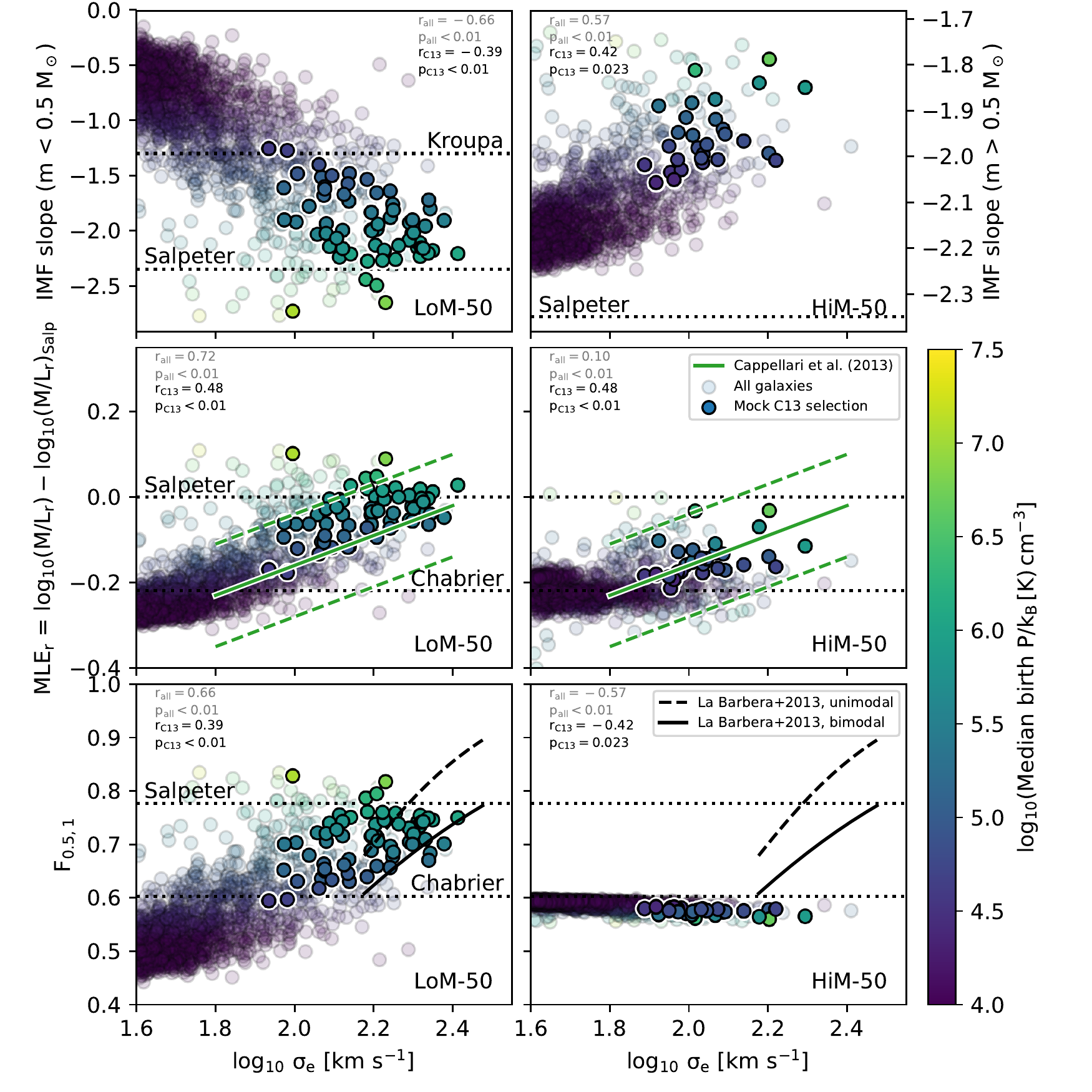}
\caption{ IMF diagnostics as a function of light-weighted stellar velocity dispersion, $\sigma_e$, all measured within the 2D, projected, $r$-band half-light radius, $r_e$, for all galaxies with $\sigma_e > 10^{1.6}\kms$ in our \lom{} (left column) and \him{} (right column) simulations at $z=0.1$. Upper row: low-mass ($m<0.5\Msun$) and high-mass ($m>0.5\Msun$) $r$-band light weighted mean IMF slope for \lom{} and \him{}, respectively. Middle row: $(M/L_r)$-excess with respect to a Salpeter IMF. Lower row: Mass fraction of stars in the IMF with $m<0.5\Msun$ relative to that for stars with $m<1\Msun$, $F_{0.5,1}$ (Equation \ref{eqn:F051}).  Expected values for fixed IMFs are indicated as dotted horizontal lines. To facilitate comparison with \citet[][C13]{Cappellari2013b}, we make a ``mock C13'' selection of early-type galaxies with $M_K < -21.5$ mag and intrinsic $u^*-r^* > 2$ (C13 cut; see text for details), indicated with opaque filled circles coloured by the light-weighted mean birth ISM pressure; points for galaxies outside this sample are translucent. The observed MLE-$\sigma_e$ trend from C13 is shown as a \CapColour{} solid line, with the intrinsic scatter shown as dashed lines. Dwarf-to-giant mass fractions derived from the correlations between IMF slope and $\sigma$ by \citet{LaBarbera2013} are shown as black-solid and -dashed lines for bimodal and unimodal IMF parameterizations, respectively. The Pearson correlation coefficient, $r$, and its $p$-value are indicated in each panel for the full sample with $\sigma > 10^{1.6} \kms$ and the C13 cut in grey and black, respectively.}
\label{fig:IMF_vs_sigma}
\end{figure*}

\begin{figure*}
  \centering
\includegraphics[width=0.45\textwidth]{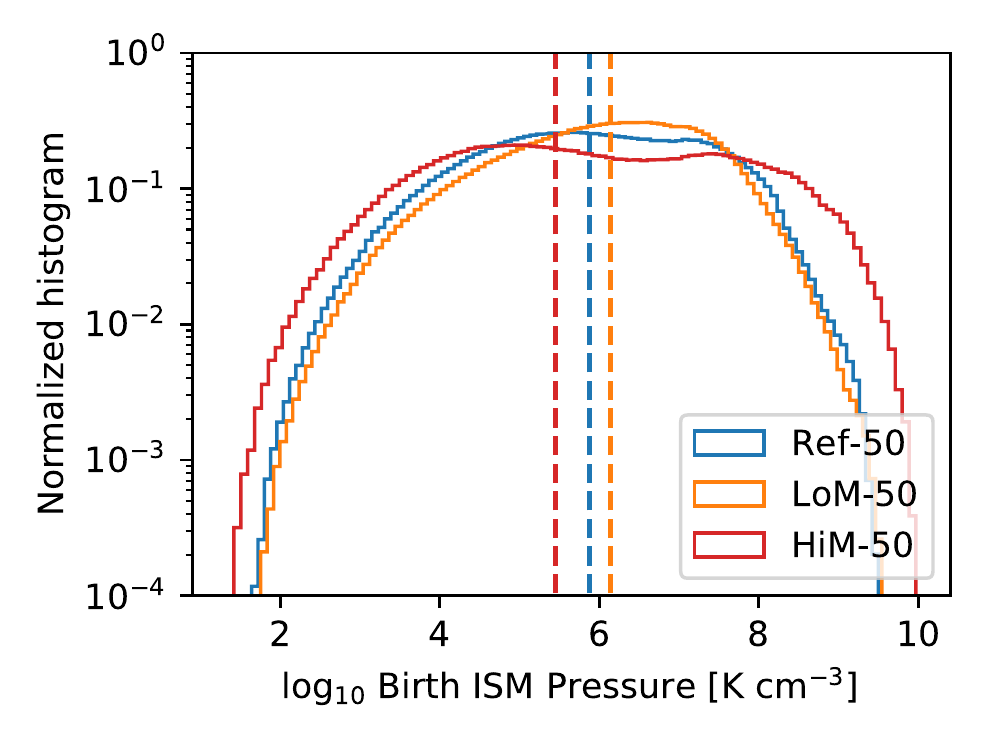}
\includegraphics[width=0.45\textwidth]{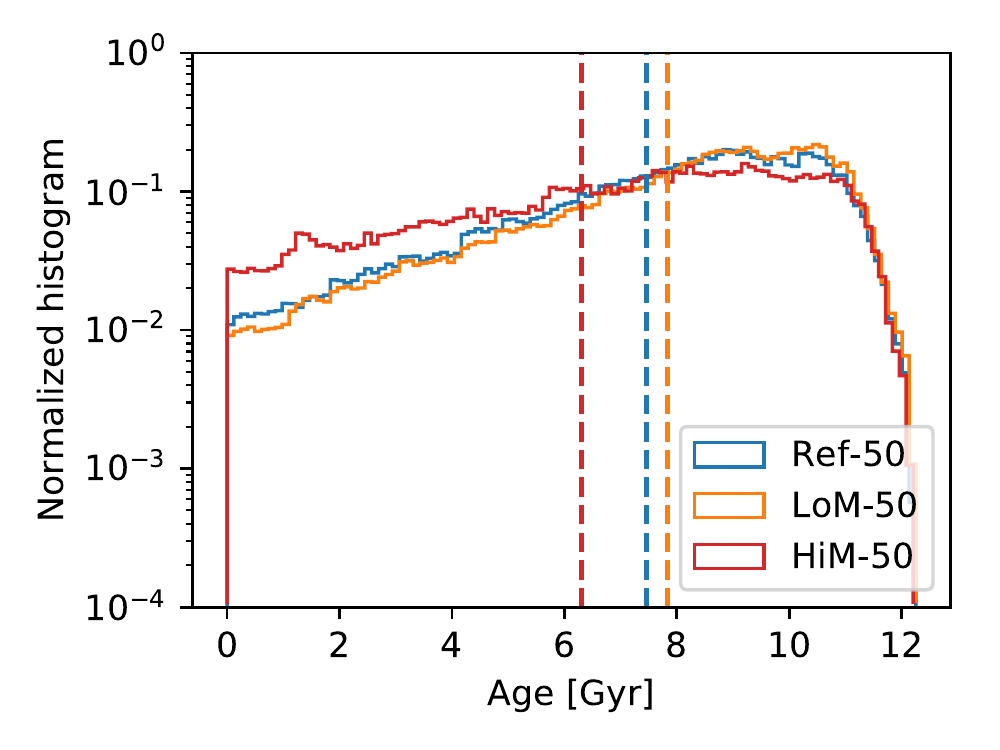}
\caption{Properties of stars within the half-light radii of galaxies with $\sigma_e > 100\kms$ in our \lom{} (orange) and \him{} (\HimColour{}) simulations compared with Ref-50 (blue). The left and right panels show birth ISM pressures and stellar ages, respectively. Vertical dashed lines denote medians. While \lom{} matches Ref-50 reasonably well, \him{} produces stars with lower median birth ISM pressures and younger ages.}
\label{fig:star_props}
\end{figure*}

\begin{figure}
  \centering
\includegraphics[width=0.45\textwidth]{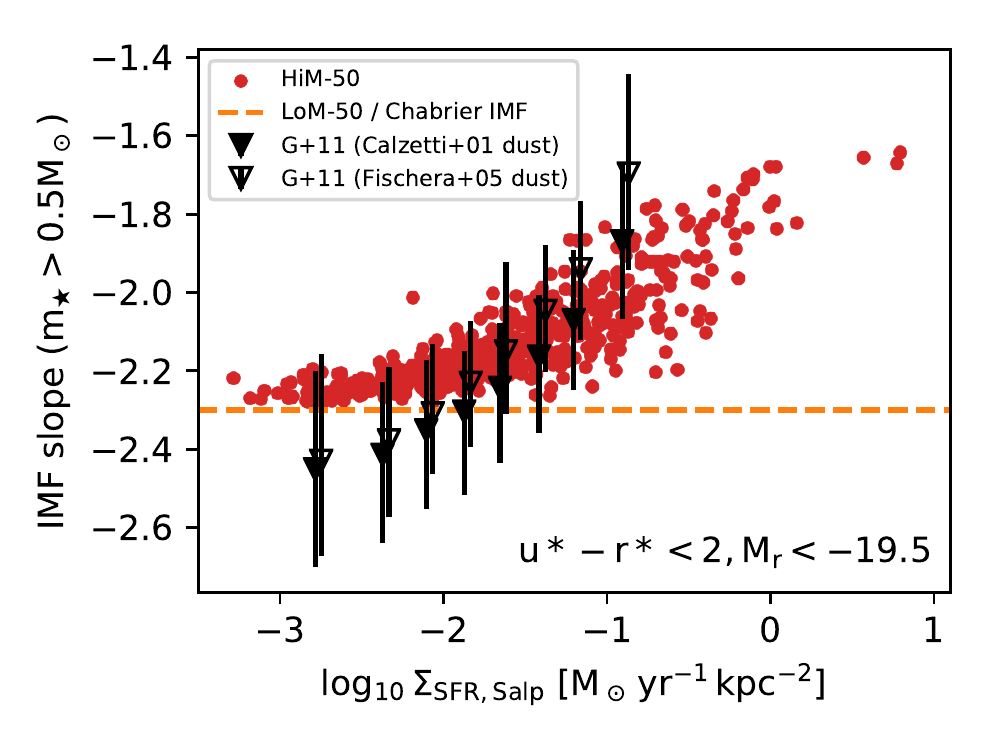}
\includegraphics[width=0.45\textwidth]{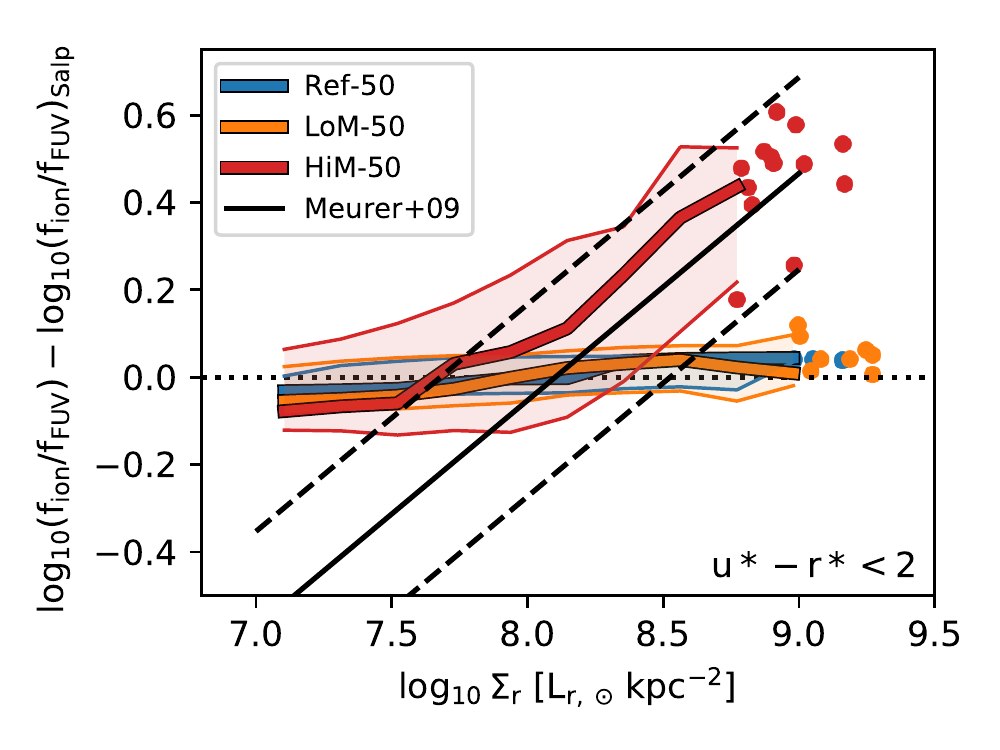}
\caption{{\bf Upper panel:} FUV-weighted high-mass ($m>0.5\Msun$) IMF slope as a function of the Salpeter-reinterpretted star formation rate surface density, $\Sigma_{\rm SFR,Salp}$, of star-forming (intrinsic $u^*-r^*<2$) galaxies at $z=0.1$. To compare with \citet{Gunawardhana2011}, we only include galaxies also with $M_r < -19.5$. The horizontal orange dashed line shows the high-mass slope for all galaxies in \lom{}, corresponding to the Kroupa/Chabrier value. Open and filled triangles show results from the GAMA survey by \citet{Gunawardhana2011} assuming \citet{Calzetti2001}/\citet{Cardelli1989} and \citet{Fischera2005} dust corrections, respectively. {\bf Lower panel:} Ratio of flux in ionizing photons to that in the GALEX FUV band, normalized to the value expected for a Salpeter IMF, as a function of $r$-band luminosity surface density for star-forming galaxies in Ref-50 (blue), \lom{} (orange), and \him{} (red). Solid thick lines indicate medians, while filled regions denote 10-90$^{\rm th}$ percentiles. The black solid line shows results from \citet{Meurer2009} for the flux in H$\alpha$ relative to FUV, normalized to the expected Salpeter value, for star-forming galaxies. Dashed black lines indicate rms residuals. The positive trends seen in observations are qualitatively reproduced for \him{} star-forming galaxies, while an IMF prescription that varies only the low-mass slope of the IMF by construction does not.}
\label{fig:highmassslope}
\end{figure}

The trend between the IMF and central stellar velocity dispersion, $\sigma_e$, is the most prevalent correlation between the IMF and galaxy properties in the observational literature. In \Fig{IMF_vs_sigma} we plot $r$-band light-weighted diagnostics related to the IMF as a function of $\sigma_e$ for galaxies in the \lom{} (left column) and \him{} (right column) simulations. As translucent coloured circles we show all galaxies with $\sigma_e > 10^{1.6}\ (\approx 40) \kms$, coloured by the light-weighted mean pressure at which the stars within each galaxy's $r_e$ were formed. 

The upper row shows the $r$-band light-weighted mean IMF slope for individual galaxies, where in the upper-left and -right panels we plot the low-mass ($m<0.5\Msun$) and high-mass ($m>0.5\Msun$) slopes, respectively. As expected, for \lom{} the IMF slope transitions from shallower than Kroupa to a steeper Salpeter-like slope with increasing $\sigma_e$, while for \him{}, the high-mass slope becomes shallower than Salpeter with increasing $\sigma_e$, reaching values up to $\approx -1.8$, comparable to the shallowest slopes inferred in local highly star-forming galaxies \citep{Gunawardhana2011}.

In the middle row we plot the resulting relation between MLE$_r$ and $\sigma_e$ for our variable IMF simulations. For both simulations, galaxies with $\sigma_e < 10^{1.8}\ (\approx 60) \kms$ lie close to the Chabrier MLE$_r$ value of $-0.22$, with MLE$_r$ increasing for higher-mass galaxies. To compare with C13, we select galaxies in a similar way to that study. The C13 sample consists of 260 early-type elliptical and lenticular galaxies selected morphologically based on whether they contain dust lanes or spiral arms, and is complete down to an absolute magnitude of $M_{\rm K} = -21.5$ mag. We mimic their selection by first taking only galaxies with $M_{\rm K} < -21.5$ mag (without any dust correction).  This cut roughly corresponds to a stellar mass $\gtrsim 10^{10.5} \Msun$ for all models (although the exact correspondence depends on the IMF assumed). Then, to select only early-type galaxies, we make a cut in intrinsic $u^*-r^* > 2.0$, which roughly separates the blue cloud from the red sequence in EAGLE \citep{Correa2017} and is similar to removing galaxies with specific star formation rate (sSFR) $ \gtrsim 10^{-1.8}$ and $10^{-1.7}\,{\rm Gyr}^{-1}$ for \lom{} and \him{}, respectively. C13 additionally remove galaxies with very young stellar populations by excluding those with an H$\beta$ stellar absorption line with equivalent width greater than $2.3 \buildrel _\circ \over {\mathrm{A}}$. \citet{McDermid2015} show that this cut corresponds roughly to an SSP age of 3.1 Gyr, which is already younger than any of our galaxies with $u^*-r^* > 2.0$. We refer to this selection as the ``mock C13'' sample.
 
The mock C13 galaxies are highlighted as the opaque coloured circles in \Fig{IMF_vs_sigma}. When selecting galaxies in this manner, both simulations produce galaxies reasonably consistent with the C13 MLE$_r-\sigma_e$ trend, with the majority of galaxies lying within the intrinsic scatter.

For \lom{}, a least absolute deviation (LAD) fit to the MLE$_r-\sigma_e$ relation for these mock C13 galaxies yields a slope of $0.23 \pm 0.07$, which agrees with the slope of $0.35 \pm 0.06$ reported by C13. However, our galaxies are offset by $\approx 0.05$ dex above the C13 trend. This small discrepancy is partly due to the fact that the LoM prescription was initially calibrated using stars within an aperture larger than $r_e$, which we show in \App{Appendix_calibration} can make a significant difference to the normalization of the MLE$_r-\sigma_e$ relation.  Indeed, with a slightly larger choice of aperture, one can decrease the normalization of the \lom{} MLE$-\sigma_e$ trend to match or even lie below the C13 trend. We caution that care with regards to aperture choices should be taken when comparing variable IMF claims between observational studies. Aperture choices vary between observational IMF studies (e.g. \citealt{McDermid2014} and \citealt{Conroy2012b} measure $M/L$ within $r_e$ and $r_e/8$, respectively) and even within them (\citealt{McDermid2014} measure other properties like age and metallicity within $r_e/8$). Consistent apertures are crucial for making fair comparisons between such studies. 

This positive offset is further increased (slightly) due to the fact that stars in \lom{} tend to form from gas at slightly higher pressures than in Ref-50, which was used for the IMF calibrations. This can be seen in \Fig{star_props}, where we show the distribution of gas birth ISM pressures and ages for stars within $r_e$ of galaxies with $\sigma_e > 100\kms$ for the two variable IMF simulations and Ref-50. This result is likely due to the weaker stellar feedback resulting from the more bottom-heavy IMF. 

For \him{}, while nearly all of the mock C13 galaxies lie within one standard deviation of the C13 MLE$_r-\sigma_e$ relation, an LAD fit is slightly shallower for the simulation, with a slope of $0.16 \pm 0.09$ (compared to the C13 slope of $0.35 \pm 0.06$). This shallower trend is the result of several factors. Firstly, we are no longer sampling as many galaxies at high-$\sigma_e$; Ref-50 had poor sampling to begin with for $\sigma_e > 10^{2.3} \kms$ due to the limited simulation volume, which is now compounded by the fact that galaxies at fixed $M_{\rm DM}$ tend to have lower $\sigma_e$ in \him{} than in Ref-50 due to the more efficient feedback. 

A second factor is the impact that the stronger stellar feedback from HiM has on the times and gas pressures at which stars form in the simulation. In \Fig{star_props}, we see that for \him{}, stars in the centres of $\sigma_e > 100\kms$ galaxies typically form at lower pressures and later times than they did in Ref-50. This behaviour is due to the stronger stellar feedback delaying star formation to later times (and thus lower pressures). Consequently, galaxies in the simulation obtained IMFs with steeper high-mass slopes than expected, as well as less time to evolve, both of which lower the MLE relative to the post-processing analysis of Ref-50 (although this is not a strong effect for mock C13 galaxies; see \App{Appendix_calibration}).

Despite the trend between MLE$_r$ and $\sigma_e$ being less clear for \him{} than \lom{}, the high-$\sigma_e$ galaxies in \him{} are certainly not inconsistent with the C13 trend, and thus represent a conservative approach to studying top-heavy IMF variations in high-mass galaxies. Indeed, we will show in Paper II that this HiM IMF prescription causes the MLE to vary much more strongly with age than with $\sigma_e$.

We note that while the MLE$_r-\sigma_e$ trends for both simulations are consistent with dynamical IMF measurements, \him{} may not be consistent with spectroscopic IMF studies that are sensitive to the present-day fraction of dwarf to giant stars, which tend to find an increasing dwarf-to-giant ratio with increasing $\sigma_e$ \citep[e.g.][]{LaBarbera2013}. To show this explicitly, following \citet{Clauwens2016}, we compute for each galaxy the fraction of the mass in stars with $m<0.5\Msun$ relative to the mass of stars with $m<1\Msun$ given its IMF. Specifically, we compute
\be
F_{0.5,1} = \frac{ \int_{0.1}^{0.5}M\Phi(M)dM}{\int_{0.1}^{1}M\Phi(M)dM}.
\label{eqn:F051}
\ee
where $\Phi(M)$ is the IMF. This upper limit of $m<1\Msun$ in the denominator roughly corresponds to the highest stellar mass expected in the old stellar populations of ETGs.  These results are plotted in the lower row of \Fig{IMF_vs_sigma}, where, while $f_{0.5,1}$ increases strongly with $\sigma_e$ for \lom{} galaxies, it is relatively constant for \him{}, remaining close to the value expected for a Chabrier IMF. This demonstrates that the increase in MLE$_r$ for high-$\sigma_e$ galaxies in \him{} is the result of excess mass in stellar remnants, rather than dwarf stars. 

We compare with the results of \citet{LaBarbera2013} by converting their IMF slopes for their 2SSP models with bimodal and unimodal IMF parameterizations to $F_{0.5,1}$. Note that we do not use their definition of $F_{0.5}$ which integrates the denominator in \Eqn{F051} to $100\Msun$, since $F_{0.5}$ is sensitive to the choice of IMF parameterization at fixed (present-day) $F_{0.5,1}$, which we show explicitly in \App{appendix_F05}. In the \lom{} case, $F_{0.5,1}$ agrees very well with the \citet{LaBarbera2013} results of increasing mass fraction of dwarf stars in high-$\sigma_e$ galaxies. \him{}, as expected, does not agree.

On the other hand, recall that the HiM prescription was motivated by the fact that highly star-forming galaxies have recently been found to have top-heavy IMFs. For example, \citet{Gunawardhana2011} have found that for local, bright ($M_r < -19.5$) star-forming galaxies, those with larger H$\alpha$-inferred star formation rate surface density are redder than expected given their SFR and a standard IMF, implying that the high-mass IMF slope may be shallower in such systems. In \Fig{highmassslope}, we compare with the results of \citet{Gunawardhana2011} by plotting the Galex FUV-weighted high-mass slope of the IMF for star-forming (intrinsic $u^*-r^*<2$) galaxies with $M_r < -19.5$ at $z=0.1$ in \him{} as a function of the star formation rate surface density, defined as $\Sigma_{\rm SFR,Salp} = {\rm SFR_{Salp}}/(2\pi r_{\rm e,FUV}^2)$, where SFR$_{\rm Salp}$ is the Salpeter-reinterpretted total star formation rate within a 3D aperture of radius 30 pkpc and $r_{\rm e,FUV}$ is the half-light radius in the FUV band. The Salpeter reinterpretation is performed by multiplying the true SFR by the ratio of the FUV flux relative to that expected for a Salpeter IMF, similar to that done by \citet{Clauwens2016}. The result from \citet{Gunawardhana2011} is shown for two assumptions for the dust corrections. The positive trend for \him{} is qualitatively consistent with the observations, albeit slightly shallower. For reference, we include as a horizontal line the high-mass slope in all \lom{} (as well as Ref-50) galaxies, corresponding to the Kroupa/Chabrier high-mass value. \lom{} is, as expected, inconsistent with the observations since the high-mass slope is not varied in that model.

Another example comes from \citet{Meurer2009}, who conclude that the increasing ratio of H$\alpha$ to FUV flux toward higher-pressure environments implies that the high-mass slope of the IMF may be becoming shallower in such environments. To compare with their data, we compute the flux of ionizing radiation, $f_{\rm ion}$, by integrating the spectra output by FSPS up to $912 \buildrel _\circ \over {\mathrm{A}}$ (as in \citealt{Clauwens2016}) and dividing by the flux in the FUV band, $f_{\rm FUV}$. Since the ionizing flux is not identical to the H$\alpha$ flux, we normalize by the value of the ratio expected for a Salpeter IMF. In the lower panel of \Fig{highmassslope}, we plot this ratio for our star-forming galaxies in Ref-50, \lom{}, and \him{} as a function of $r$-band luminosity surface density, $\Sigma_r$. We compare with the corresponding relation from \citet{Meurer2009} shown as a solid black line. Ref-50 and \lom{} show a constant ``Salpeter''-like $f_{\rm ion}/f_{\rm FUV}$ at all $\Sigma_r$. For $\Sigma_r>8\,{\rm L}_{r,\odot}\,{\rm kpc}^{-2}$, \him{} galaxies increase in $f_{\rm ion}/f_{\rm FUV}$ with increasing  $\Sigma_r$, in agreement with the observations. At lower $\Sigma_r$ the relation flattens to the Salpeter value since in HiM we do not vary the IMF high-mass slope to values steeper than Salpeter. This result implies that the IMF high-mass slope may need to become steeper than Salpeter at low pressures to conform with these observations. However, it is not clear that the MLE-$\sigma_e$ correlation would then still match the observed C13 trend, as this would increase the MLE at low pressure (or low $\sigma_e$), weakening the otherwise positive MLE$-\sigma_e$ correlation.

Since strongly star-forming galaxies are the progenitors of present-day high-mass ETGs, it is not clear how to reconcile the observed top-heavy IMF in strongly star-forming galaxies with the observed bottom-heavy IMF seen in present-day ETGs as constrained by the dwarf-to-giant ratio (lower row of \Fig{IMF_vs_sigma}. These variable IMF simulations will thus be extremely useful in testing these different, possibly conflicting, IMF variation scenarios with galaxy formation models.

\subsection{Subgrid calibration diagnostics}
\label{sec:calibration_diagnostics}
The success of the EAGLE model stems in part from calibrating the subgrid feedback parameters to match key observables (the GSMF, sizes, and BH masses) that are difficult to predict from first principles in cosmological simulations. Thus, a first check to see if the variable IMF simulations are reasonable is to verify that they also reproduce these calibration diagnostics. 

\begin{figure*}
  \centering 
\includegraphics[width=0.95\textwidth]{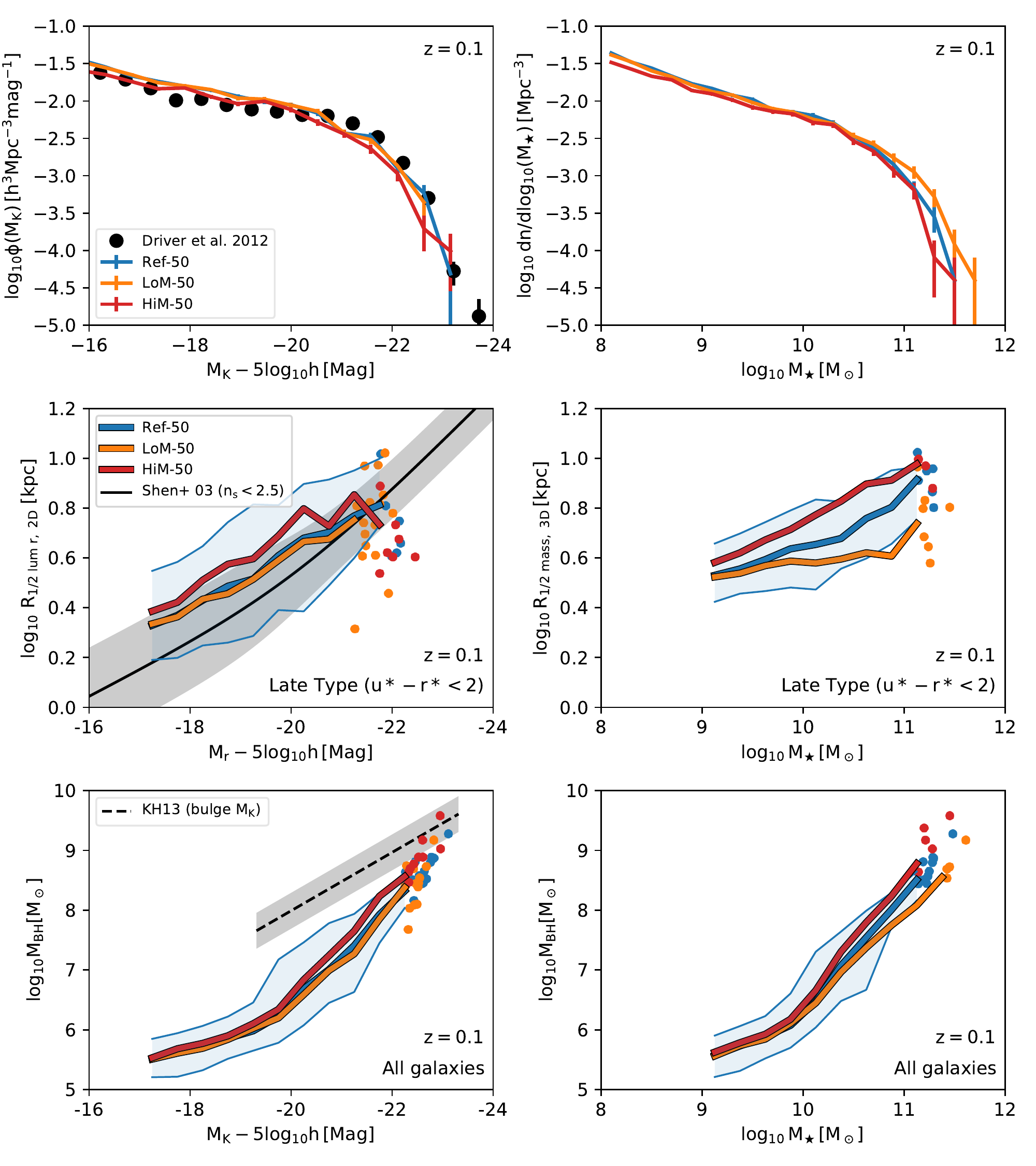}
\caption{{\bf Left column:} Subgrid calibration diagnostics for the \lom{} (orange), \him{} (\HimColour{}) simulations, compared to the reference model (Ref-50; blue) and observations for galaxies at $z=0.1$. To remain consistent with S15, masses and luminosities (sizes) are measured  within a 30 kpc 3D (2D) aperture.  Top-left panel: K-band galaxy luminosity ($M_{\rm K}$) function. Observational data from the GAMA survey are shown as black points with $1\sigma$ error bars \citep{Driver2012}. Middle-left panel: Projected K-band stellar half-light radius as a function of $M_{\rm K}$ for star-forming galaxies (intrinsic $u^*-r^* < 2.0$) with more than 600 star particles. Filled regions show the 10 to 90th percentile range for Ref-50; the other curves have similar scatter. Individual points are shown for bins containing fewer than 10 galaxies. Sersic $r$-band half-light radii from SDSS are shown for galaxies with Sersic indices $n_S < 2.5$ as a black solid line \citep[][the grey shaded region indicates $1\sigma$ scatter]{Shen2003}.  Lower-left panel: As in the middle-left panel but showing the black hole mass$-$ galaxy $M_{\rm K}$ relation. For reference we show the observed relation with bulge luminosity (converted to AB magnitudes) from \citet{Kormendy2013} as a dashed black line with intrinsic scatter indicated with the grey filled region.  {\bf Right column:} Physical quantities corresponding to the subgrid calibration diagnostics shown in the left column. Top-right panel: Galaxy stellar mass function.  Middle-right and bottom-right panels: 3D half-mass radius and $\MBH$, respectively, as a function of $\Mstar$.  Galaxies in \lom{} with $\Mstar > 10^{10.5} \Msun$ have higher masses at fixed number density and smaller physical sizes at fixed mass than in Ref-50 due to an excess (dark) mass of dwarf stars in the central regions. Both \lom{} and \him{} produce galaxies with reasonably realistic luminosities, half-light radii, and BH masses. }
\label{fig:calibration_diagnostics}
\end{figure*}

\subsubsection{Observable diagnostics}
\label{sec:calibration_diagnostics_observable}
Since physical quantities like the GSMF and galaxy half-mass radii can only be derived from observables once an IMF is assumed, we must compare the ``observable'' versions of the calibration diagnostics with the reference model and observations. The left column of \Fig{calibration_diagnostics} shows the galaxy $K$-band luminosity function, 2D projected $r$-band half-light radii for late-type galaxies, and BH masses as a function of luminosity for Ref-50 (blue), \lom{} (orange) and \him{} (\HimColour{}).

We show the $z=0.1$ $K$-band luminosity functions of our simulations in the top-left panel of \Fig{calibration_diagnostics}, and compare them to observational data from the GAMA survey \citep{Driver2012} down to $M_K - 5\log_{10}h = -16$, corresponding to galaxies with $\approx 100$ stellar particles, the resolution limit of the GSMF (S15). Both variable IMF runs agree well with Ref-50, with \him{} slightly under-predicting the luminosity function relative to Ref-50 by $\lesssim 0.1$ dex at all luminosities. This under-prediction is likely caused by the stronger stellar feedback in \him{}.

In the reference model, the density-dependence of the stellar feedback strength was calibrated to broadly match the observed sizes of late-type galaxies. This calibration was necessary to prevent the formation of overly compact galaxies due to artificial thermal losses in high-density environments due to the limited resolution of the simulation. It is thus important to verify that our variable IMF simulations also reproduce the sizes of such galaxies. In the middle-left panel of \Fig{calibration_diagnostics}, we plot the $r$-band 2D projected half-light radii as a function of $M_r$ for all star-forming galaxies at $z=0.1$ with at least 600 bound stellar particles, corresponding to $\Mstar \approx 10^9 \Msun$ or $M_r-5\log_{10}h<-17$. It was shown in S15 that galaxy sizes are converged down to this limit. To compare with observed sizes of late-type galaxies, we plot only star-forming galaxies defined as those with intrinsic (dust-free) $u^*-r^*<2$.  
The variable IMF simulations match Ref-50 well, with \him{} producing larger galaxies by $\approx 0.1$ dex, which is less than the typical discrepancies between observational studies of galaxy sizes \citep[e.g][]{Shen2003, Baldry2012}. These slightly larger sizes may be caused by the fact that galaxies in \him{} typically form at later times, from higher angular momentum gas, giving them larger sizes for the same amount of stellar mass formed. This effect is compounded by the fact that stars in the central regions of these galaxies typically formed at higher pressures than those in the outskirts, and thus with more top-heavy IMFs, yielding dimmer present-day stellar populations and  boosting the half-light radii to larger values.

For reference, we include the observed relation between Petrosian $r$-band half-light radius and $r$-band absolute magnitude for SDSS galaxies with Sersic indices $n<2.5$ \citep{Shen2003}. Such indices correspond to surface density profiles typical of discy, star-forming galaxies. Note that a fairer comparison would be to also select simulated galaxies with $n<2.5$.  However, \citet{Shen2003} showed that the size-luminosity relation is not significantly affected for different reasonable choices of morphological discriminator. Indeed, S15 found that such an $n<2.5$ cut selects 94 per cent of EAGLE galaxies with more than 600 star particles. We have checked that selecting all galaxies instead of only those with $u^*-r^*<2$ makes little difference to this plot, so we expect that an $n<2.5$ cut for the simulated galaxies would not change our result. While \lom{} and Ref-50 agree relatively well with the data, \him{} systematically over-predicts galaxy sizes relative to SDSS by $\approx 0.2$ dex, which differs from the observations at the $1\sigma$ level. Despite this good agreement for late type galaxies, we will show in \Sec{ETG_sizes} that the same is not true for ETGs in \him{}.

The efficiency of AGN feedback was calibrated to match the normalization of the observed $\MBH-\Mstar$ relation by \citet{Booth2009} as part of the OWLS project \citep{Schaye2010}. Since it also gave good results for the much higher-resolution EAGLE simulations, this efficiency was adopted for the reference model. The lower-left panel of \Fig{calibration_diagnostics} shows $\MBH$ as a function of $M_{\rm K}$. Note that we use the actual $\MBH$ values from the simulation rather than attempting to reinterpret them observationally. Both of the variable IMF simulations agree with Ref-50 extremely well for $M_{K} > -20$, while for more luminous objects there are slight variations of up to $\approx 0.1$ and 0.3 dex above the Ref-50 relation for \lom{} and \him{}, respectively. These variations are much smaller than the scatter in the observed $\MBH-\Mstar$ relation and are an acceptable match to Ref-50. Note that the BH masses agree much more poorly if the KS law is recalibrated but feedback is not made self-consistent (see discussion in \Sec{mods} and \App{Appendix_self_consistency}). 

The observed relation between $\MBH$ and bulge $K$-band luminosity from \citet{Kormendy2013} for classical bulges and elliptical galaxies is also shown in the lower left panel of \Fig{calibration_diagnostics}. Since we plot total rather than bulge $K$-band luminosity, we expect our results to fall to the right of the observed relation, which is indeed the case. At the brightest end, where most galaxies are elliptical, all of our simulations converge to the normalization of the observed relation.

We conclude that the luminosities, sizes, and BH masses of galaxies in the variable IMF simulations are reasonable, and match the Ref-50 run quite well. This is encouraging, since it means that we do not need to recalibrate the simulations to obtain an acceptable match to the observed luminosity function, galaxy sizes and BH masses, even when including self-consistent feedback.

\subsubsection{Physical diagnostics}
\label{sec:calibration_diagnostics_physical}

While the observable calibration diagnostics are consistent with Ref-50, the same is not necessarily true for the associated physical quantities. In the right-hand column of \Fig{calibration_diagnostics}, we plot the physical quantities corresponding to the diagnostics in the left-hand column, namely: the GSMF, 3D half-mass radii, and $\MBH$ as a function of $\Mstar$ for galaxies in our variable IMF simulations and Ref-50. We cannot compare with observations here, as these typically assume a universal IMF when deriving physical quantities. While it is possible in principle to reinterpret these physical quantities assuming a Chabrier IMF, thus allowing a comparison with observations, we choose to show the true physical quantities here to highlight the importance of IMF assumptions in the translation between observables and true physical quantities.

The GSMF is shown in the top-right panel. At low masses ($\lesssim 10^{10} \Msun$), the variable IMF runs and Ref-50 match very well, owing to the fact that since these galaxies tend to have lower stellar birth ISM pressures, they form stars with an IMF similar to the Chabrier IMF used in the reference model. At higher masses ($\Mstar \gtrsim 10^{11} \Msun$), \him{} is consistent with Ref-50 but \lom{} predicts larger masses at fixed number density by $\approx 0.12$ dex. Since the luminosity function in \lom{} traces Ref-50 nearly perfectly, this difference in their GSMFs is purely due to the increased stellar mass required to produce a fixed luminosity for a bottom-heavy IMF, due to an excess mass of (dim) dwarf stars. Thus, this difference represents the error one incurs when converting K-band luminosity to stellar mass assuming a Chabrier IMF for a galaxy with an intrinsically bottom-heavy IMF. These results are consistent with the recent work of \citet{Clauwens2016} and \citet{Bernardi2018a}, who investigated the effect that bottom-heavy IMF variations in high-mass ETGs would have on the GSMF derived from SDSS galaxies.

Intrinsic sizes are also significantly affected by a variable IMF. In the middle-right panel of \Fig{calibration_diagnostics} we show the 3D half-mass radii of our galaxies as a function of $\Mstar$. As in the middle-left panel of \Fig{calibration_diagnostics}, we show only late-type galaxies. At fixed $\Mstar$, galaxies in \lom{} and \him{} are smaller and larger, respectively, than Ref-50 by $\approx 0.2$ dex at the highest masses. For \lom{}, the smaller physical sizes are due to the excess mass in dwarf stars in the central regions, while high-mass galaxies in \him{} are larger due to a deficit of low- and intermediate-mass stars in these regions which is not quite balanced by the excess mass in stellar remnants, as well as the stronger stellar feedback that tends to yield larger galaxies \citep[e.g.][]{Sales2010, Crain2015}. 

The $\MBH-\Mstar$ relation is plotted in the bottom-right panel of \Fig{calibration_diagnostics}. The relation for \lom{} is shallower than for Ref-50 owing to the increased $\Mstar$ of the most-massive galaxies. The discrepancy for \him{} is similar to that for the $\MBH-M_K$ relation, and is reasonably consistent with Ref-50. 

As a check, we reinterpreted $\Mstar$ and the half-mass radii of these galaxies assuming a Chabrier IMF by multiplying their K-band luminosities by the $M/L$ ratio they would have had if they had evolved with a Chabrier IMF. We refer to this reinterpreted mass as $\MstarChab$. Doing so puts the masses, sizes, and BH masses of the \lom{} galaxies into excellent agreement with Ref-50 by decreasing the inferred masses of high-mass galaxies by a factor $\approx 2$, but has little effect for \him{} (not shown).

\begin{figure}
  \centering
\includegraphics[width=0.45\textwidth]{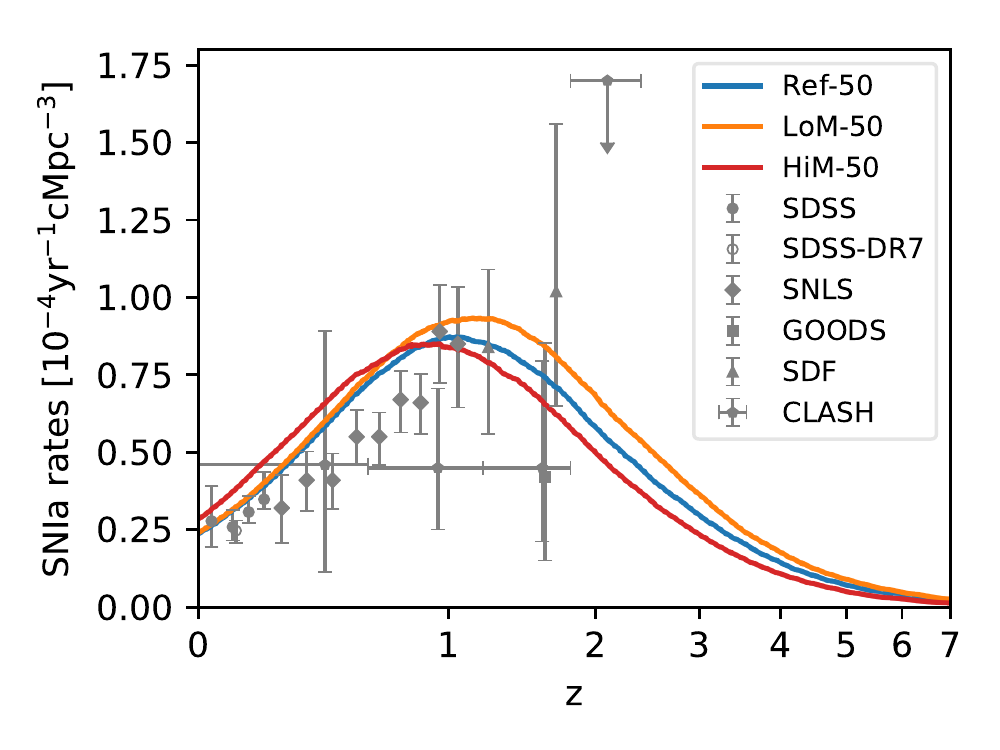}
\caption{Cosmic supernova Ia rate density as a function of redshift in the \lom{} (orange), \him{} (\HimColour{}) and Ref-50 (blue) simulations. Grey data points show observations compiled by and classified as the ``most accurate and precise measurements'' by \citet{Graur2014}: SDSS Stripe 82 \citep{Dilday2010}, SDSS-DR7 \citep{Graur2013}, SNLS \citep{Perrett2012}, GOODS \citep{Dahlen2008}, SDF \citep{Graur2014}, and CLASH \citep{Graur2014}. The 1$\sigma$ systematic and statistical uncertainties are indicated with error bars. The SNIa rates are not strongly affected in the variable IMF simulations and match observations about as well as the reference model.  }
\label{fig:snia_evolution}
\end{figure}

Finally, we investigate the SNIa rate evolution in the simulations. As mentioned in \Sec{mods}, the predicted rates do not explicitly depend on the IMF, but they are affected by the star formation history of the simulations. \Fig{snia_evolution} compares the evolution of the SNIa rate density with observations compiled by \citet{Graur2014}. The rates are not strongly affected in either variable IMF simulation, where the agreement with observations is about as good as it is for Ref-50. This is encouraging, since a strong deviation from the observed rates would require a recalibration of the empirical time-delay function. All differences here are due to the effect that the IMF has on the star formation history, which is slightly delayed in \him{} relative to Ref-50.

\section{Effect of variable IMFs on galaxy properties}
\label{sec:IMF_effect}

In this section we investigate the effect that our variable IMF prescriptions have on the predicted properties of galaxies in the (50 Mpc)$^3$ self-consistent variable IMF simulations relative to Ref-50 and observations. Specifically, Sections \ref{sec:alpha_enhancement}, \ref{sec:metallicities}, \ref{sec:SFRs}, and \ref{sec:ETG_sizes} discuss the effect on alpha enhancement, metal abundances, star formation rates, and ETG sizes, respectively. All results are shown for galaxies at $z=0.1$; properties at higher redshift will be discussed in a future work. 

\subsection{Alpha enhancement}
\label{sec:alpha_enhancement}

\begin{figure}
  \centering
\includegraphics[width=0.5\textwidth]{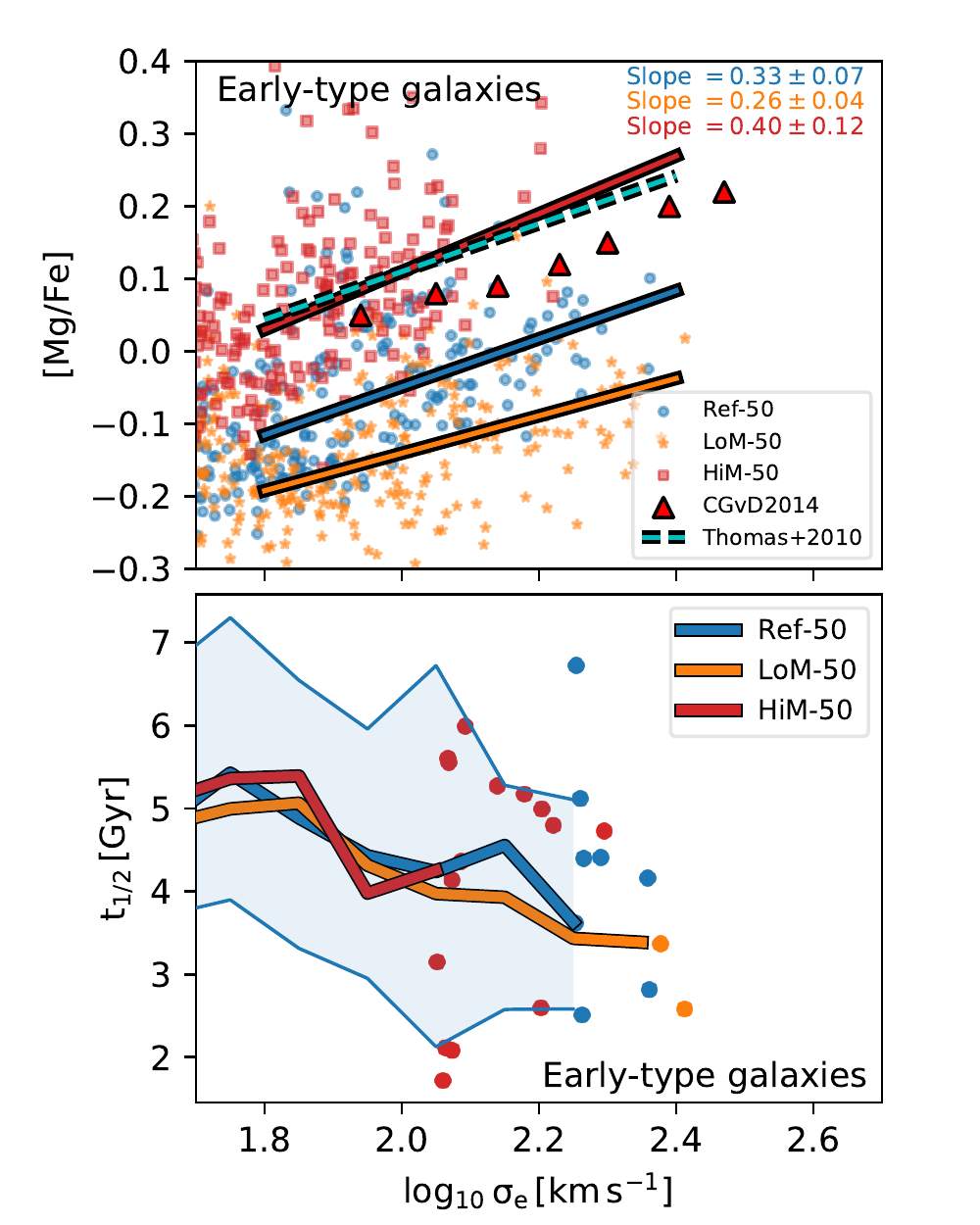}
\caption{{\bf Top panel:} Stellar [Mg/Fe] as a function of stellar velocity dispersion for galaxies at $z=0.1$ in \lom{} (orange stars) and \him{} (\HimColour{} squares) compared to Ref-50 (blue circles).  All quantities are SDSS $r$-band light-weighted and measured within the projected half-light radius, $r_e$. To facilitate a fairer comparison with observations, included are only early-type galaxies ($u^*-r^*>2.0$). The observed trend for ETGs from \citet{Thomas2010} is shown as a dashed cyan line, while the observed trend for quiescent galaxies from \citet{Conroy2014} is shown as red triangles.  Least absolute deviation fits for early-type galaxies with $\sigma_e > 10^{1.8}\kms$ in each simulation are shown as solid coloured lines, with slopes labeled.  The [Mg/Fe] abundances in \lom{} (\him{}) are normalized lower (higher) relative to Ref-50 for high-mass galaxies. While the slopes of the [Mg/Fe]-$\sigma$ relation for both Ref-50 and \him{} are consistent with observations, that for \lom{} is somewhat shallower. {\bf Bottom panel:} Median formation time of stars within $r_e$ for the same galaxy samples as the upper panel. The positive [Mg/Fe]$-\sigma_e$ correlation reflects the star formation histories of early-type galaxies.}
\label{fig:alpha_abundances}
\end{figure}

One of the strongest effects that the variable IMF has in the simulations is on the abundances of $\alpha$-elements in high-$\sigma_e$ galaxies. In the upper panel of \Fig{alpha_abundances} we show [Mg/Fe] as a function of $\sigma_e$ for early-type galaxies (defined as those with $u^*-r^*>2.0$) in our variable IMF simulations.

\citet{Segers2016} have already shown that the trend of $\alpha$-enhancement with stellar velocity dispersion in the reference simulations agrees well with observations of quiescent ETGs although the normalization of the relation is too low by about a factor of 2. We note that nucleosynthetic yields are uncertain at the factor of 2 level \citep[e.g.][]{Wiersma2009b}, while uncertainties in stellar population modelling lead to systematic errors in the observed values of [$\alpha$/Fe] of $\sim 0.1$ dex. Thus, the normalization of the trend is not nearly as constraining as the slope. Least absolute deviation fits to these galaxies above $\sigma_e > 10^{1.8} \kms$ yield slopes of $0.33 \pm 0.07$, $0.26 \pm 0.04$, and $0.40 \pm 0.12$ for Ref-50, \lom{}, and \him{}, respectively. While the slopes for Ref-50 and \him{} agree with the observed values of $0.33 \pm 0.01$ \citep{Thomas2010} and $0.33 \pm 0.03$ \citep{Conroy2014}, that for \lom{} is somewhat shallower.

We have investigated the cause of the difference in the normalization and slopes of these trends, and find that the culprit is the difference in the metal yields in the different simulations. In high-mass galaxies, \lom{} produces fewer massive stars per unit stellar mass formed and thus less Mg is synthesized for future generations of stars, leading to reduced [Mg/Fe]. On the other hand, \him{} produces many more massive stars per unit stellar mass formed, increasing the Mg yields. By independently switching on the different effects of the variable IMF in smaller (25 Mpc)$^3$ boxes, we confirmed that indeed the yields, rather than the feedback or the KS law re-normalization, drive these differences. While the differences in the normalization are not large, it is concerning that the [Mg/Fe] values in \lom{} fall so far below those from observations, especially since, following  \citet{Portinari1998}, the Mg yields have already been doubled in the reference model with respect to the standard yields.

It is interesting that our IMF variations have such little effect on the slope of the [Mg/Fe]$-\sigma_e$ relation. Recent studies of IMF variations in SAMs have concluded that a top-heavy IMF in rapidly star-forming environments (which occurs in the high-redshift progenitors of present-day high-mass ETGs) may be necessary to produce a positive correlation in this relation, a result of higher Mg abundances due to a larger number of SNII \citep{Gargiulo2015b}. In the EAGLE model, a variable IMF is not necessary to reproduce this slope, as the positive trend comes from the quenching of star formation via AGN feedback, preventing much of the Fe from type Ia SNe from being incorporated into future stellar populations \citep{Segers2016}. It is thus encouraging that the slope does not become even steeper with a top-heavy IMF in EAGLE. Although the difference in Mg abundance between \him{} and Ref-50 increases with $\sigma_e$, the same is true for the Fe abundances, maintaining the slope of the [Mg/Fe]$-\sigma_e$ relation. 

To help support this quenching scenario, in the lower panel of \Fig{alpha_abundances} we plot the median formation time, $t_{1/2}$, of stars within the 2d projected half-light radius, $r_e$, of the same early-type galaxies shown in the upper panel. Here we see a trend of decreasing $t_{1/2}$ with increasing $\sigma_e$ for all three simulations, in qualitative agreement with recent results from the ATLAS$^{3D}$ survey \citep{McDermid2015}. These results support the idea that short star formation histories leads to higher [Mg/Fe] ratios in early-type galaxies.

This quenching scenario is also supported by semi-analytic modelling performed by \citet{DeLucia2017}, although they find that abrupt quenching of high-mass galaxies prevents them from reaching high enough metallicities at $z=0$, requiring a variable IMF to match both the mass-metallicity relation and the alpha-enhancement of high-mass galaxies simultaneously \citep[see also][]{Arrigoni2010}. We will show in the next section that our \him{} simulation does indeed match the slope of the observed mass-metallicity relation better than the reference model.

\subsection{Metallicities}
\label{sec:metallicities}

\begin{figure}
  \centering
\includegraphics[width=0.45\textwidth]{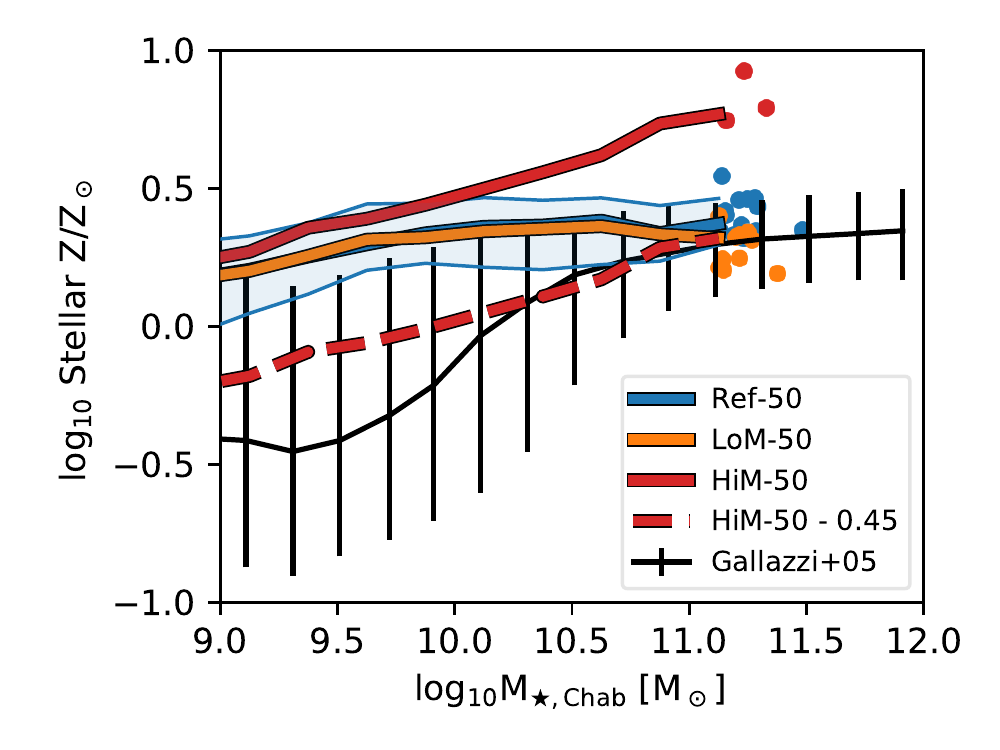}
\includegraphics[width=0.45\textwidth]{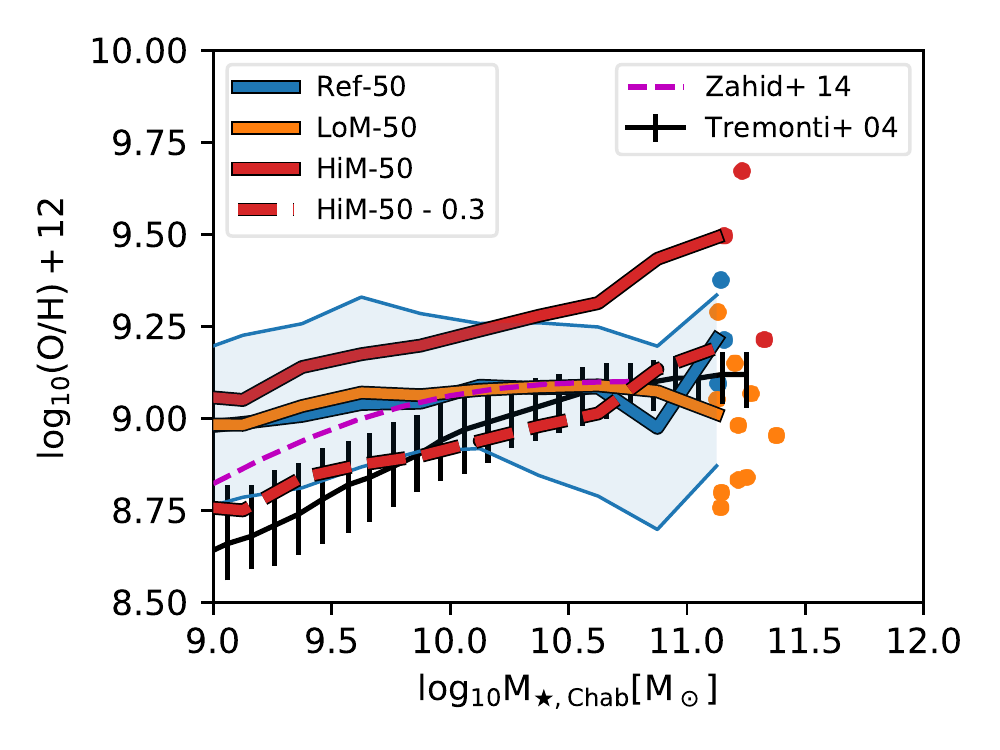}
\caption{Metal abundances in \lom{} (orange) and \him{} (\HimColour{}) compared to Ref-50 (blue) as a function of $\MstarChab$ for all galaxies at $z=0.1$. Medians are indicated by solid lines and filled regions show the 10th to 90th percentile ranges. Abundances are measured in projection within one effective radius, $r_e$, while $\MstarChab$ is measured within a 30 kpc 3D aperture. Top: Stellar metallicity. Observations reported by \citet{Gallazzi2005} are shown as the black solid line with error bars indicating the scatter.  Bottom: Gas-phase oxygen abundance for star-forming gas.  For each galaxy, abundances are measured for all bound star-forming gas particles. The fit to observations from \citet{Zahid2014} is plotted as a magenta dashed line. Observations from \citet{Tremonti2004} are shown as black error bars. To ease comparison of the slopes with observations, the thick, red dashed curve lowers the trend for \him{} in both panels to match the normalization of the observations at $\MstarChab \approx 10^{11}\Msun$. Stellar and gas-phase metallicities in high-mass galaxies in \lom{} are consistent with Ref-50 but rise $\approx 0.4$ dex higher for \him{} at the highest masses, in tension with the flattening in the observed relation above $10^{10}\Msun$. }
\label{fig:metals}
\end{figure}

We also investigate the effect of the variable IMFs on the stellar mass-metallicity relation where, in the top panel of \Fig{metals}, we plot the stellar metallicity, $Z$, measured within $r_e$, as a function of Chabrier-reinterpreted stellar mass, $\MstarChab$, for all galaxies with $\MstarChab > 10^8 \Msun$. Interestingly, the total metallicities are largely unchanged in the \lom{} run relative to Ref-50, with only a slight ($< 0.1$ dex) decrease in $Z$ for $\MstarChab > 10^{11}\Msun$.  However, for \him{}, the mass-metallicity relation is much steeper, with $Z/Z_\odot \approx 0.4$ dex higher in high-mass ($\MstarChab \sim 10^{11} \Msun$) galaxies. This difference comes from the much higher production of metals per unit stellar mass formed from an IMF with a shallow high-mass slope. 

We compare the simulated trends with the observed relation for SDSS galaxies from \citet{Gallazzi2005}. For $\MstarChab > 10^{10.5}\Msun$, the mass-metallicity relation is observed to flatten off, as is also seen in Ref-50 and \lom{}, albeit at lower mass than observed. \him{}, on the other hand, shows no sign of a saturating metallicity at high mass, which may be in tension with observations. \citet{Zahid2014} argue that this saturation occurs when the gas-phase abundances are high enough that, during star-formation, the metal mass that is removed from the ISM and locked up into low-mass stars is comparable to that produced and released back into the ISM by high-mass stars. \him{} keeps increasing in metallicity because for a top-heavy IMF, many more supernovae are produced per dwarf star formed, producing more metals than are being locked up, delaying saturation to higher metallicities. Note, however, that \citet{DeRossi2017} have shown that the mass-metallicity relation in EAGLE saturates in part due to AGN feedback quenching star formation and ejecting metal-rich gas out of high-mass galaxies.  This implies that the \citet{Zahid2014} explanation is at least incomplete. Indeed, we will show in \Sec{SFRs} that galaxies with $\Mstar \gtrsim 10^{11}\Msun$ have typically higher SFRs in \him{} than in Ref-50, possibly contributing to their higher metallicities.

Given the large uncertainties in the simulated nucleosynthetic yields \citep{Wiersma2009b} and the calibration of the metallicity indicators applied to observations \citep[e.g. ][]{Kewley2008}, the slope of the mass-metallicity relation is more constraining than its normalization. Because our (50 Mpc)$^3$ simulations only contain galaxies up to $\MstarChab \sim 10^{11}\Msun$, it is not clear if the lack of flattening at the high-mass end is actually inconsistent with the observations. To facilitate visual comparison of the slopes, we lower the normalization of the \him{} relation until the high-mass median value matches the value from \citet{Gallazzi2005}, which we show as a dashed-\HimColour{} line in \Fig{metals}. Indeed, the reduced \him{} trend agrees with \citet{Gallazzi2005} better than is the case for Ref-50 or \lom{}, which begin to flatten at lower stellar masses than observed galaxies. This result is also in qualitative agreement with some SAMs that find that the slope of the mass-metallicity relation and the alpha-enhancement can be simultaneously reproduced with similar top-heavy IMF variations \citep{DeLucia2017}. The better agreement may also be related to the stellar feedback, as S15 found that the higher efficiency of star formation required to match the GSMF in the high-resolution simulation Recal-L025N0752 resulted in stronger outflows that decreased the metallicity of the ISM enough to better match the the observed mass-metallicity relations. Nevertheless, this rescaling is inconsistent with the radiative cooling rates in the simulation, so yields in \him{} would need to be rescaled in the simulation input parameters to verify that the better agreement of \him{} persists.

It is unclear how stellar metallicities derived from spectroscopic observations depend on the assumed IMF. Because the observationally-inferred gas-phase metallicity may be less sensitive to the assumed IMF, we plot in the lower panel of \Fig{metals} the mass-weighted gas-phase oxygen abundances for star-forming gas in our variable IMF simulations as a function of $\MstarChab$ \citep[although gas-phase abundances may still be affected by IMF assumptions in observational studies, see e.g.][]{Paalvast2017}. We compare our simulations with the observations of \citet{Tremonti2004} and the fit to observations of \citet{Zahid2014} for $z=0.1$. Note that we again only focus on the slope, rather than the absolute values, as a function of mass. Our trends here are qualitatively similar to those found for the stellar metallicities in the top panel of \Fig{metals}, with perhaps more scatter. \him{} continues increasing toward high masses while metallicities in Ref-50 and \lom{} tend to flatten off above $10^{10} \Msun$. Lowering the \him{} trend by 0.3 dex brings it into reasonable agreement with \citet{Tremonti2004}, and yields a better match than Ref-50 or \lom{}. Again, self-consistent rescaling of the yields would be required to confirm this better agreement. We conclude that none of the models are ruled out by the observed trends between $\alpha$-enhancement or metallicity and mass. For \him{} the absence of any flattening in the mass-metallicity relation is in tension with observations but larger volume simulations are required to judge the severity of the problem.

\subsection{Star formation}
\label{sec:SFRs}

\begin{figure*}
  \centering
\includegraphics[width=0.95\textwidth]{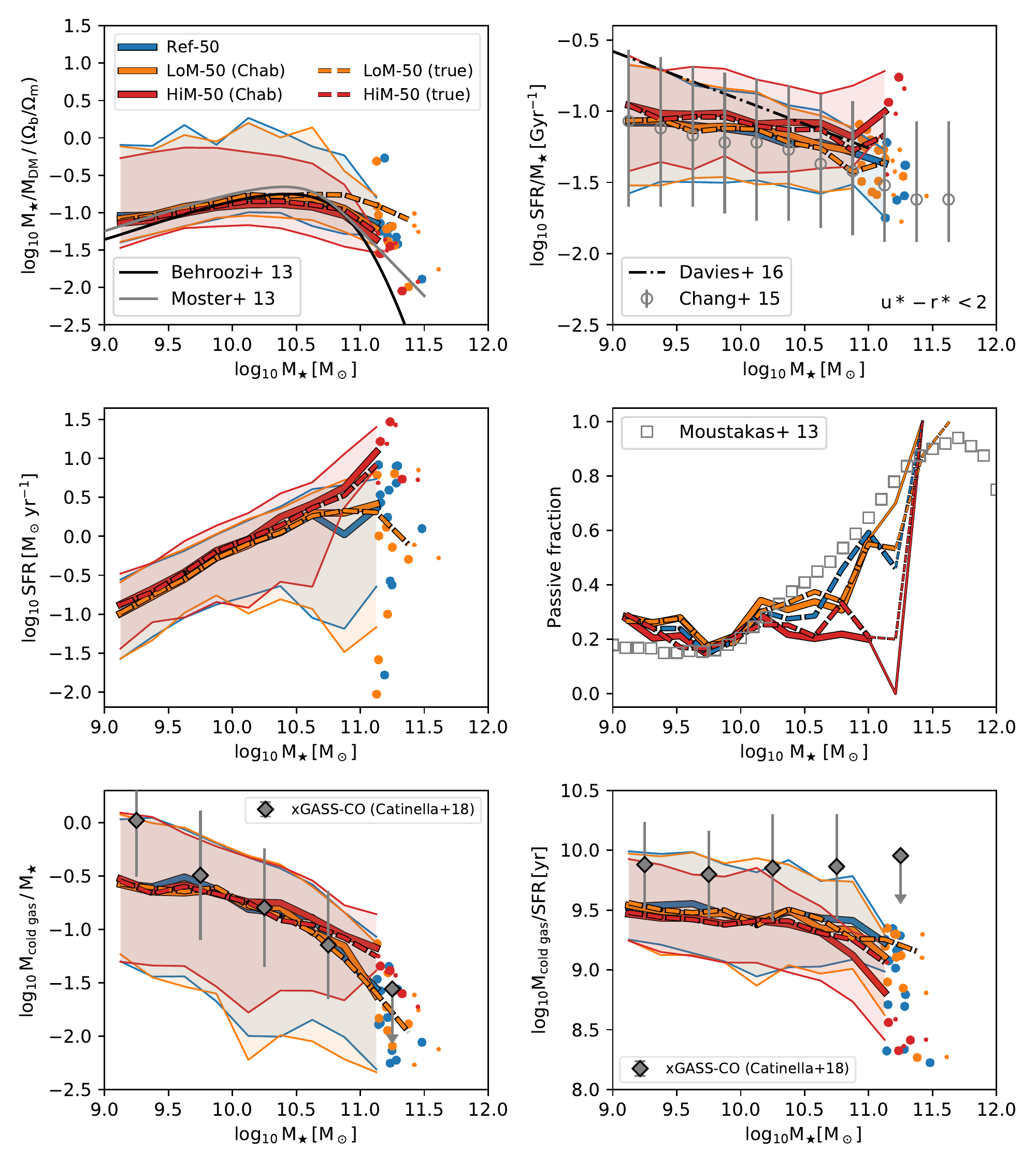}
\caption{ Star formation properties of all galaxies with $\MstarChab > 10^{9}\Msun$ at $z=0.1$ as a function of stellar mass in our \lom{} (orange) and \him{} (\HimColour{}) simulations, compared with Ref-50 (blue). Solid lines show results when reinterpreting the $\Mstar$ and SFR assuming a Chabrier IMF, while dashed lines show true values. In reading order from top-left to bottom-right: Galaxy formation efficiency, specific star formation rate (sSFR) for star-forming galaxies, SFR for all galaxies, passive fractions, neutral hydrogen fractions, and gas consumption timescale. Star-forming galaxies are defined as those with intrinsic $u^*-r^*<2$. Medians are indicated as thick lines and shaded regions mark the 10-90$^{\rm th}$ percentiles for the Chabrier-inferred lines. For bins with fewer than 10 galaxies, we show individual galaxies, where large and small dots refer to Chabrier-inferred and true values, respectively; for passive fractions these bins are indicated with thinner lines. Thin solid black and grey lines show abundance matching results from \citet{Behroozi2013} and \citet{Moster2013}, respectively. Grey circles show the median sSFR for SDSS+WISE galaxies from $z=0$ to $0.2$ with $\rm{sSFR} > 10^{-2}\,{\rm Gyr}^{-1}$, with error bars showing the 10-90$^{\rm th}$ percentiles \citep{Chang2015}. The black dash-dotted line in the upper right panel shows a fit to the observed sSFRs for star-forming galaxies from $z=0$ to $0.1$ from the GAMA survey \citep{Davies2016}. Passive fractions for SDSS galaxies are shown as grey squares \citep{Moustakas2013}. Median neutral hydrogen fractions and gas consumption timescales from the xGASS and xCOLDGASS surveys are shown as grey diamonds \citep{Catinella2018}; error bars denote 1$\sigma$ scatter while arrows indicate upper limits on the median. The results for \lom{} are consistent with Ref-50, but for \him{} the SFRs are too high for $\MstarChab > 10^{11}\Msun$, resulting in lower passive fractions at high mass. These higher SFRs in \him{} are due to a higher gas fraction in high-mass galaxies.}
\label{fig:star_formation}
\end{figure*}

As mentioned in earlier sections, star formation rates can be strongly affected by the IMF, both the SFRs inferred from (virtual) observations as well as the true SFRs in the simulations due to the IMF's effect on metallicity and stellar feedback. \Fig{star_formation} details the effect of these variable IMF prescriptions on star formation in the simulations as a function of stellar mass. We show both the true quantities (dashed lines) and the corresponding values after interpreting the $\Mstar$ and SFR from the K-band and FUV light, respectively, under the assumption of a Chabrier IMF (solid lines). The reinterpreted SFR, $\SFRChab$, is computed by multiplying the actual SFR by the ratio between the Galex FUV-band luminosity of the stars and that they would have had if evolved with a Chabrier IMF, as also done by \citet{Clauwens2016}. We note that this modifies the SFR by only $\sim 10$ per cent for most galaxies, leading to nearly identical trends of true and Chabrier-reinterpreted SFR with $\Mstar$. Note as well that only the Chabrier-interpreted values (solid lines) should be compared with the observations.

The top-left panel of \Fig{star_formation} shows the ``galaxy formation efficiency'' of galaxies for our variable IMF simulations, defined as the ratio $\MstarChab/\MDM$ normalized to the cosmic baryon fraction. Both variable IMF simulations agree with Ref-50 (solid thick lines), as well as with abundance matching results from \citet{Behroozi2013} and \citet{Moster2013} shown as thin solid lines. For \lom{}, the true efficiency is shifted toward higher values for $\MstarChab > 10^{11}\Msun$ (dashed orange line), where the excess stellar mass originates from dwarf stars. Interestingly, the true efficiencies for \him{} galaxies are quite close to their Chabrier-interpreted values. This implies that any correction to the observed galaxy formation efficiencies to account for IMF variations depends sensitively on how such variations are parameterized.

The specific star formation rates (sSFR), measured within 3D spherical apertures of radius 30 kpc, are shown in the upper right panel of \Fig{star_formation} for star-forming galaxies, defined as those with $u^*-r^*<2$. In all simulations, the sSFRs are consistent for $\MstarChab < 10^{10.5} \Msun$. At higher masses, sSFRs turn upwards for \him{}, while they decrease for \lom{} and Ref-50. The decrease in sSFR is consistent with the observed relations for local star-forming galaxies from the GAMA survey \citep{Davies2016} and SDSS+WISE \citep{Chang2015}\footnote{We show results for galaxies from their online catalogue with FLAG=1, ${\rm sSFR} > 10^{-2}\,{\rm Gyr}^{-1}$, and $\Mstar$ above their mass completeness limit.}, while the upturn at high mass in \him{} is in tension with the observations. This upturn is not as strong for the true sSFR (red-dashed line), implying that the discrepancy for \him{} is partially due to reinterpreting the SFRs assuming a Chabrier IMF.

These higher sSFRs in \him{} are due to a lack of quenching at high masses. This can be seen in the middle left panel of \Fig{star_formation}, where we show total $\SFRChab$ as a function of $\MstarChab$ for all galaxies at $z=0.1$. Here it can be seen that in both Ref-50 and \lom{}, the positive trend between $\SFRChab$ and $\MstarChab$ flattens at $\MstarChab > 10^{10.5}\Msun$ with a large amount of scatter toward low $\SFRChab$ as galaxies are quenched. However, in \him{}, the trend becomes even steeper with less scatter at high mass. This behaviour leads to a lower passive fraction at high mass for \him{} galaxies, as shown in the middle right panel of \Fig{star_formation}, where we define the passive fraction as the fraction of galaxies with intrinsic $u^*-r^*>2$ in bins of $\Mstar$. These low passive fractions in \him{} contrast with the increasing passive fraction with increasing $\Mstar$ seen in observations of SDSS galaxies \citep{Moustakas2013}, as well as Ref-50 and \lom{}. 

In the lower-left panel of \Fig{star_formation}, we show that these higher SFRs in \him{} galaxies are a consequence of a larger cold gas fraction. We compute cold gas masses within apertures of 70 pkpc following \citet{Crain2017}, where for each gas particle, we compute the mass of neutral hydrogen following the prescription of \citet{Rahmati2013}, accounting for self-shielding and assuming a \citet{Haardt2001} ionizing UV background.\footnote{As in \citet{Catinella2018}, to account for helium we multiply the neutral hydrogen mass by 1.3 to obtain total cold gas mass.} While the ratio between star-forming gas mass and $\MstarChab$ decreases steeply with increasing $\MstarChab$ in Ref-50 and \lom{}, this fraction drops less steeply in \him{}, lying $\approx 0.5\,$dex above Ref-50 at $\MstarChab \gtrsim 10^{11}\Msun$. The higher cold gas fractions in \him{} galaxies at $z=0.1$ are likely due to burstier stellar feedback ejecting more gas out of galaxies at high-$z$. This effect causes a delay in the peak of star formation -- we will investigate the time-dependent properties of galaxies in these variable IMF simulations in a future paper.

We compare our results in the lower-left panel of \Fig{star_formation} with the median cold gas fractions of galaxies from the mass-selected xGASS and xCOLD GASS surveys \citep{Catinella2018}. These observations match all of our simulations well for $\MstarChab > 10^{9.5} \Msun$ but are too high in the lowest-mass bin. This low-mass tension is likely due to the fact that atomic hydrogen masses are not converged in the ``intermediate'' resolution EAGLE model, especially in the range $\MstarChab \sim 10^{9-10}\Msun$ where they are lower by nearly an order of magnitude relative to higher-resolution models \citep[see][]{Crain2017}. Thus, the use of higher-resolution simulations may resolve this tension with observations for low-mass galaxies. 

The SFR change is not due to the re-normalization of the star formation law (see \Sec{mods}). To show this, we plot the gas consumption timescale, parametrized by the cold gas mass divided by the SFR, in the lower-right panel of \Fig{star_formation} for all galaxies in each of our simulations. In all of our simulations the true value of this timescale (dashed lines) is quite constant and is barely affected by the variable IMF (by at most 0.1 dex). We thus conclude that a higher cold gas mass fraction, rather than the renormalization of the star formation law, is responsible for the higher SFRs in high-mass \him{} galaxies. We also compare these results with the observed xGASS-CO sample, who also find a nearly constant relation. There is, however, a systematic $\approx 0.4$ dex offset toward lower gas consumption timescales relative to the observations. The reason for this offset is unclear, but is consistent with the expected systematic uncertainties associated with different SFR calibrators (compare results from \citealt{Davies2016} and \citealt{Chang2015} in the upper right panel of \Fig{star_formation}).

\subsection{ETG galaxy sizes}
\label{sec:ETG_sizes}

\begin{figure*}
  \centering 
\includegraphics[width=0.45\textwidth]{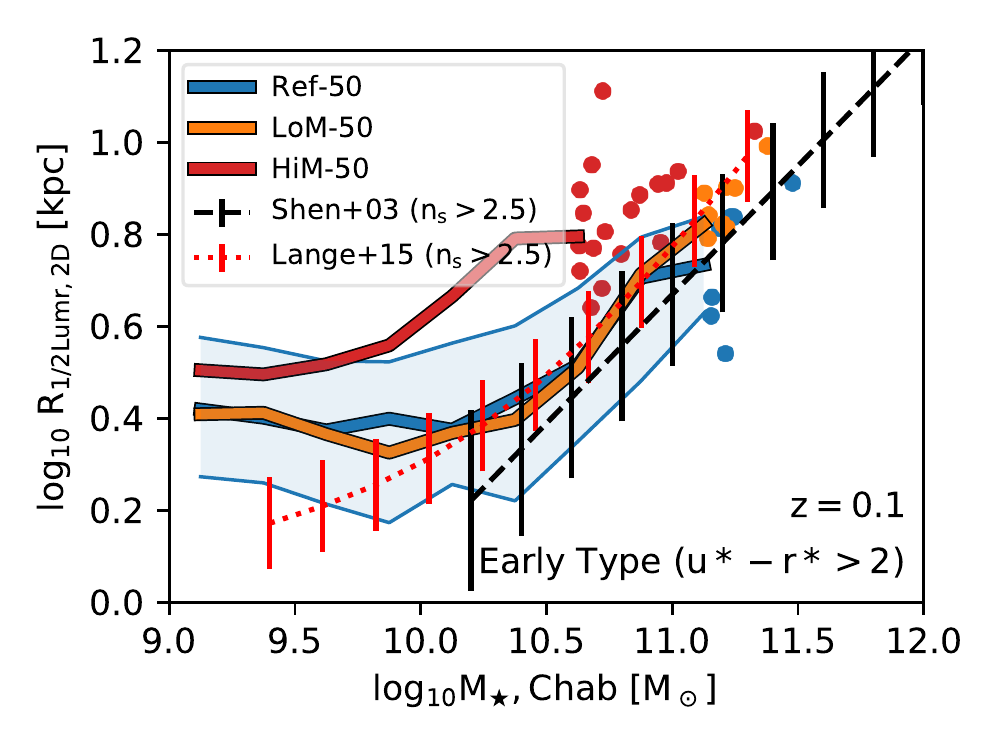}
\includegraphics[width=0.45\textwidth]{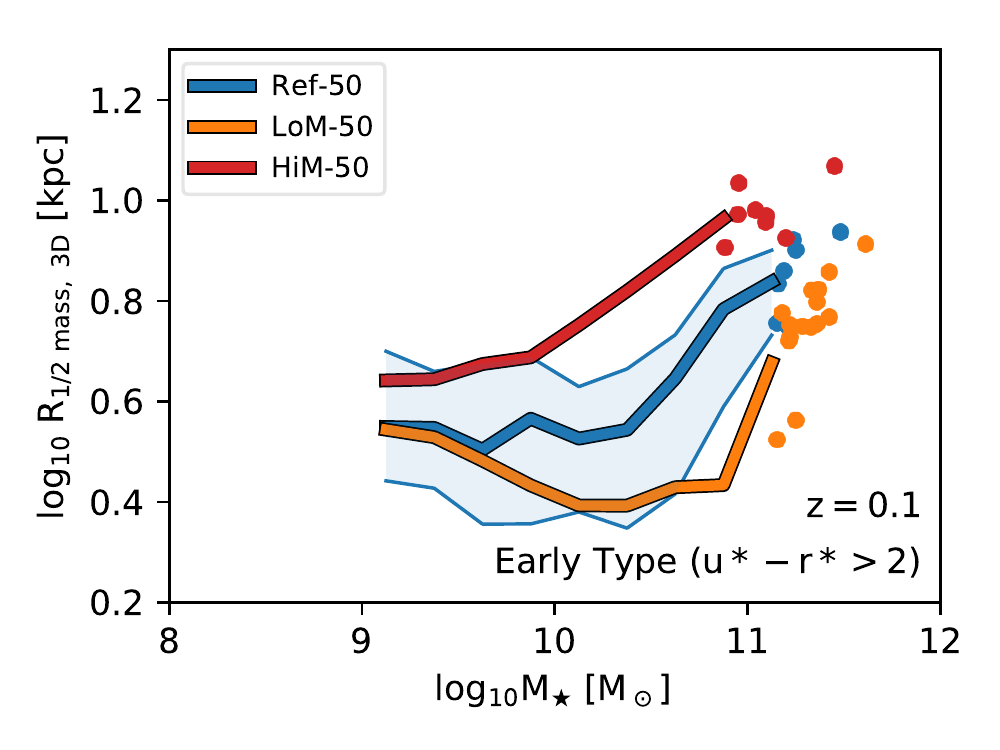}
\caption{ Size-$\Mstar$ relation for early-type galaxies, defined as those with intrinsic $u^*-r^*>2$, in Ref-50 (blue), \lom{} (orange), and \him{} (\HimColour{}) at $z=0.1$. Left panel: 2D projected $r$-band half-light radius as a function of Chabrier-reinterpreted stellar mass, $\MstarChab$. The black-dashed and red-dotted lines with error bars show the observed trend from SDSS and GAMA, respectively, for galaxies with Sersic index $n_s > 2.5$ \citep{Shen2003, Lange2015}. For the GAMA data we assume errors of 0.1 dex based on visual inspection of their 25-75$^{\rm th}$ percentiles, while for the SDSS data we assume the errors are the same as those for their same relation for late-type galaxies. Right panel: Physical 3D half-mass radius as a function of $\Mstar$. While the half-light radii of ETGs in Ref-50 and \lom{} are consistent with observations, ETGs in \him{} are larger at fixed $\MstarChab$ by $\approx 0.2-0.4$ dex. As is the case for late-type galaxies (see middle row of \Fig{calibration_diagnostics}), \lom{} ETGs are physically smaller by $\approx 0.2-0.3$ dex at fixed $\Mstar$ relative to Ref-50.}
\label{fig:sizes_ETGs}
\end{figure*}

We showed in \Sec{calibration_diagnostics_observable} that the observed sizes of late-type galaxies are reproduced in our variable IMF simulations. We now investigate if this agreement persists for early-type galaxies as well. In the left panel of \Fig{sizes_ETGs}, we plot the half-light radius, $r_e$, as a function of $\MstarChab$ for non-star-forming galaxies, defined as those with $u^*-r^*>2$. While the relation for \lom{} agrees well with Ref-50, \him{} ETGs are strongly offset to larger sizes at fixed $\MstarChab$. 

The reasons for this offset in \him{} are the same as those responsible for the smaller $\approx 0.1$ dex offset seen for late-type galaxies in \Sec{calibration_diagnostics_observable}. Stronger stellar feedback, later formation times, and the fact that the luminosities of the central, top-heavy regions of the galaxies are dimmer relative to a Chabrier IMF than the less top-heavy outskirts, all inflate the sizes at fixed $\MstarChab$ relative to Ref-50. These effects are exacerbated in ETGs due to the fact that they are older, since old, top-heavy stellar populations are much dimmer than those with a Chabrier IMF, causing ETGs to shift toward lower $\MstarChab$ at fixed $r_e$. 

For comparison, we plot the observed relations from SDSS \citep{Shen2003} and the GAMA survey \citep{Lange2015} for $z<0.1$ galaxies with Sersic index $n_s > 2.5$. We multiply the \citet{Shen2003} sizes by a factor 1.075 to convert from $z$- to $r$-band half-light radii \citep[see][]{Kelvin2012, Lange2015}. Both \lom{} and Ref-50 match the observed relations well for $\MstarChab > 10^{10}\Msun$, but \him{} ETGs are too large by $\approx 0.2-0.3$ dex, in tension with the observations.

In the right panel of \Fig{sizes_ETGs} we plot the true 3D half mass radius as a function of true $\Mstar$ for the same galaxies. As was the case for star-forming galaxies in \Fig{calibration_diagnostics}, the physical sizes of high-mass ETGs in \lom{} are smaller than in Ref-50 by $\approx 0.2-0.3$ dex at fixed $\Mstar$ due to higher mass fractions of dwarf stars in their central regions. The positive offset in \him{} galaxies seen in the observable diagnostics is also present in the physical ones since, as with the light, the masses of the central, old stellar populations with top-heavy IMFs are lower than those with a Chabrier IMF, pushing these galaxies to lower $\Mstar$ and larger half-mass radii.

\section{Summary and Conclusions}
\label{sec:conclusions}

We have modified the reference EAGLE cosmological, hydrodynamical simulations to self-consistently include a prescription for a stellar initial mass function (IMF) that varies per-star particle as a function of the ISM pressure at which it was formed. Two prescriptions are explored: in both cases we begin with the Kroupa double-power law IMF and vary the slope either below or above $0.5 \Msun$ (hereafter referred to as LoM and HiM, respectively; see \Fig{IMF}). For each prescription the dependence of the slope on birth ISM pressure was calibrated such that the observed \citet[][hereafter C13]{Cappellari2013b} trend of increasing excess stellar mass-to-light ($M/L$) ratio with central stellar velocity dispersion, $\sigma_e$, is roughly recovered. For LoM this recovery is accomplished by an increase in the fraction of low-mass stars, while for HiM the increasing mass fraction in stellar remnants and decreased luminosity are responsible. 

The calibration of the pressure-dependence of the IMF was performed by post-processing the (100 Mpc)$^3$ reference EAGLE simulation. From this post-processing procedure, we have found that:

\begin{itemize}
\item
In order to reproduce the observed trends between the stellar $M/L$ ratio excess (MLE) relative to that expected for a Salpeter IMF (eq. \ref{eqn:MLE}), and central stellar velocity dispersion ($\sigma_e$), LoM and HiM must respectively become more bottom- and top-heavy in higher-pressure environments (although this may also be possible with a ``top-light'' HiM prescription, see \Sec{calibration}). Since ISM pressures typically decrease with the age of the universe \citep[e.g.][]{Crain2015}, these IMF prescriptions are implicitly time-dependent.
\item
The MLE is only an excellent proxy for the IMF (i.e. independent of age and metallicity) when the high-mass slope is close to the ``reference" IMF (which here is Salpeter). For IMFs with shallower high-mass slopes, the MLE becomes strongly age-dependent for stars less than a few Gyr old (\Fig{IMF}).\\
\end{itemize}

We ran two new (50 Mpc)$^3$ simulations with the same physics and resolution as the reference EAGLE model, but now each including one of our variable IMF prescriptions, which we refer to as \lom{} and \him{}, respectively. These simulations use variable nucleosynthetic yields, a star formation law, and stellar feedback consistent with their locally varying IMFs. Our conclusions are as follows:
\begin{itemize}
\item
Both variable IMF simulations are broadly consistent with the observed trend between MLE and $\sigma_e$ of C13 (\Fig{IMF_vs_sigma}). However, the trend in \him{} is less clear than for \lom{} due to a lack of high-$\sigma_e$ galaxies in the former, likely caused by burstier stellar feedback.
\item
Galaxies in \lom{} are consistent with the increasing fraction of dwarf stars toward higher $\sigma_e$ in early-type galaxies inferred by spectroscopic IMF studies, while those in \him{} do not show such a trend (\Fig{IMF_vs_sigma}).  On the other hand, star-forming galaxies in \him{} show increasing ratios of ionizing flux to FUV flux with increasing $r$-band surface brightness, in agreement with recent observations, while \lom{} galaxies show no trend (\Fig{highmassslope}). It is unclear how to reconcile these apparently conflicting observations.
\item
Relative to Ref-50, stellar ages and birth ISM pressures are largely unchanged in \lom{}, while \him{} produces younger stars at lower birth ISM pressures on average. This change may be due to the stronger stellar feedback from a top-heavy IMF (\Fig{star_props}).
\item
Observational proxies for the EAGLE subgrid calibration diagnostics (galaxy $K$-band luminosity function, $r$-band half-light radii ($r_e$) of late-type galaxies (LTGs), and black hole masses) are consistent with the reference model for \lom{}. The same is true for \him{}, except that the $r_e$ of LTGs in \him{} are larger, but only by $\approx 0.1$ dex, at fixed $r$-band luminosity (left column of Fig. \ref{fig:calibration_diagnostics}). 
\item
Stellar masses and LTG half-mass radii are larger and smaller, respectively, by $\approx 0.1-0.2$ dex in \lom{} relative to Ref-50 (right column of Fig. \ref{fig:calibration_diagnostics}). This difference is due to an excess of (dim) dwarf stars that increase the mass (but not the light) in the central regions of these galaxies as a result of the bottom-heavy IMF.  \\
\end{itemize}

We also investigated the effect that the IMF has on predicted global galaxy scaling relations. Our results are as follows:
\begin{itemize}
\item
While the slopes of the [Mg/Fe]-$\sigma_e$ relation in the Ref-50 and \him{} simulations for high-mass ($\sigma_e > 60\kms$), early-type ($u-r>2$) galaxies ($0.33 \pm 0.07$ and $0.40 \pm 0.12$, respectively) are consistent with the observed relation, \lom{} produces a relation with slightly (but significantly) shallower slope of $0.26 \pm 0.04$, compared to observed values $0.33 \pm 0.01$ \citep{Thomas2010} and $0.33 \pm 0.03$ \citep{Conroy2014} (\Fig{alpha_abundances}). The normalization of the [Mg/Fe]$-\sigma_e$ relation in \him{} is $\approx 0.15$ dex higher than in Ref-50, bringing it into agreement with observations, while it is $\approx 0.05-0.1$ dex lower in \lom{}, $\approx 0.2$ dex below the observed relation. However, given the large systematic uncertainties in the normalization for both observations and simulations, these differences in normalization are not significant. 
\item
Stellar and gas-phase metallicities in \lom{} are consistent with Ref-50, but in \him{} both quantities increase steeply with $\MstarChab$, the stellar mass inferred under the assumption of a Chabrier IMF, above $10^{10}\Msun$ with no sign of flattening at higher mass. This is contrary to the flattening seen in Ref-50 and \lom{}, and is possibly inconsistent with similar flattening seen in observations.
\item
The relations between $\MstarChab/\MDM$ and $\MstarChab$ for galaxies in \lom{} and \him{} are consistent with Ref-50. Adopting true $\Mstar$ in place of $\MstarChab$ increases $\MstarChab/\MDM$ by $\approx 0.3$ dex for \lom{} galaxies with $\Mstar > 10^{10.5}\Msun$, but has little effect for \him{} galaxies.
Specific SFRs of galaxies with $\MstarChab \gtrsim 10^{10.5}\Msun$ are higher in \him{} than in Ref-50, resulting in a lower passive fraction that does not rise with stellar mass up to at least $\MstarChab = 10^{11}\Msun$, in tension with observations (\Fig{star_formation}). This higher SFR is a result of a higher star-forming gas fraction in high-mass galaxies, likely due to the burstier feedback being more efficient at ejecting gas from galaxies at early times, delaying star-formation to lower $z$.
\item
While the half-light radii of early-type galaxies in \lom{} are consistent with Ref-50 and observations, those in \him{} with $\MstarChab > 10^{10} \Msun$ are about a factor of 2 larger at fixed $\MstarChab$, inconsistent with observations. The larger half-light radii are likely due to stronger stellar feedback and the stronger dimming of old stellar populations with a top-heavy IMF relative to a Chabrier IMF, increasing $r_e$ and decreasing $\MstarChab$, respectively.

\end{itemize}

The results of this project are intended to aid in the interpretation of evidence for IMF variations in real galaxies, especially in terms of {\it how} the IMF varies, either at the high-mass or the low-mass end. While a high-mass slope variation cannot be definitively ruled out by the present analysis, the model in which the IMF varies at the low-mass end (as well as the reference model with a Chabrier IMF) produces galaxies that match observations much more closely than one in which the high-mass slope is varied.  

It is intriguing that observations that are sensitive to the low-mass and high-mass slopes of the IMF prefer LoM and HiM, respectively, with neither model matching all observations simultaneously (Figs. \ref{fig:IMF_vs_sigma} and \ref{fig:highmassslope}). This result could indicate that IMF variations are more complex than those explored in this paper, or that systematic uncertainties in models used to constrain the IMF observationally are underestimated. 

This paper lays the groundwork for further analysis of the predicted trends between the IMF (parameterized by the MLE) and galaxy properties, which will be explored in a series of upcoming papers (Papers II and III). Paper II will investigate the trends between the MLE and global properties across the galaxy population, uncovering the observable properties that are predicted to correlate with the MLE most strongly. In Paper III we delve into spatially-resolved properties of individual galaxies, exploring how IMF variations affect radial gradients in $M/L$, metal abundances, and MLE to further expose the differences in the predictions due to the non-universality of the IMF. Paper III will also investigate the time dependence of the IMF and its effect on the evolution of galaxies in our simulations.

\section*{Acknowledgements}

 We are grateful to the anonymous referee for constructive feedback which increased the overall quality of the paper. We thank Fabio Fontanot for useful comments. C.B. thanks Bart Clauwens, Madusha Gunawardhana, Padraig Alton, Richard Bower, Tom Theuns, Marijn Franx, and Matthieu Schaller for insightful discussions regarding this project. 
 This work used the DiRAC Data Centric system at Durham University, operated by the Institute for Computational Cosmology on behalf of the STFC DiRAC HPC Facility (www.dirac.ac.uk). This equipment was funded by BIS National E-infrastructure capital grant ST/K00042X/1, STFC capital grants ST/H008519/1 and ST/K00087X/1, STFC DiRAC Operations grant ST/K003267/1 and Durham University. DiRAC is part of the National E-Infrastructure. RAC is a Royal Society University Research Fellows. We also gratefully acknowledge PRACE for awarding us access to the resource Curie based in France at Tr$\grave{\rm e}$s Grand Centre de Calcul. This work was sponsored by the Dutch National Computing Facilities Foundation (NCF) for the use of supercomputer facilities, with financial support from the Netherlands Organization for Scientific Research (NWO). This research made use of {\sc astropy}, a community-developed core {\sc python} package for Astronomy \citep{Astropy2013}.




\bibliographystyle{mnras} 
\bibliography{IMF} 

\begin{thebibliography}{}
\makeatletter
\relax
\def\mn@urlcharsother{\let\do\@makeother \do\$\do\&\do\#\do\^\do\_\do\%\do\~}
\def\mn@doi{\begingroup\mn@urlcharsother \@ifnextchar [ {\mn@doi@}
  {\mn@doi@[]}}
\def\mn@doi@[#1]#2{\def\@tempa{#1}\ifx\@tempa\@empty \href
  {http://dx.doi.org/#2} {doi:#2}\else \href {http://dx.doi.org/#2} {#1}\fi
  \endgroup}
\def\mn@eprint#1#2{\mn@eprint@#1:#2::\@nil}
\def\mn@eprint@arXiv#1{\href {http://arxiv.org/abs/#1} {{\tt arXiv:#1}}}
\def\mn@eprint@dblp#1{\href {http://dblp.uni-trier.de/rec/bibtex/#1.xml}
  {dblp:#1}}
\def\mn@eprint@#1:#2:#3:#4\@nil{\def\@tempa {#1}\def\@tempb {#2}\def\@tempc
  {#3}\ifx \@tempc \@empty \let \@tempc \@tempb \let \@tempb \@tempa \fi \ifx
  \@tempb \@empty \def\@tempb {arXiv}\fi \@ifundefined
  {mn@eprint@\@tempb}{\@tempb:\@tempc}{\expandafter \expandafter \csname
  mn@eprint@\@tempb\endcsname \expandafter{\@tempc}}}

\bibitem[\protect\citeauthoryear{Alton, Smith  \& Lucey}{Alton
  et~al.}{2017}]{Alton2017}
Alton P.~D.,  Smith R.~J.,   Lucey J.~R.,  2017, \mn@doi [\mnras]
  {10.1093/mnras/stx464}, 468, 1594

\bibitem[\protect\citeauthoryear{Arrigoni, Trager, Somerville  \&
  Gibson}{Arrigoni et~al.}{2010}]{Arrigoni2010}
Arrigoni M.,  Trager S.~C.,  Somerville R.~S.,   Gibson B.~K.,  2010, \mn@doi
  [\mnras] {10.1111/j.1365-2966.2009.15924.x}, 402, 173

\bibitem[\protect\citeauthoryear{{Astropy Collaboration}}{{Astropy
  Collaboration}}{2013}]{Astropy2013}
{Astropy Collaboration} 2013, \mn@doi [\aap] {10.1051/0004-6361/201322068},
  558, A33

\bibitem[\protect\citeauthoryear{Auger, Treu, Gavazzi, Bolton, Koopmans  \&
  Marshall}{Auger et~al.}{2010}]{Auger2010}
Auger M.~W.,  Treu T.,  Gavazzi R.,  Bolton A.~S.,  Koopmans L.~V.,   Marshall
  P.~J.,  2010, \mn@doi [\apjl] {10.1088/2041-8205/721/2/L163}, 721, L163

\bibitem[\protect\citeauthoryear{Baldry et~al.,}{Baldry
  et~al.}{2012}]{Baldry2012}
Baldry I.~K.,  et~al., 2012, \mn@doi [\mnras]
  {10.1111/j.1365-2966.2012.20340.x}, 421, 621

\bibitem[\protect\citeauthoryear{Barnab{\`{e}}, Spiniello, Koopmans, Trager,
  Czoske  \& Treu}{Barnab{\`{e}} et~al.}{2013}]{Barnabe2013}
Barnab{\`{e}} M.,  Spiniello C.,  Koopmans L.~V.,  Trager S.~C.,  Czoske O.,
  Treu T.,  2013, \mn@doi [\mnras] {10.1093/mnras/stt1727}, 436, 253

\bibitem[\protect\citeauthoryear{Bastian, Covey  \& Meyer}{Bastian
  et~al.}{2010}]{Bastian2010}
Bastian N.,  Covey K.~R.,   Meyer M.~R.,  2010, \mn@doi [\araa]
  {10.1146/annurev-astro-082708-101642}, 48, 339

\bibitem[\protect\citeauthoryear{Bate}{Bate}{2009}]{Bate2009}
Bate M.~R.,  2009, \mn@doi [\mnras] {10.1111/j.1365-2966.2008.14165.x}, 392,
  1363

\bibitem[\protect\citeauthoryear{Bate \& Bonnell}{Bate \&
  Bonnell}{2005}]{Bate2005}
Bate M.~R.,  Bonnell I.~A.,  2005, \mn@doi [\mnras]
  {10.1111/j.1365-2966.2004.08593.x}, 356, 1201

\bibitem[\protect\citeauthoryear{Baugh, Lacey, Frenk, Granato, Silva, Bressan,
  Benson  \& Cole}{Baugh et~al.}{2005}]{Baugh2005}
Baugh C.~M.,  Lacey C.~G.,  Frenk C.~S.,  Granato G.~L.,  Silva L.,  Bressan
  A.,  Benson A.~J.,   Cole S.,  2005, \mn@doi [\mnras]
  {10.1111/j.1365-2966.2004.08553.x/abs/}, 356, 1191

\bibitem[\protect\citeauthoryear{Behroozi, Wechsler  \& Conroy}{Behroozi
  et~al.}{2013}]{Behroozi2013}
Behroozi P.~S.,  Wechsler R.~H.,   Conroy C.,  2013, \mn@doi [\apj]
  {10.1088/0004-637X/770/1/57}, 770

\bibitem[\protect\citeauthoryear{Bekki}{Bekki}{2013}]{Bekki2013b}
Bekki K.,  2013, \mn@doi [\mnras] {10.1093/mnras/stt1735}, 436, 2254

\bibitem[\protect\citeauthoryear{Bernardi et~al.,}{Bernardi
  et~al.}{2018}]{Bernardi2018a}
Bernardi M.,  et~al., 2018, \mn@doi [\mnras] {10.1093/mnras/stx3171}, 475, 757

\bibitem[\protect\citeauthoryear{Blancato, Genel  \& Bryan}{Blancato
  et~al.}{2017}]{Blancato2017}
Blancato K.,  Genel S.,   Bryan G.,  2017, \mn@doi [\apj]
  {10.3847/1538-4357/aa7b84}, 845, 21

\bibitem[\protect\citeauthoryear{Bondi \& Hoyle}{Bondi \&
  Hoyle}{1944}]{Bondi1944}
Bondi H.,  Hoyle F.,  1944, \mn@doi [\mnras] {10.1093/mnras/104.5.273}, 104,
  273

\bibitem[\protect\citeauthoryear{Booth \& Schaye}{Booth \&
  Schaye}{2009}]{Booth2009}
Booth C.~M.,  Schaye J.,  2009, \mn@doi [\mnras]
  {10.1111/j.1365-2966.2009.15043.x}, 398, 53

\bibitem[\protect\citeauthoryear{Bower, Schaye, Frenk, Theuns, Schaller, Crain
  \& McAlpine}{Bower et~al.}{2017}]{Bower2017}
Bower R.~G.,  Schaye J.,  Frenk C.~S.,  Theuns T.,  Schaller M.,  Crain R.~A.,
   McAlpine S.,  2017, \mn@doi [\mnras] {10.1093/mnras/stw2735}, 465, 32

\bibitem[\protect\citeauthoryear{Bruzual \& Charlot}{Bruzual \&
  Charlot}{2003}]{Bruzual2003}
Bruzual G.,  Charlot S.,  2003, \mn@doi [\mnras]
  {10.1046/j.1365-8711.2003.06897.x}, 344, 1000

\bibitem[\protect\citeauthoryear{Calzetti}{Calzetti}{2001}]{Calzetti2001}
Calzetti D.,  2001, \mn@doi [\pasp] {10.1086/324269}, 113, 1449

\bibitem[\protect\citeauthoryear{Camps \& Baes}{Camps \&
  Baes}{2015}]{Camps2015}
Camps P.,  Baes M.,  2015, \mn@doi [Astron. Comput.]
  {10.1016/j.ascom.2014.10.004}, 9, 20

\bibitem[\protect\citeauthoryear{Cappellari et~al.,}{Cappellari
  et~al.}{2013}]{Cappellari2013b}
Cappellari M.,  et~al., 2013, \mn@doi [\mnras] {10.1093/mnras/stt644}, 432,
  1862

\bibitem[\protect\citeauthoryear{Cardelli, Clayton  \& Mathis}{Cardelli
  et~al.}{1989}]{Cardelli1989}
Cardelli J.,  Clayton G.,   Mathis J.,  1989, in Allamandola L.,  Tielens A.,
  eds,  IAU Symposium Vol. 135, Interstellar Dust. p.~5, \url
  {http://adsabs.harvard.edu/abs/1989IAUS..135P...5C}

\bibitem[\protect\citeauthoryear{Catinella et~al.,}{Catinella
  et~al.}{2018}]{Catinella2018}
Catinella B.,  et~al., 2018, \mn@doi [\mnras] {10.1093/mnras/sty089}, 476, 875

\bibitem[\protect\citeauthoryear{Cenarro, Gorgas, Vazdekis, Cardiel  \&
  Peletier}{Cenarro et~al.}{2003}]{Cenarro2003}
Cenarro A.~J.,  Gorgas J.,  Vazdekis A.,  Cardiel N.,   Peletier R.~F.,  2003,
  \mn@doi [\mnras] {10.1046/j.1365-8711.2003.06360.x}, 339, L12

\bibitem[\protect\citeauthoryear{Chabrier}{Chabrier}{2003}]{Chabrier2003a}
Chabrier G.,  2003, \mn@doi [\apj] {10.1086/374879}, 586, 1

\bibitem[\protect\citeauthoryear{Chang, Wel, Cunha  \& Rix}{Chang
  et~al.}{2015}]{Chang2015}
Chang Y.~Y.,  Wel A. V.~D.,  Cunha E.~D.,   Rix H.~W.,  2015, \mn@doi [\apjs]
  {10.1088/0067-0049/219/1/8}, 219

\bibitem[\protect\citeauthoryear{Charlot \& Fall}{Charlot \&
  Fall}{2000}]{Charlot2000}
Charlot S.,  Fall S.~M.,  2000, \mn@doi [\apj] {10.1086/309250}, 539, 718

\bibitem[\protect\citeauthoryear{Clauwens, Schaye  \& Franx}{Clauwens
  et~al.}{2015}]{Clauwens2015}
Clauwens B.,  Schaye J.,   Franx M.,  2015, \mn@doi [\mnras]
  {10.1093/mnras/stv603}, 449, 4091

\bibitem[\protect\citeauthoryear{Clauwens, Schaye  \& Franx}{Clauwens
  et~al.}{2016}]{Clauwens2016}
Clauwens B.,  Schaye J.,   Franx M.,  2016, \mn@doi [\mnras]
  {10.1093/mnras/stw1808}, 462, 2832

\bibitem[\protect\citeauthoryear{Collier, Smith  \& Lucey}{Collier
  et~al.}{2018}]{Collier2018}
Collier W.~P.,  Smith R.~J.,   Lucey J.~R.,  2018, \mn@doi [\mnras]
  {10.1093/mnras/sty1188}

\bibitem[\protect\citeauthoryear{Conroy \& Gunn}{Conroy \&
  Gunn}{2010}]{Conroy2010}
Conroy C.,  Gunn J.~E.,  2010, \mn@doi [\apj] {10.1088/0004-637X/712/2/833},
  712, 833

\bibitem[\protect\citeauthoryear{Conroy \& van Dokkum}{Conroy \& van
  Dokkum}{2012}]{Conroy2012b}
Conroy C.,  van Dokkum P.~G.,  2012, \mn@doi [\apj]
  {10.1088/0004-637X/760/1/71}, 760, 71

\bibitem[\protect\citeauthoryear{Conroy, Gunn  \& White}{Conroy
  et~al.}{2009}]{Conroy2009}
Conroy C.,  Gunn J.~E.,   White M.,  2009, \mn@doi [\apj]
  {10.1088/0004-637X/699/1/486}, 699, 486

\bibitem[\protect\citeauthoryear{Conroy, Graves  \& van Dokkum}{Conroy
  et~al.}{2014}]{Conroy2014}
Conroy C.,  Graves G.~J.,   van Dokkum P.~G.,  2014, \mn@doi [\apj]
  {10.1088/0004-637X/780/1/33}, 780, 33

\bibitem[\protect\citeauthoryear{Conroy, van Dokkum  \& Villaume}{Conroy
  et~al.}{2017}]{Conroy2017}
Conroy C.,  van Dokkum P.,   Villaume A.,  2017, \mn@doi [\apj]
  {10.3847/1538-4357/aa6190}, 837, 8

\bibitem[\protect\citeauthoryear{Correa, Schaye, Clauwens, Bower, Crain,
  Schaller, Theuns  \& Thob}{Correa et~al.}{2017}]{Correa2017}
Correa C.~A.,  Schaye J.,  Clauwens B.,  Bower R.~G.,  Crain R.~A.,  Schaller
  M.,  Theuns T.,   Thob A.~C.,  2017, \mn@doi [\mnras]
  {10.1093/mnrasl/slx133}, 472, L45

\bibitem[\protect\citeauthoryear{Crain et~al.,}{Crain et~al.}{2015}]{Crain2015}
Crain R.~A.,  et~al., 2015, \mn@doi [\mnras] {10.1093/mnras/stv725}, 450, 1937

\bibitem[\protect\citeauthoryear{Crain et~al.,}{Crain et~al.}{2017}]{Crain2017}
Crain R.~A.,  et~al., 2017, \mn@doi [\mnras] {10.1093/mnras/stw2586}, 464, 4204

\bibitem[\protect\citeauthoryear{Dahlen \& al.}{Dahlen \&
  al.}{2008}]{Dahlen2008}
Dahlen T.,  al. E.,  2008, \mn@doi [\apj] {10.1086/587978}, 681, 462

\bibitem[\protect\citeauthoryear{{Dalla Vecchia} \& Schaye}{{Dalla Vecchia} \&
  Schaye}{2012}]{DallaVecchia2012}
{Dalla Vecchia} C.,  Schaye J.,  2012, \mn@doi [\mnras]
  {10.1111/j.1365-2966.2012.21704.x}, 426, 140

\bibitem[\protect\citeauthoryear{Davies et~al.,}{Davies
  et~al.}{2016}]{Davies2016}
Davies L.~J.,  et~al., 2016, \mn@doi [\mnras] {10.1093/mnras/stw1342}, 461, 458

\bibitem[\protect\citeauthoryear{Davis \& McDermid}{Davis \&
  McDermid}{2017}]{Davis2017}
Davis T.~A.,  McDermid R.~M.,  2017, \mn@doi [\mnras] {10.1093/mnras/stw2366},
  464, 453

\bibitem[\protect\citeauthoryear{Davis, Efstathiou, Frenk  \& White}{Davis
  et~al.}{1985}]{Davis1985}
Davis M.,  Efstathiou G.,  Frenk C.~S.,   White S. D.~M.,  1985, \mn@doi [\apj]
  {10.1086/163168}, 292, 371

\bibitem[\protect\citeauthoryear{{De Lucia}, Fontanot  \& Hirschmann}{{De
  Lucia} et~al.}{2017}]{DeLucia2017}
{De Lucia} G.,  Fontanot F.,   Hirschmann M.,  2017, \mn@doi [\mnras]
  {10.1093/mnrasl/slw242}, 466, L88

\bibitem[\protect\citeauthoryear{{De Rossi}, Bower, Font, Schaye  \&
  Theuns}{{De Rossi} et~al.}{2017}]{DeRossi2017}
{De Rossi} M.~E.,  Bower R.~G.,  Font A.~S.,  Schaye J.,   Theuns T.,  2017,
  \mn@doi [\mnras] {10.1093/mnras/stx2158}, 472, 3354

\bibitem[\protect\citeauthoryear{Dilday et~al.,}{Dilday
  et~al.}{2010}]{Dilday2010}
Dilday B.,  et~al., 2010, \mn@doi [\apj] {10.1088/0004-637X/713/2/1026}, 713,
  1026

\bibitem[\protect\citeauthoryear{Dolag, Borgani, Murante  \& Springel}{Dolag
  et~al.}{2009}]{Dolag2009}
Dolag K.,  Borgani S.,  Murante G.,   Springel V.,  2009, \mn@doi [\mnras]
  {10.1111/j.1365-2966.2009.15034.x}, 399, 497

\bibitem[\protect\citeauthoryear{Driver et~al.,}{Driver
  et~al.}{2012}]{Driver2012}
Driver S.~P.,  et~al., 2012, \mn@doi [\mnras]
  {10.1111/j.1365-2966.2012.22036.x}, 427, 3244

\bibitem[\protect\citeauthoryear{Dutton, Mendel  \& Simard}{Dutton
  et~al.}{2012}]{Dutton2012}
Dutton A.~A.,  Mendel J.~T.,   Simard L.,  2012, \mn@doi [\mnras]
  {10.1111/j.1745-3933.2012.01230.x}, 422, L33

\bibitem[\protect\citeauthoryear{Ferreras, {La Barbera}, de~la Rosa, Alexandre,
  de Carvalho, Falc{\'{o}}n-Barroso  \& Ricciardelli}{Ferreras
  et~al.}{2013}]{Ferreras2013}
Ferreras I.,  {La Barbera} F.,  de~la Rosa I.~G.,  Alexandre V.,  de Carvalho
  R.~R.,  Falc{\'{o}}n-Barroso J.,   Ricciardelli E.,  2013, \mn@doi [\mnras]
  {10.1093/mnrasl/sls014}, 429, L15

\bibitem[\protect\citeauthoryear{Fischera \& Dopita}{Fischera \&
  Dopita}{2005}]{Fischera2005}
Fischera J.,  Dopita M.,  2005, \mn@doi [\apj] {10.1086/426185}, 619, 340

\bibitem[\protect\citeauthoryear{Fontanot}{Fontanot}{2014}]{Fontanot2014}
Fontanot F.,  2014, \mn@doi [\mnras] {10.1093/mnras/stu1078}, 442, 3138

\bibitem[\protect\citeauthoryear{Fontanot, {De Lucia}, Hirschmann, Bruzual,
  Charlot  \& Zibetti}{Fontanot et~al.}{2017}]{Fontanot2017}
Fontanot F.,  {De Lucia} G.,  Hirschmann M.,  Bruzual G.,  Charlot S.,
  Zibetti S.,  2017, \mn@doi [\mnras] {10.1093/mnras/stw2612}, 464, 3812

\bibitem[\protect\citeauthoryear{Fontanot, {De Lucia}, Xie, Hirschmann, Bruzual
   \& Charlot}{Fontanot et~al.}{2018}]{Fontanot2018}
Fontanot F.,  {De Lucia} G.,  Xie L.,  Hirschmann M.,  Bruzual G.,   Charlot
  S.,  2018, \mn@doi [\mnras] {10.1093/mnras/stx3323}, 475, 2467

\bibitem[\protect\citeauthoryear{Gallazzi, Charlot, Brinchmann, White  \&
  Tremonti}{Gallazzi et~al.}{2005}]{Gallazzi2005}
Gallazzi A.,  Charlot S.,  Brinchmann J.,  White S.~D.,   Tremonti C.~A.,
  2005, \mn@doi [\mnras] {10.1111/j.1365-2966.2005.09321.x}, 362, 41

\bibitem[\protect\citeauthoryear{Gargiulo et~al.,}{Gargiulo
  et~al.}{2015}]{Gargiulo2015b}
Gargiulo I.~D.,  et~al., 2015, \mn@doi [\mnras] {10.1093/mnras/stu2272}, 446,
  3820

\bibitem[\protect\citeauthoryear{Graur \& Maoz}{Graur \&
  Maoz}{2013}]{Graur2013}
Graur O.,  Maoz D.,  2013, \mn@doi [\mnras] {10.1093/mnras/sts718}, 430, 1746

\bibitem[\protect\citeauthoryear{Graur et~al.,}{Graur et~al.}{2014}]{Graur2014}
Graur O.,  et~al., 2014, \mn@doi [\apj] {10.1088/0004-637X/783/1/28}, 783, 28

\bibitem[\protect\citeauthoryear{Gunawardhana et~al.,}{Gunawardhana
  et~al.}{2011}]{Gunawardhana2011}
Gunawardhana M. L.~P.,  et~al., 2011, \mn@doi [\mnras]
  {10.1111/j.1365-2966.2011.18800.x}, 415, 1647

\bibitem[\protect\citeauthoryear{Guszejnov, Hopkins  \& Ma}{Guszejnov
  et~al.}{2017}]{Guszejnov2017}
Guszejnov D.,  Hopkins P.~F.,   Ma X.,  2017, \mn@doi [\mnras]
  {10.1093/mnras/stx2067}, 472, 2107

\bibitem[\protect\citeauthoryear{Gutcke \& Springel}{Gutcke \&
  Springel}{2017}]{Gutcke2017}
Gutcke T.~A.,  Springel V.,  2017, preprint (arXiv:1710.04222)

\bibitem[\protect\citeauthoryear{Haardt \& Madau}{Haardt \&
  Madau}{2001}]{Haardt2001}
Haardt F.,  Madau P.,  2001, in Neumann D.~M.,  Tran J. T.~V.,  eds, Clusters
  of Galaxies and the High Redshift Universe Observed in X-rays. Savoie,
  France, p.~64, \url {http://adsabs.harvard.edu/abs/2001cghr.confE..64H}

\bibitem[\protect\citeauthoryear{Haas, Schaye, Booth, Vecchia, Springel, Theuns
   \& Wiersma}{Haas et~al.}{2013}]{Haas2013}
Haas M.~R.,  Schaye J.,  Booth C.~M.,  Vecchia C.~D.,  Springel V.,  Theuns T.,
    Wiersma R. P.~C.,  2013, \mn@doi [\mnras] {10.1093/mnras/stt1487}, 435,
  2931

\bibitem[\protect\citeauthoryear{Habergham, Anderson  \& James}{Habergham
  et~al.}{2010}]{Habergham2010}
Habergham S.~M.,  Anderson J.~P.,   James P.~A.,  2010, \mn@doi [\apj]
  {10.1088/0004-637X/717/1/342}, 717, 342

\bibitem[\protect\citeauthoryear{Hennebelle \& Chabrier}{Hennebelle \&
  Chabrier}{2013}]{Hennebelle2013}
Hennebelle P.,  Chabrier G.,  2013, \mn@doi [\apj]
  {10.1088/0004-637X/770/2/150}, 770, 150

\bibitem[\protect\citeauthoryear{Hopkins}{Hopkins}{2012}]{Hopkins2012}
Hopkins P.~F.,  2012, \mn@doi [\mnras] {10.1111/j.1365-2966.2012.20731.x}, 423,
  2037

\bibitem[\protect\citeauthoryear{Jappsen, Klessen, Larson, Li  \& {Mac
  Low}}{Jappsen et~al.}{2005}]{Jappsen2005}
Jappsen A.-K.,  Klessen R.~S.,  Larson R.~B.,  Li Y.,   {Mac Low} M.-M.,  2005,
  \mn@doi [\aap] {10.1051/0004-6361:20042178}, 435, 611

\bibitem[\protect\citeauthoryear{Kelvin et~al.,}{Kelvin
  et~al.}{2012}]{Kelvin2012}
Kelvin L.~S.,  et~al., 2012, \mn@doi [\mnras]
  {10.1111/j.1365-2966.2012.20355.x}, 421, 1007

\bibitem[\protect\citeauthoryear{{Kennicutt, Jr.}}{{Kennicutt,
  Jr.}}{1998}]{KennicuttJr.1998}
{Kennicutt, Jr.} R.~C.,  1998, \mn@doi [\apj] {10.1086/305588}, 498, 541

\bibitem[\protect\citeauthoryear{Kewley \& Ellison}{Kewley \&
  Ellison}{2008}]{Kewley2008}
Kewley L.~J.,  Ellison S.~L.,  2008, \mn@doi [\apj] {10.1086/587500}, 681, 1183

\bibitem[\protect\citeauthoryear{Kormendy \& Ho}{Kormendy \&
  Ho}{2013}]{Kormendy2013}
Kormendy J.,  Ho L.~C.,  2013, \mn@doi [\araa]
  {10.1146/annurev-astro-082708-101811}, 51, 511

\bibitem[\protect\citeauthoryear{Kroupa}{Kroupa}{2001}]{Kroupa2001}
Kroupa P.,  2001, \mn@doi [\mnras] {10.1046/j.1365-8711.2001.04022.x}, 322, 231

\bibitem[\protect\citeauthoryear{Kroupa \& Weidner}{Kroupa \&
  Weidner}{2003}]{Kroupa2003}
Kroupa P.,  Weidner C.,  2003, \mn@doi [\apj] {10.1086/379105}, 598, 1076

\bibitem[\protect\citeauthoryear{Krumholz}{Krumholz}{2011}]{Krumholz2011}
Krumholz M.~R.,  2011, \mn@doi [\apj] {10.1088/0004-637X/743/2/110}, 743, 110

\bibitem[\protect\citeauthoryear{{La Barbera}, Ferreras, Vazdekis, de~la Rosa,
  de Carvalho, Trevisan, Falc{\'{o}}n-Barroso  \& Ricciardelli}{{La Barbera}
  et~al.}{2013}]{LaBarbera2013}
{La Barbera} F.,  Ferreras I.,  Vazdekis A.,  de~la Rosa I.~G.,  de Carvalho
  R.~R.,  Trevisan M.,  Falc{\'{o}}n-Barroso J.,   Ricciardelli E.,  2013,
  \mn@doi [\mnras] {10.1093/mnras/stt943}, 433, 3017

\bibitem[\protect\citeauthoryear{{La Barbera}, Ferreras  \& Vazdekis}{{La
  Barbera} et~al.}{2015}]{LaBarbera2015}
{La Barbera} F.,  Ferreras I.,   Vazdekis A.,  2015, \mn@doi [\mnras]
  {10.1093/mnrasl/slv029}, 449, L137

\bibitem[\protect\citeauthoryear{{La Barbera}, Vazdekis, Ferreras, Pasquali,
  Cappellari, Mart{\'{i}}n-Navarro, Sch{\"{o}}nebeck  \&
  Falc{\'{o}}n-Barroso}{{La Barbera} et~al.}{2016}]{LaBarbera2016}
{La Barbera} F.,  Vazdekis A.,  Ferreras I.,  Pasquali A.,  Cappellari M.,
  Mart{\'{i}}n-Navarro I.,  Sch{\"{o}}nebeck F.,   Falc{\'{o}}n-Barroso J.,
  2016, \mn@doi [\mnras] {10.1093/mnras/stv2996}, 457, 1468

\bibitem[\protect\citeauthoryear{Lange et~al.,}{Lange et~al.}{2015}]{Lange2015}
Lange R.,  et~al., 2015, \mn@doi [\mnras] {10.1093/mnras/stu2467}, 447, 2603

\bibitem[\protect\citeauthoryear{Lejeune, Cuisinier  \& Buser}{Lejeune
  et~al.}{1997}]{Lejeune1997}
Lejeune T.,  Cuisinier F.,   Buser R.,  1997, \mn@doi [\aaps]
  {10.1051/aas:1997373}, 125, 229

\bibitem[\protect\citeauthoryear{Lejeune, Cuisinier  \& Buser}{Lejeune
  et~al.}{1998}]{Lejeune1998}
Lejeune T.,  Cuisinier F.,   Buser R.,  1998, \mn@doi [\aaps]
  {10.1051/aas:1998405}, 130, 65

\bibitem[\protect\citeauthoryear{Li et~al.,}{Li et~al.}{2017}]{Li2017}
Li H.,  et~al., 2017, \mn@doi [\apj] {10.3847/1538-4357/aa662a}, 838, 77

\bibitem[\protect\citeauthoryear{Lupton, Blanton, Fekete, Hogg, O'Mullane,
  Szalay  \& Wherry}{Lupton et~al.}{2004}]{Lupton2004}
Lupton R.~H.,  Blanton M.~R.,  Fekete G.,  Hogg D.~W.,  O'Mullane W.,  Szalay
  A.~S.,   Wherry N.,  2004, \mn@doi [\pasp] {10.1086/382245}, 116, 133

\bibitem[\protect\citeauthoryear{Lyubenova et~al.,}{Lyubenova
  et~al.}{2016}]{Lyubenova2016}
Lyubenova M.,  et~al., 2016, \mn@doi [\mnras] {10.1093/mnras/stw2434}, 463,
  3220

\bibitem[\protect\citeauthoryear{Marigo \& Girardi}{Marigo \&
  Girardi}{2007}]{Marigo2007}
Marigo P.,  Girardi L.,  2007, \mn@doi [\aap] {10.1051/0004-6361:20066772},
  469, 239

\bibitem[\protect\citeauthoryear{Marigo, Girardi, Bressan, Groenewegen, Silva
  \& Granato}{Marigo et~al.}{2008}]{Marigo2008}
Marigo P.,  Girardi L.,  Bressan A.,  Groenewegen M. A.~T.,  Silva L.,
  Granato G.~L.,  2008, \mn@doi [\aap] {10.1051/0004-6361:20078467}, 482, 883

\bibitem[\protect\citeauthoryear{Mart{\'{i}}n-Navarro}{Mart{\'{i}}n-Navarro}{2016}]{Martin-Navarro2016}
Mart{\'{i}}n-Navarro I.,  2016, \mn@doi [\mnras] {10.1093/mnrasl/slv181}, 456,
  L104

\bibitem[\protect\citeauthoryear{Mart{\'{i}}n-Navarro, {La Barbera}, Vazdekis,
  Falc{\'{o}}n-Barroso  \& Ferreras}{Mart{\'{i}}n-Navarro
  et~al.}{2015a}]{Martin-Navarro2015b}
Mart{\'{i}}n-Navarro I.,  {La Barbera} F.,  Vazdekis A.,  Falc{\'{o}}n-Barroso
  J.,   Ferreras I.,  2015a, \mn@doi [\mnras] {10.1093/mnras/stu2480}, 447,
  1033

\bibitem[\protect\citeauthoryear{Mart{\'{i}}n-Navarro
  et~al.,}{Mart{\'{i}}n-Navarro et~al.}{2015b}]{Martin-Navarro2015c}
Mart{\'{i}}n-Navarro I.,  et~al., 2015b, \mn@doi [\apj]
  {10.1088/2041-8205/806/2/L31}, 806, L31

\bibitem[\protect\citeauthoryear{McDermid et~al.,}{McDermid
  et~al.}{2014}]{McDermid2014}
McDermid R.~M.,  et~al., 2014, \mn@doi [\apjl] {10.1088/2041-8205/792/2/L37},
  792, 5

\bibitem[\protect\citeauthoryear{McDermid et~al.,}{McDermid
  et~al.}{2015}]{McDermid2015}
McDermid R.~M.,  et~al., 2015, \mn@doi [\mnras] {10.1093/mnras/stv105}, 448,
  3484

\bibitem[\protect\citeauthoryear{Meurer et~al.,}{Meurer
  et~al.}{2009}]{Meurer2009}
Meurer G.~R.,  et~al., 2009, \mn@doi [\apj] {10.1088/0004-637X/695/1/765}, 695,
  765

\bibitem[\protect\citeauthoryear{Moster, Naab  \& White}{Moster
  et~al.}{2013}]{Moster2013}
Moster B.~P.,  Naab T.,   White S.~D.,  2013, \mn@doi [\mnras]
  {10.1093/mnras/sts261}, 428, 3121

\bibitem[\protect\citeauthoryear{Moustakas et~al.,}{Moustakas
  et~al.}{2013}]{Moustakas2013}
Moustakas J.,  et~al., 2013, \mn@doi [\apj] {10.1088/0004-637X/767/1/50}, 767

\bibitem[\protect\citeauthoryear{Narayanan \& Dav{\'{e}}}{Narayanan \&
  Dav{\'{e}}}{2012}]{Narayanan2012}
Narayanan D.,  Dav{\'{e}} R.,  2012, \mn@doi [\mnras]
  {10.1111/j.1365-2966.2012.21159.x}, 423, 3601

\bibitem[\protect\citeauthoryear{Narayanan \& Dav{\'{e}}}{Narayanan \&
  Dav{\'{e}}}{2013}]{Narayanan2013}
Narayanan D.,  Dav{\'{e}} R.,  2013, \mn@doi [\mnras] {10.1093/mnras/stt1548},
  436, 2892

\bibitem[\protect\citeauthoryear{Newman, Smith, Conroy, Villaume  \& van
  Dokkum}{Newman et~al.}{2017}]{Newman2017}
Newman A.~B.,  Smith R.~J.,  Conroy C.,  Villaume A.,   van Dokkum P.,  2017,
  \mn@doi [\apj] {10.3847/1538-4357/aa816d}, 845, 18

\bibitem[\protect\citeauthoryear{Oldham \& Auger}{Oldham \&
  Auger}{2018}]{Oldham2018}
Oldham L.,  Auger M.,  2018, \mn@doi [\mnras] {10.1093/mnras/stx2969}, 474,
  4169

\bibitem[\protect\citeauthoryear{Paalvast \& Brinchmann}{Paalvast \&
  Brinchmann}{2017}]{Paalvast2017}
Paalvast M.,  Brinchmann J.,  2017, \mn@doi [\mnras] {10.1093/mnras/stx1271},
  470, 1612

\bibitem[\protect\citeauthoryear{Perrett et~al.,}{Perrett
  et~al.}{2012}]{Perrett2012}
Perrett K.,  et~al., 2012, \mn@doi [\aj] {10.1088/0004-6256/144/2/59}, 144, 59

\bibitem[\protect\citeauthoryear{{Planck Collaboration}}{{Planck
  Collaboration}}{2014}]{Planck2014}
{Planck Collaboration} 2014, \mn@doi [\aap] {10.1051/0004-6361/201321529}, 571,
  A1

\bibitem[\protect\citeauthoryear{Portinari, Chiosi  \& Bressan}{Portinari
  et~al.}{1998}]{Portinari1998}
Portinari L.,  Chiosi C.,   Bressan A.,  1998, \aap, 334, 38

\bibitem[\protect\citeauthoryear{Posacki, Cappellari, Treu, Pellegrini  \&
  Ciotti}{Posacki et~al.}{2015}]{Posacki2015}
Posacki S.,  Cappellari M.,  Treu T.,  Pellegrini S.,   Ciotti L.,  2015,
  \mn@doi [\mnras] {10.1093/mnras/stu2098}, 446, 493

\bibitem[\protect\citeauthoryear{Rahmati, Pawlik, Rai{\v{c}}evic̀  \&
  Schaye}{Rahmati et~al.}{2013}]{Rahmati2013}
Rahmati A.,  Pawlik A.~H.,  Rai{\v{c}}evic̀ M.,   Schaye J.,  2013, \mn@doi
  [\mnras] {10.1093/mnras/stt066}, 430, 2427

\bibitem[\protect\citeauthoryear{Rosani, Pasquali, {La Barbera}, Ferreras  \&
  Vazdekis}{Rosani et~al.}{2018}]{Rosani2018}
Rosani G.,  Pasquali A.,  {La Barbera} F.,  Ferreras I.,   Vazdekis A.,  2018,
  \mn@doi [\mnras] {10.1093/mnras/sty528}, 476, 5233

\bibitem[\protect\citeauthoryear{Rosas-Guevara et~al.,}{Rosas-Guevara
  et~al.}{2015}]{Rosas-Guevara2015}
Rosas-Guevara Y.~M.,  et~al., 2015, \mn@doi [\mnras] {10.1093/mnras/stv2056},
  454, 1038

\bibitem[\protect\citeauthoryear{Sales, Navarro, Schaye, Vecchia, Springel  \&
  Booth}{Sales et~al.}{2010}]{Sales2010}
Sales L.~V.,  Navarro J.~F.,  Schaye J.,  Vecchia C.~D.,  Springel V.,   Booth
  C.~M.,  2010, \mn@doi [\mnras] {10.1111/j.1365-2966.2010.17391.x}, 409, 1541

\bibitem[\protect\citeauthoryear{Salpeter}{Salpeter}{1955}]{Salpeter1955}
Salpeter E.~E.,  1955, \mn@doi [\apj] {10.1086/145971}, 121, 161

\bibitem[\protect\citeauthoryear{Sarzi, Spiniello, {La Barbera},
  Krajnovi{\'{c}}  \& van~den Bosch}{Sarzi et~al.}{2018}]{Sarzi2018}
Sarzi M.,  Spiniello C.,  {La Barbera} F.,  Krajnovi{\'{c}} D.,   van~den Bosch
  R.,  2018, \mn@doi [\mnras] {10.1093/mnras/sty1092}

\bibitem[\protect\citeauthoryear{Schaller, {Dalla Vecchia}, Schaye, Bower,
  Theuns, Crain, Furlong  \& McCarthy}{Schaller et~al.}{2015}]{Schaller2015b}
Schaller M.,  {Dalla Vecchia} C.,  Schaye J.,  Bower R.~G.,  Theuns T.,  Crain
  R.~A.,  Furlong M.,   McCarthy I.~G.,  2015, \mn@doi [\mnras]
  {10.1093/mnras/stv2169}, 454, 2277

\bibitem[\protect\citeauthoryear{Schaye}{Schaye}{2004}]{Schaye2004}
Schaye J.,  2004, \mn@doi [\apj] {10.1086/421232}, 609, 667

\bibitem[\protect\citeauthoryear{Schaye \& {Dalla Vecchia}}{Schaye \& {Dalla
  Vecchia}}{2008}]{Schaye2008}
Schaye J.,  {Dalla Vecchia} C.,  2008, \mn@doi [\mnras]
  {10.1111/j.1365-2966.2007.12639.x}, 383, 1210

\bibitem[\protect\citeauthoryear{Schaye et~al.,}{Schaye
  et~al.}{2010}]{Schaye2010}
Schaye J.,  et~al., 2010, \mn@doi [\mnras] {10.1111/j.1365-2966.2009.16029.x},
  402, 1536

\bibitem[\protect\citeauthoryear{Schaye et~al.,}{Schaye
  et~al.}{2015}]{Schaye2015}
Schaye J.,  et~al., 2015, \mn@doi [\mnras] {10.1093/mnras/stu2058}, 446, 521

\bibitem[\protect\citeauthoryear{Segers, Schaye, Bower, Crain, Schaller  \&
  Theuns}{Segers et~al.}{2016}]{Segers2016}
Segers M.~C.,  Schaye J.,  Bower R.~G.,  Crain R.~A.,  Schaller M.,   Theuns
  T.,  2016, \mn@doi [\mnras] {10.1093/mnrasl/slw111}, 461, L102

\bibitem[\protect\citeauthoryear{Shen, Mo, White, Blanton, Kauffmann, Voges,
  Brinkmann  \& Csabai}{Shen et~al.}{2003}]{Shen2003}
Shen S.,  Mo H.~J.,  White S. D.~M.,  Blanton M.~R.,  Kauffmann G.,  Voges W.,
  Brinkmann J.,   Csabai I.,  2003, \mn@doi [\mnras]
  {10.1046/j.1365-8711.2003.06740.x}, 343, 978

\bibitem[\protect\citeauthoryear{Smith}{Smith}{2014}]{Smith2014}
Smith R.~J.,  2014, \mn@doi [\mnras] {10.1093/mnrasl/slu082}, 443, L69

\bibitem[\protect\citeauthoryear{Smith, Alton, Lucey, Conroy  \& Carter}{Smith
  et~al.}{2015}]{Smith2015b}
Smith R.~J.,  Alton P.,  Lucey J.~R.,  Conroy C.,   Carter D.,  2015, \mn@doi
  [\mnras] {10.1093/mnrasl/slv132}, 454, L71

\bibitem[\protect\citeauthoryear{Sonnenfeld, Treu, Marshall, Suyu, Gavazzi,
  Auger  \& Nipoti}{Sonnenfeld et~al.}{2015}]{Sonnenfeld2015}
Sonnenfeld A.,  Treu T.,  Marshall P.~J.,  Suyu S.~H.,  Gavazzi R.,  Auger
  M.~W.,   Nipoti C.,  2015, \mn@doi [\apj] {10.1088/0004-637X/800/2/94}, 800

\bibitem[\protect\citeauthoryear{Sonnenfeld, Nipoti  \& Treu}{Sonnenfeld
  et~al.}{2017}]{Sonnenfeld2017}
Sonnenfeld A.,  Nipoti C.,   Treu T.,  2017, \mn@doi [\mnras]
  {10.1093/mnras/stw2919}, 465, 2397

\bibitem[\protect\citeauthoryear{Spiniello, Koopmans, Trager, Czoske  \&
  Treu}{Spiniello et~al.}{2011}]{Spiniello2011}
Spiniello C.,  Koopmans L.~V.,  Trager S.~C.,  Czoske O.,   Treu T.,  2011,
  \mn@doi [\mnras] {10.1111/j.1365-2966.2011.19458.x}, 417, 3000

\bibitem[\protect\citeauthoryear{Spiniello, Trager, Koopmans  \&
  Chen}{Spiniello et~al.}{2012}]{Spiniello2012}
Spiniello C.,  Trager S.~C.,  Koopmans L.~V.,   Chen Y.~P.,  2012, \mn@doi
  [\apjl] {10.1088/2041-8205/753/2/L32}, 753, 32

\bibitem[\protect\citeauthoryear{Spiniello, Trager, Koopmans  \&
  Conroy}{Spiniello et~al.}{2014}]{Spiniello2014}
Spiniello C.,  Trager S.,  Koopmans L.~V.,   Conroy C.,  2014, \mn@doi [\mnras]
  {10.1093/mnras/stt2282}, 438, 1483

\bibitem[\protect\citeauthoryear{Springel}{Springel}{2005}]{Springel2005}
Springel V.,  2005, \mn@doi [\mnras] {10.1111/j.1365-2966.2005.09655.x}, 364,
  1105

\bibitem[\protect\citeauthoryear{Springel, White, Tormen  \&
  Kauffmann}{Springel et~al.}{2001}]{Springel2001}
Springel V.,  White S. D.~M.,  Tormen G.,   Kauffmann G.,  2001, \mn@doi
  [\mnras] {10.1046/j.1365-8711.2001.04912.x}, 328, 726

\bibitem[\protect\citeauthoryear{Springel, {Di Matteo}  \& Hernquist}{Springel
  et~al.}{2005}]{Springel2005a}
Springel V.,  {Di Matteo} T.,   Hernquist L.,  2005, \mn@doi [\mnras]
  {10.1111/j.1365-2966.2005.09238.x}, 361, 776

\bibitem[\protect\citeauthoryear{Thomas, Maraston, Schawinski, Sarzi  \&
  Silk}{Thomas et~al.}{2010}]{Thomas2010}
Thomas D.,  Maraston C.,  Schawinski K.,  Sarzi M.,   Silk J.,  2010, \mn@doi
  [\mnras] {10.1111/j.1365-2966.2010.16427.x}, 404, 1775

\bibitem[\protect\citeauthoryear{Thomas et~al.,}{Thomas
  et~al.}{2011}]{Thomas2011}
Thomas J.,  et~al., 2011, \mn@doi [\mnras] {10.1111/j.1365-2966.2011.18725.x},
  415, 545

\bibitem[\protect\citeauthoryear{Tortora, Romanowsky  \& Napolitano}{Tortora
  et~al.}{2013}]{Tortora2013}
Tortora C.,  Romanowsky A.~J.,   Napolitano N.~R.,  2013, \mn@doi [\apj]
  {10.1088/0004-637X/765/1/8}, 765

\bibitem[\protect\citeauthoryear{Tremonti et~al.,}{Tremonti
  et~al.}{2004}]{Tremonti2004}
Tremonti C.~A.,  et~al., 2004, \mn@doi [\apj] {10.1086/423264}, 613, 898

\bibitem[\protect\citeauthoryear{Treu, Auger, Koopmans, Gavazzi, Marshall  \&
  Bolton}{Treu et~al.}{2010}]{Treu2010}
Treu T.,  Auger M.~W.,  Koopmans L. V.~E.,  Gavazzi R.,  Marshall P.~J.,
  Bolton A.~S.,  2010, \mn@doi [\apj] {10.1088/0004-637X/709/2/1195}, 709, 1195

\bibitem[\protect\citeauthoryear{{Van Dokkum} \& Conroy}{{Van Dokkum} \&
  Conroy}{2010}]{vanDokkum2010}
{Van Dokkum} P.~G.,  Conroy C.,  2010, \mn@doi [\nat] {10.1038/nature09578},
  468, 940

\bibitem[\protect\citeauthoryear{Vazdekis, Casuso, Peletier  \&
  Beckman}{Vazdekis et~al.}{1996}]{Vazdekis1996}
Vazdekis A.,  Casuso E.,  Peletier R.~F.,   Beckman J.~E.,  1996, \mn@doi
  [\apjs] {10.1086/192340}, 106, 307

\bibitem[\protect\citeauthoryear{Vazdekis, S{\'{a}}nchez-Bl{\'{a}}zquez,
  Falc{\'{o}}n-Barroso, Cenarro, Beasley, Cardiel, Gorgas  \&
  Peletier}{Vazdekis et~al.}{2010}]{Vazdekis2010}
Vazdekis A.,  S{\'{a}}nchez-Bl{\'{a}}zquez P.,  Falc{\'{o}}n-Barroso J.,
  Cenarro A.~J.,  Beasley M.~A.,  Cardiel N.,  Gorgas J.,   Peletier R.~F.,
  2010, \mn@doi [\mnras] {10.1111/j.1365-2966.2010.16407.x}, 404, 1639

\bibitem[\protect\citeauthoryear{Vazdekis, Ricciardelli, Cenarro,
  Rivero-Gonz{\'{a}}lez, D{\'{i}}az-Garc{\'{i}}a  \&
  Falc{\'{o}}n-Barroso}{Vazdekis et~al.}{2012}]{Vazdekis2012}
Vazdekis A.,  Ricciardelli E.,  Cenarro A.~J.,  Rivero-Gonz{\'{a}}lez J.~G.,
  D{\'{i}}az-Garc{\'{i}}a L.~A.,   Falc{\'{o}}n-Barroso J.,  2012, \mn@doi
  [\mnras] {10.1111/j.1365-2966.2012.21179.x}, 424, 157

\bibitem[\protect\citeauthoryear{Westera, Lejeune, Buser, Cuisinier  \&
  A.}{Westera et~al.}{2002}]{Westera2002}
Westera P.,  Lejeune T.,  Buser R.,  Cuisinier F.,   A. G.~B.,  2002, \mn@doi
  [\aap] {10.1051/0004-6361:20011493}, 381, 524

\bibitem[\protect\citeauthoryear{Wiersma, Schaye  \& Smith}{Wiersma
  et~al.}{2009a}]{Wiersma2009a}
Wiersma R. P.~C.,  Schaye J.,   Smith B.~D.,  2009a, \mn@doi [\mnras]
  {10.1111/j.1365-2966.2008.14191.x}, 393, 99

\bibitem[\protect\citeauthoryear{Wiersma, Schaye, Theuns, {Dalla Vecchia}  \&
  Tornatore}{Wiersma et~al.}{2009b}]{Wiersma2009b}
Wiersma R. P. C.~R.,  Schaye J.,  Theuns T.,  {Dalla Vecchia} C.,   Tornatore
  L.,  2009b, \mn@doi [\mnras] {10.1111/j.1365-2966.2009.15331.x}, 399, 574

\bibitem[\protect\citeauthoryear{Zahid, Dima, Kudritzki, Kewley, Geller, Hwang,
  Silverman  \& Kashino}{Zahid et~al.}{2014}]{Zahid2014}
Zahid H.~J.,  Dima G.~I.,  Kudritzki R.-P.,  Kewley L.~J.,  Geller M.~J.,
  Hwang H.~S.,  Silverman J.~D.,   Kashino D.,  2014, \mn@doi [\apj]
  {10.1088/0004-637X/791/2/130}, 791, 130

\bibitem[\protect\citeauthoryear{Zhang, Romano, Ivison, Papadopoulos  \&
  Matteucci}{Zhang et~al.}{2018}]{Zhang2018}
Zhang Z.-Y.,  Romano D.,  Ivison R.~J.,  Papadopoulos P.~P.,   Matteucci F.,
  2018, \mn@doi [\nat] {10.1038/s41586-018-0196-x}, 558, 260

\bibitem[\protect\citeauthoryear{van Dokkum, Conroy, Villaume, Brodie  \&
  Romanowsky}{van Dokkum et~al.}{2017}]{vanDokkum2017}
van Dokkum P.,  Conroy C.,  Villaume A.,  Brodie J.,   Romanowsky A.,  2017,
  \mn@doi [\apj] {10.3847/1538-4357/aa7135}, 841, 23

\makeatother
\end{thebibliography}


\renewcommand\thefigure{\thesection.\arabic{figure}} 

\appendix

\renewcommand\thefigure{\thesection.\arabic{figure}}    
\section{ Aperture effects and IMF calibration details.}
\label{sec:Appendix_calibration}
\setcounter{figure}{0}  

\begin{figure*}
\includegraphics[width=\textwidth]{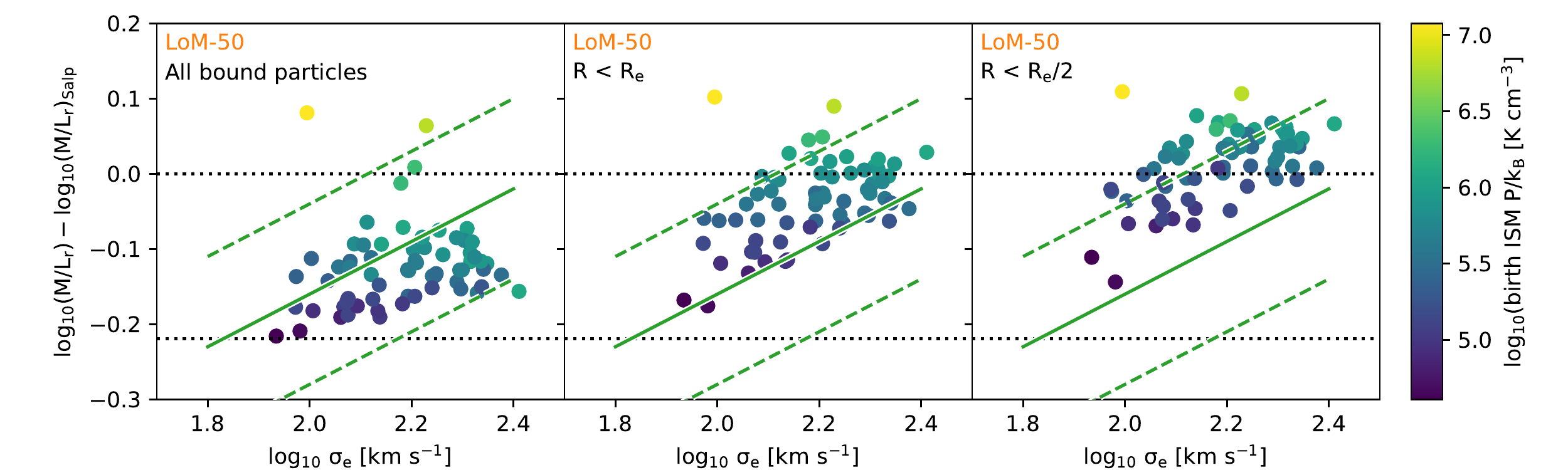}
\includegraphics[width=\textwidth]{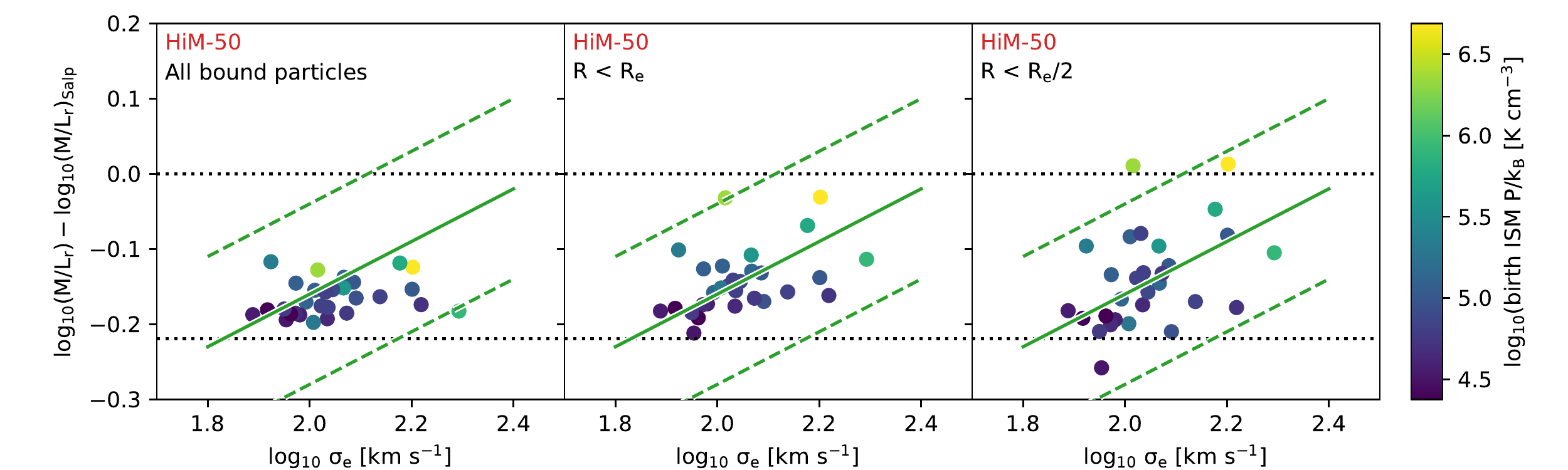}
\caption{Effect of aperture choice on IMF measurement. Mass-to-light excess (MLE) over Salpeter as a function of stellar velocity dispersion for mock C13 galaxies in our \lom{} (top row) and \him{} (bottom row) simulations at $z=0.1$. From left to right, the MLE is measured for all bound stars, stars within 1 $r_e$, and stars within 0.5 $r_e$, respectively. For reference, in all panels we plot the MLE$_r-\sigma_e$ relation of \citet{Cappellari2013b} with 1$\sigma$ scatter as \CapColour{}-solid and -dashed lines, respectively. Horizontal dashed lines indicate MLE$_r$ for Salpeter and Chabrier IMFs from top to bottom, respectively. The value of MLE is sensitive to the precise choice of aperture.}
\label{fig:IMF_vs_sigma_aperture}
\end{figure*}

In \Fig{IMF_vs_sigma_aperture}, we show the effect of aperture choice on the MLE-$\sigma_e$ relation for mock C13 galaxies in \lom{} and \him{}. Comparing the left and right columns, we see that the global IMF, measured over all stellar particles in each galaxy, underestimates the MLE values when measured within $r_e$, by $\approx 0.1$ dex. Choosing an even smaller aperture of $r_e/2$ increases the difference further to $\approx 0.1$ dex. We did not measure the IMF within $r_e/8$, as for many galaxies this is below the resolution limit of the simulations.

\begin{figure*}
\includegraphics[width=0.45\textwidth]{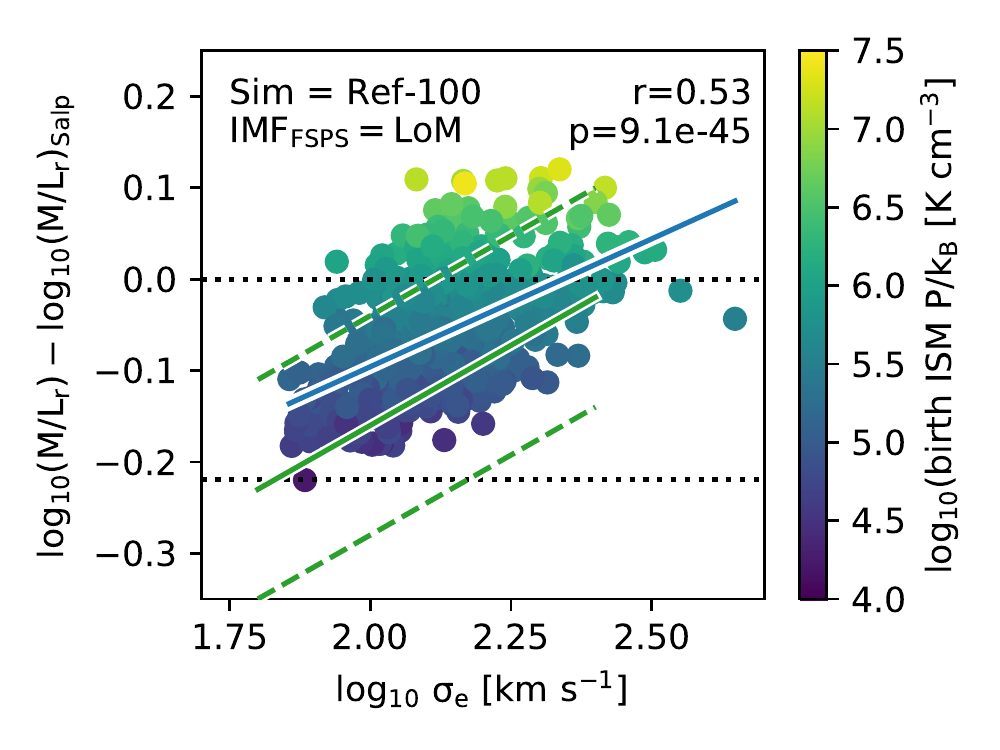}
\includegraphics[width=0.45\textwidth]{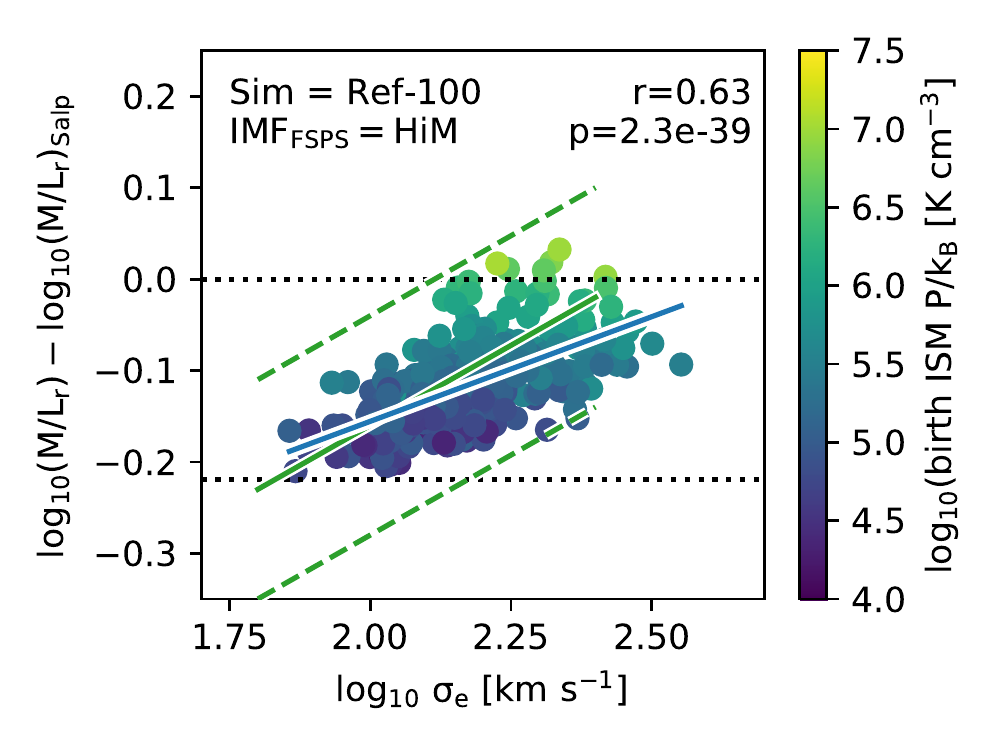}

\includegraphics[width=0.45\textwidth]{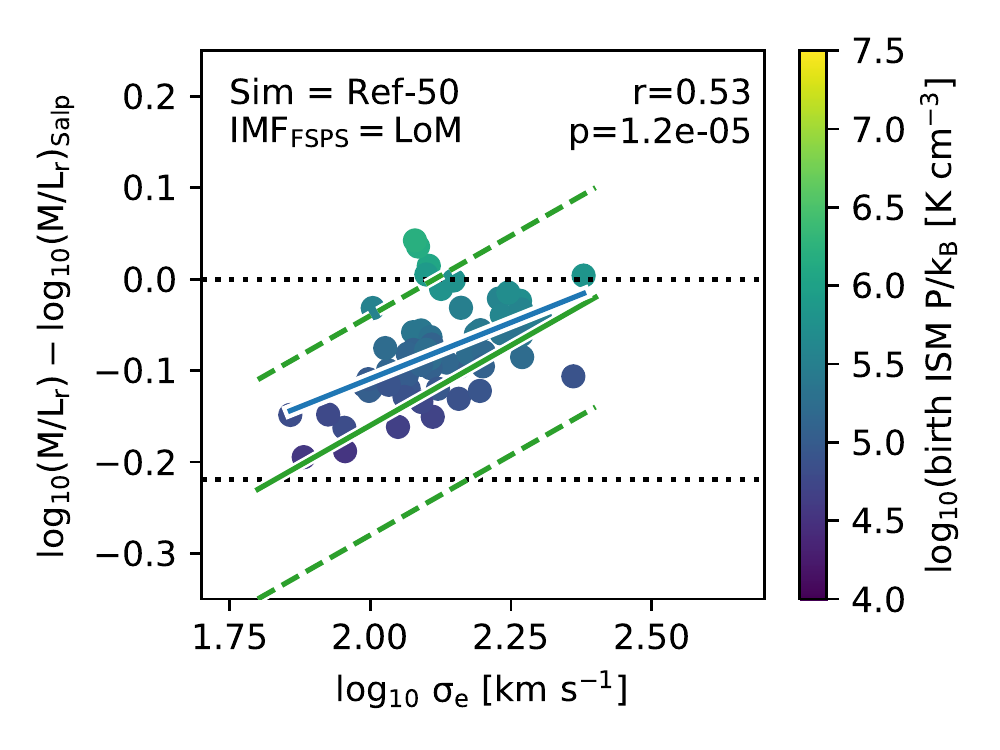}
\includegraphics[width=0.45\textwidth]{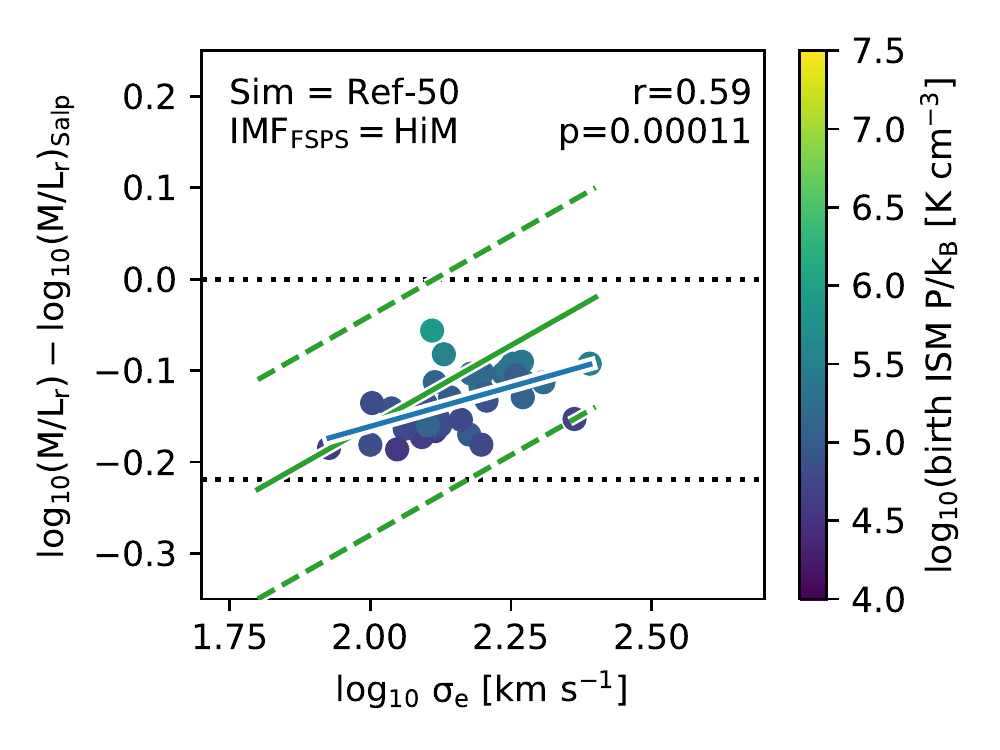}

\includegraphics[width=0.45\textwidth]{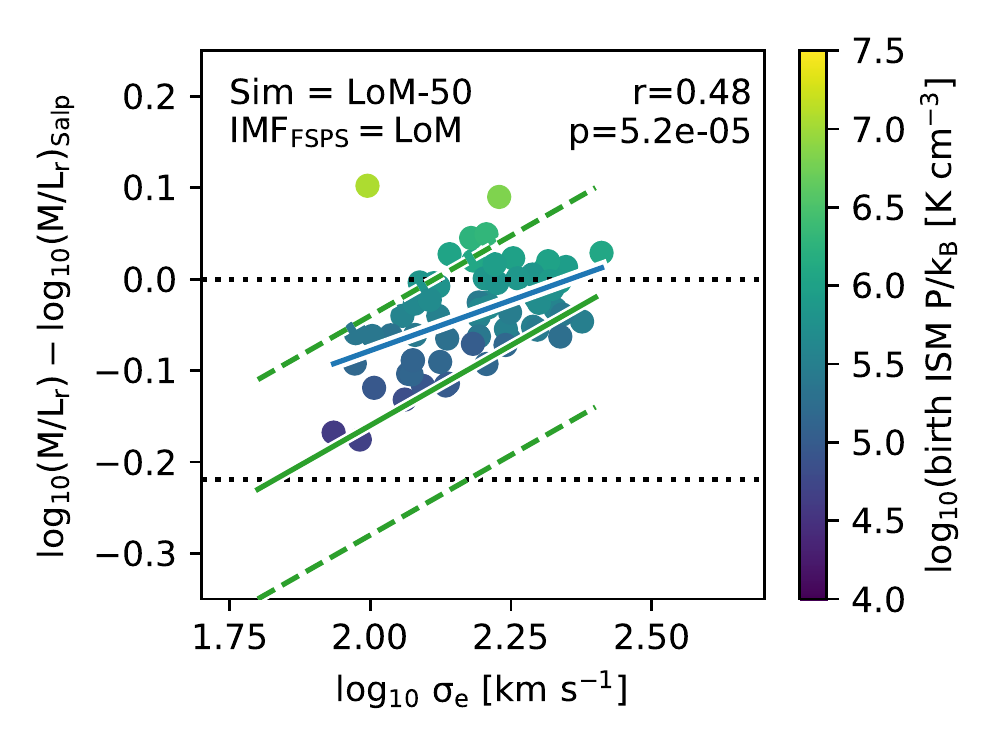}
\includegraphics[width=0.45\textwidth]{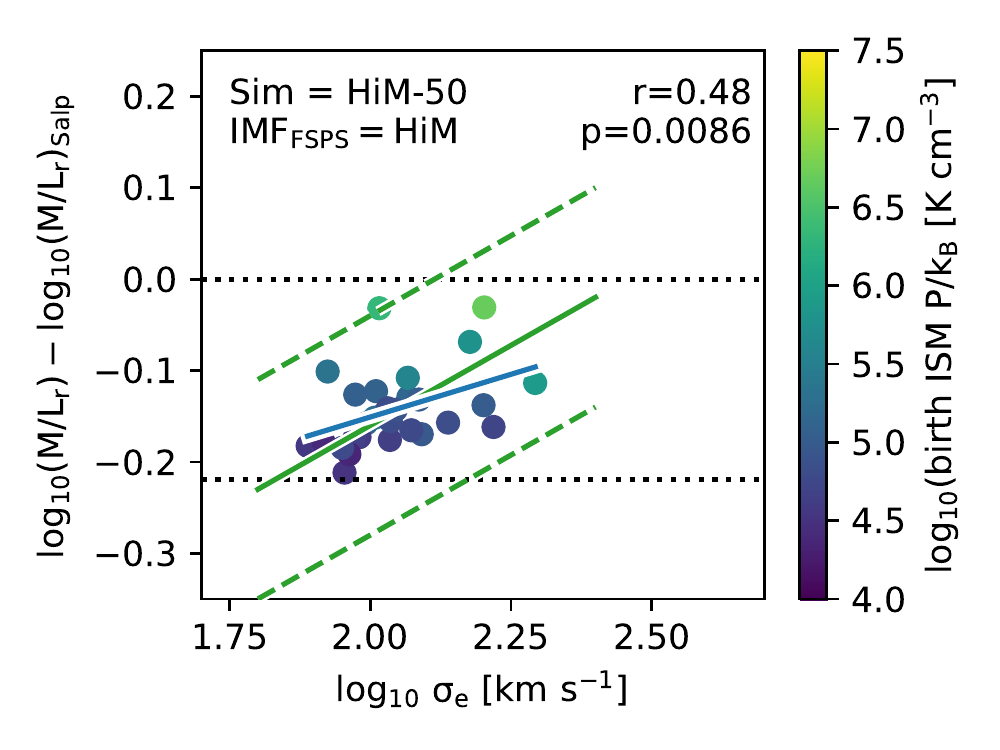}
\caption{Variable IMF prescription calibration. MLE$_r$ as a function of $\sigma_e$ for galaxies consistent with a C13 selection. All quantities are measured within $r_e$. Simulations in the left and right columns assume a LoM and HiM IMF variation prescription, respectively. Top row: Post-processed Ref-100 simulation. Middle row: Post-processed Ref-50 simulation. Bottom row: Self-consistent \lom{} (left) and \him{} (right) simulations. Least-absolute deviation fits are shown as blue solid lines, and the Spearman $r$ coefficient and its associated $p$-value are indicated in the upper right corner of each panel. The calibrated correlation between the MLE$_r$ and $\sigma_e$ is preserved in the \lom{} simulation, but is reduced in \him{}.}
\label{fig:IMF_calibration}
\end{figure*}

\Fig{IMF_calibration} outlines the process of calibrating the variable IMF prescriptions LoM and HiM for mock C13 galaxies in the left and right columns, respectively, by showing the MLE-$\sigma_e$ relation for different steps in the process. The top row shows the Ref-100 simulation (with the same resolution as Ref-50 but with a volume (100 Mpc)$^3$), where the $z=0.1$ masses and luminosities of the stars were recomputed in post-processing assuming they evolved according to either the LoM or HiM variable IMF prescription (upper-left and -right panels, respectively). The positive trend up to high $\sigma_e$ is clear in both cases. While the best-fit relation in \lom{} is offset from the C13 relation, the slopes are consistent. As mentioned in \Sec{IMF_vs_sigma}, this offset is due to the fact that this IMF variation prescription was originally calibrated using apertures larger than $r_e$. The smaller aperture size in this plot excludes stars with Chabrier-like IMFs, increasing the MLE for the entire relation. For \him{}, the normalization of the MLE$_r-\sigma_e$ relation is much closer to the C13 relation, but the slope is slightly shallower.  Since these calibrations were done by eye, a perfect match to the C13 slope is not expected. Indeed, the agreement with C13 is still very good, since most of our points lie within the 1$\sigma$ scatter of their relation.

The middle row shows the same IMF prescriptions applied to the Ref-50 simulation. Here we are missing the high-$\sigma_e$ galaxies due to the smaller box size, but the positive trend is still significant in both cases. Finally, the bottom row shows the results for \lom{} and \him{}, which were run with the LoM and HiM IMF prescriptions self-consistently included. The trend for LoM is preserved in \lom{}, with a slightly higher normalization. As mentioned in \Sec{IMF_vs_sigma}, this is due to the typically larger birth ISM densities at which stars are born in \lom{} relative to Ref-50. On the other hand, the trend in \him{} is slightly weaker due to smaller values of $\sigma_e$ as well as younger ages. Better statistics at high-$\sigma_e$ may be required to determine if galaxies in \him{} are inconsistent with the C13 trend.

\begin{figure*}
\includegraphics[width=\textwidth]{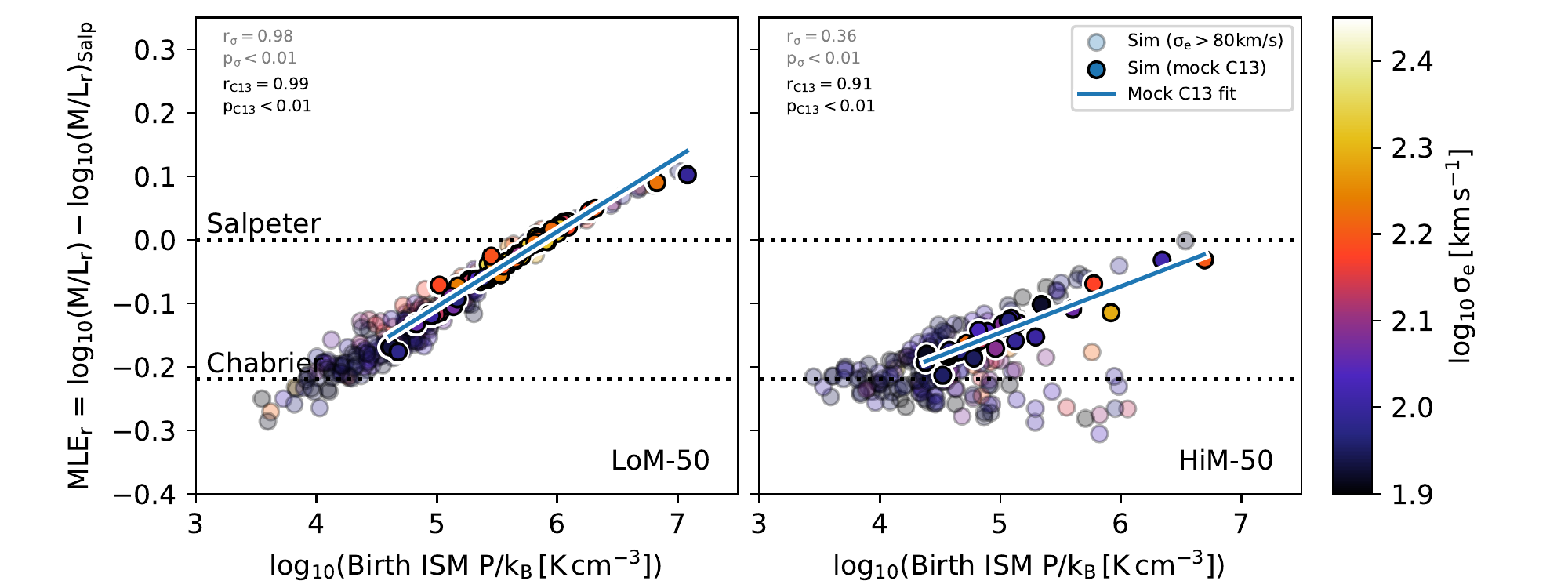}

\caption{ MLE$_r$ as a function of birth ISM pressure for \lom{} and \him{} (left and right panels, respectively), coloured by central stellar velocity dispersion, $\sigma_e$. All quantities are $r$-band light weighted, measured within 2D projected $r_e$. Galaxies with $\sigma_e > 80\kms$ are shown as translucent circles, while those that fall under our ``mock C13'' selection criteria are shown as opaque circles. }
\label{fig:IMF_vs_pressure}
\end{figure*}

\Fig{IMF_vs_pressure} shows the MLE are a function of the median birth pressure of stars within $r_e$ of galaxies at $z=0.1$ in \lom{} (left) and \him{} (right). MLE correlates extremely well with pressure in \lom{}, but more weakly in \him{}. The larger scatter in \him{} is due to the age-dependence of MLE for a shallow high-mass IMF slope.

\section{ Self-consistency tests}
\label{sec:Appendix_self_consistency}
\setcounter{figure}{0}  

\begin{figure}
  \centering 
\includegraphics[width=0.45\textwidth]{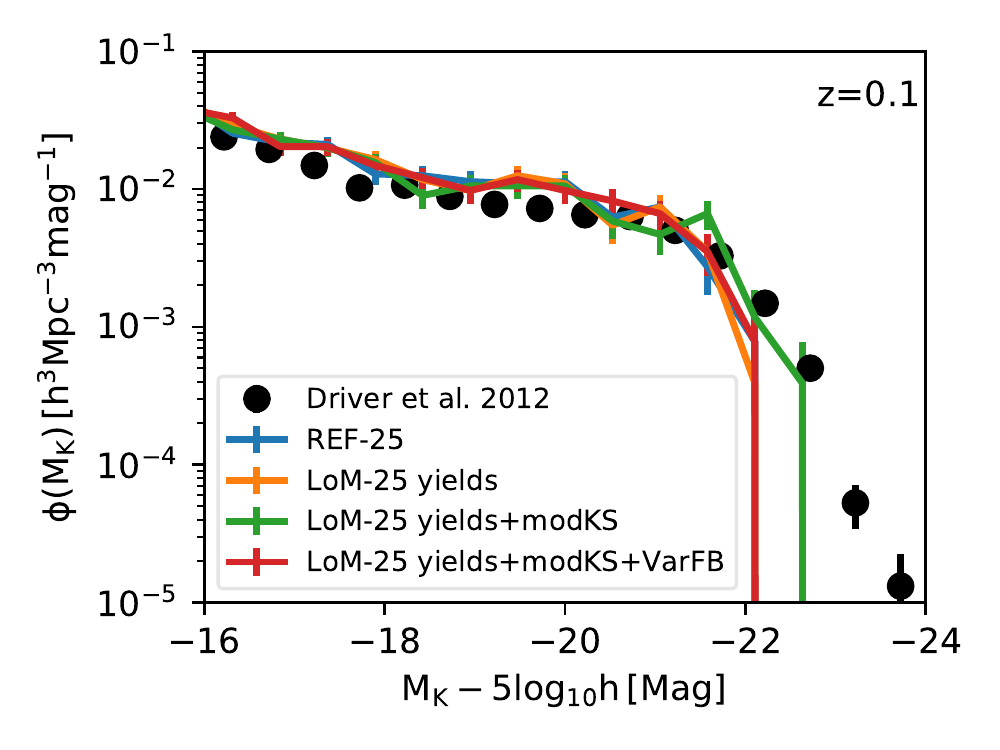}
\includegraphics[width=0.45\textwidth]{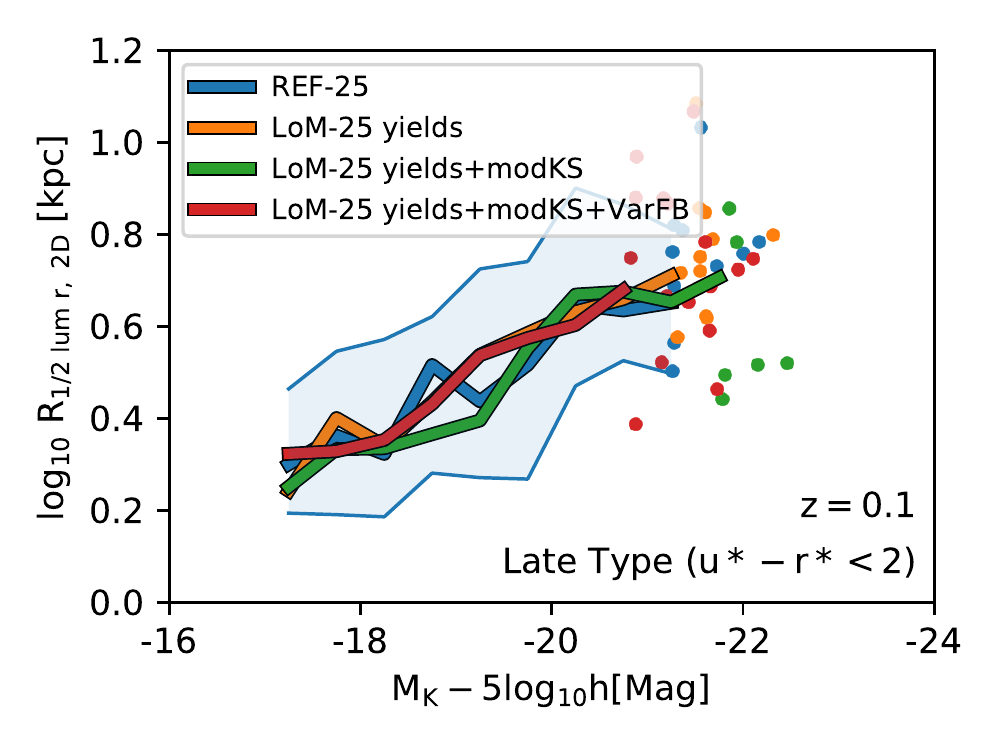}
\includegraphics[width=0.45\textwidth]{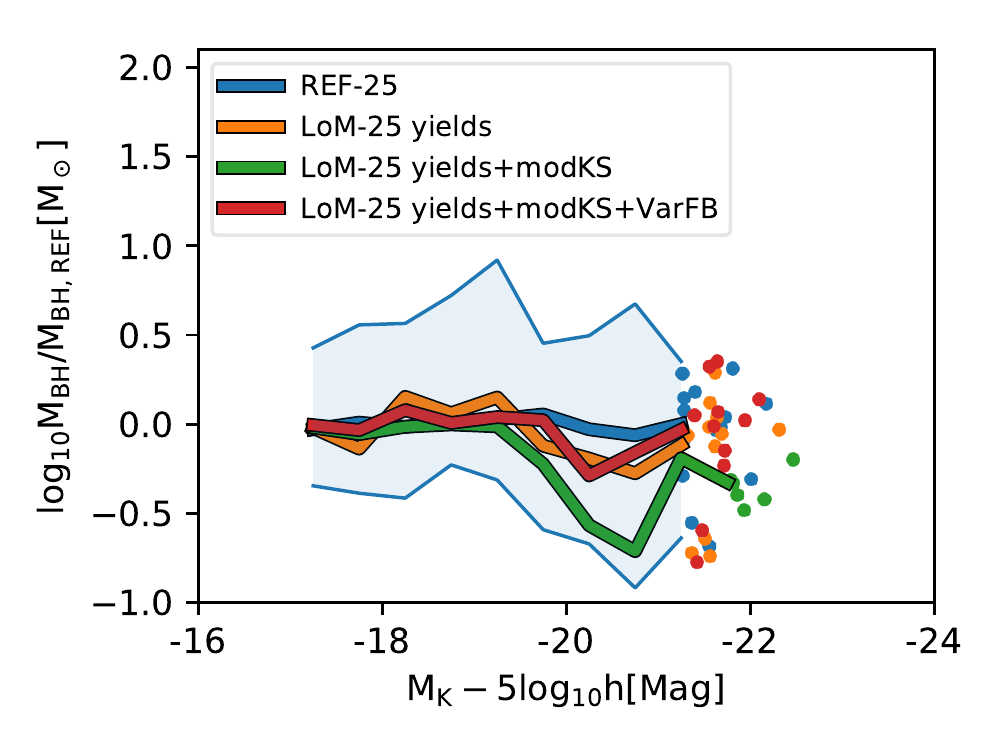}
\caption{As in the left column of \Fig{calibration_diagnostics} but for simulations with the same model as \lom{} but for a (25 Mpc)$^3$ volume (rather than (50 Mpc)$^3$). In blue we show the reference 25 Mpc box (Ref-25), while other colours show simulations with LoM, adding new effects of the variable IMF one at a time. In orange we show the effect of only changing the stellar yields while keeping everything else fixed at the reference (Chabrier) prescription. In green we self-consistently modify the star-formation law. In red we further make the stellar feedback self-consistent with the variable IMF, this being our fiducial LoM model. The left panel shows the K-band luminosity function. The middle panel shows half light radius as a function of K-band absolute magnitude. To clarify the deviations of $\MBH$ from the Ref-25 model, the right panel shows the median $\MBH$ relative to the median for Ref-25, rather than absolute $\MBH$ as in \Fig{calibration_diagnostics}. The "yields" model and our fiducial, self-consistent model match the calibrated values of the Ref model very closely, while the model "yields+modKS" deviates with higher luminosities and lower BH masses.}
\label{fig:calibration_25_LoM}
\end{figure}

Making the simulations completely self-consistent with the variable IMF while simultaneously ensuring that the subgrid calibration diagnostics remained consistent with the reference model was a challenging and painstaking process. Due to the non-linear process of galaxy formation and its connection with the IMF, it can be unclear which effect of the variable IMF is responsible for changes in the galaxy properties. Thus, we investigated the effect of a variable IMF in smaller, (25 Mpc)$^3$ boxes, adding new effects of the variable IMFs one at a time until they became self-consistent. \Fig{calibration_25_LoM} shows this process for \lom{}, and its effects on the subgrid physics calibration diagnostics: the K-band luminosity ($M_K$) function, the $r_e-M_K$ relation, and the $\MBH-M_K$ relation. To emphasize discrepancies in $\MBH$, we plot medians relative to the Ref-25 simulation in the right panel.

First we allowed the IMF to affect only the yields in the simulation, while keeping the feedback and star formation law consistent with that used for the reference (Chabrier IMF) model. This model we refer to as ``LoM-25 yields'', represented by the orange curves in \Fig{calibration_25_LoM}. For all calibration diagnostics, this model agrees very well with Ref-25, which is not surprising since metallicities are not strongly affected in the \lom{} runs.

We next added the effect of modifying the physical star-formation law such that the observed star formation law, i.e., intrinsic UV surface brightness as a function of gas surface density, was preserved, shown as the green curve. In this case the simulation produced a larger number of high-mass galaxies, while reducing $\MBH$ by over 0.5 dex for these galaxies, without strongly affecting the sizes. This result can be explained by self-regulation of stellar and BH growth. Since the feedback per stellar mass formed was still set as ``Chabrier'', the increased normalization of the SF law (i.e. the smaller gas consumption timescale) would result in stronger stellar feedback at fixed gas surface density. Thus, galaxies naturally decrease the gas density (and thus the SFR) until the feedback returns to the value appropriate for self-regulation (i.e. outflows roughly balance inflows). This lower gas density reduces the ability of gas to accrete onto BHs, reducing their accretion rates and thus their final $z=0.1$ masses. With lower BH masses, AGN feedback is suppressed, causing stellar feedback to compensate, resulting in higher $\Mstar$ and thus brighter $M_K$.

To alleviate this issue, we made the feedback self-consistent, shown by the red curves in \Fig{calibration_25_LoM}. With decreased stellar feedback per unit stellar mass, the effect of the modified star-formation law is canceled out such that the amount of feedback at fixed gas surface density is more consistent with the reference model. This model is much more consistent with Ref-25 and with the calibration data, and is our fiducial model.

\begin{figure}
  \centering 
\includegraphics[width=0.45\textwidth]{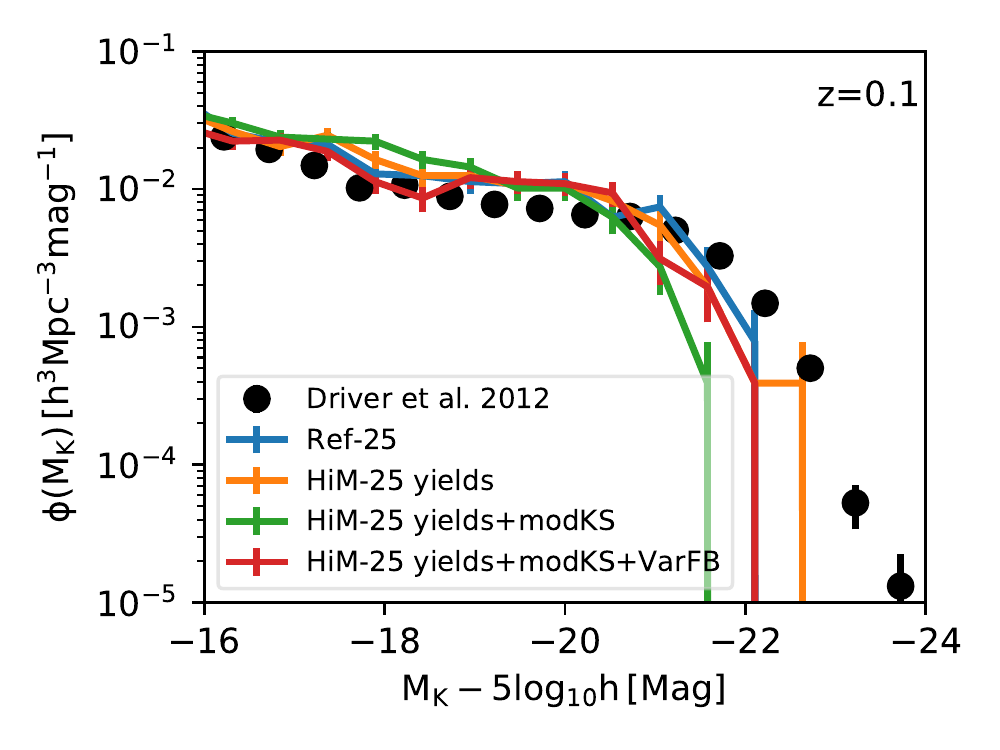}
\includegraphics[width=0.45\textwidth]{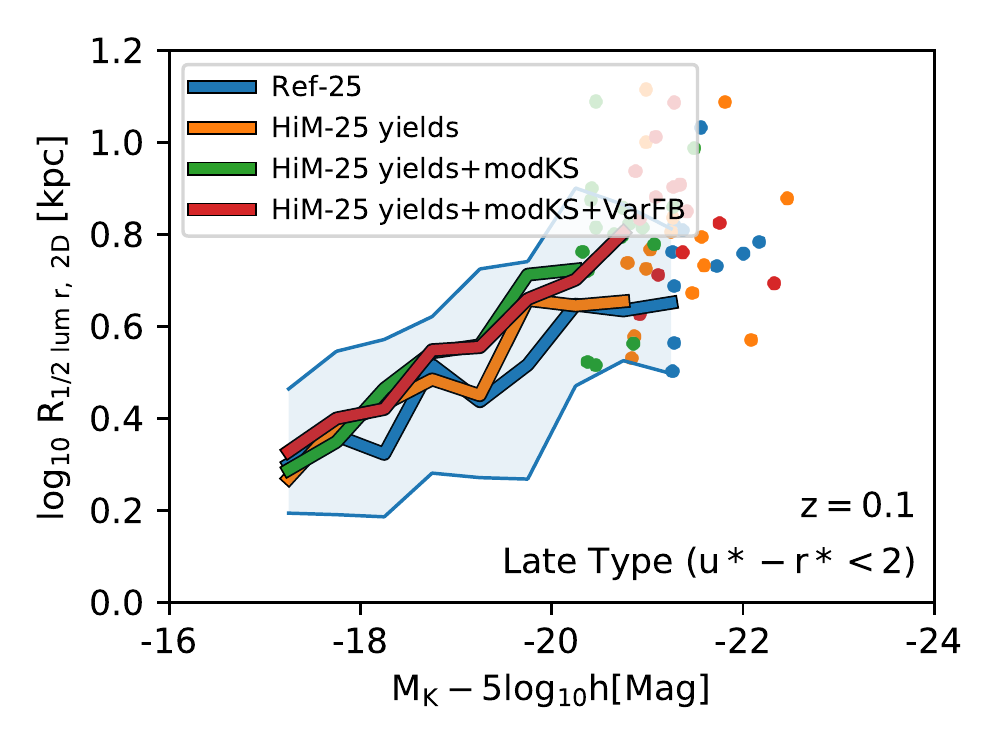}
\includegraphics[width=0.45\textwidth]{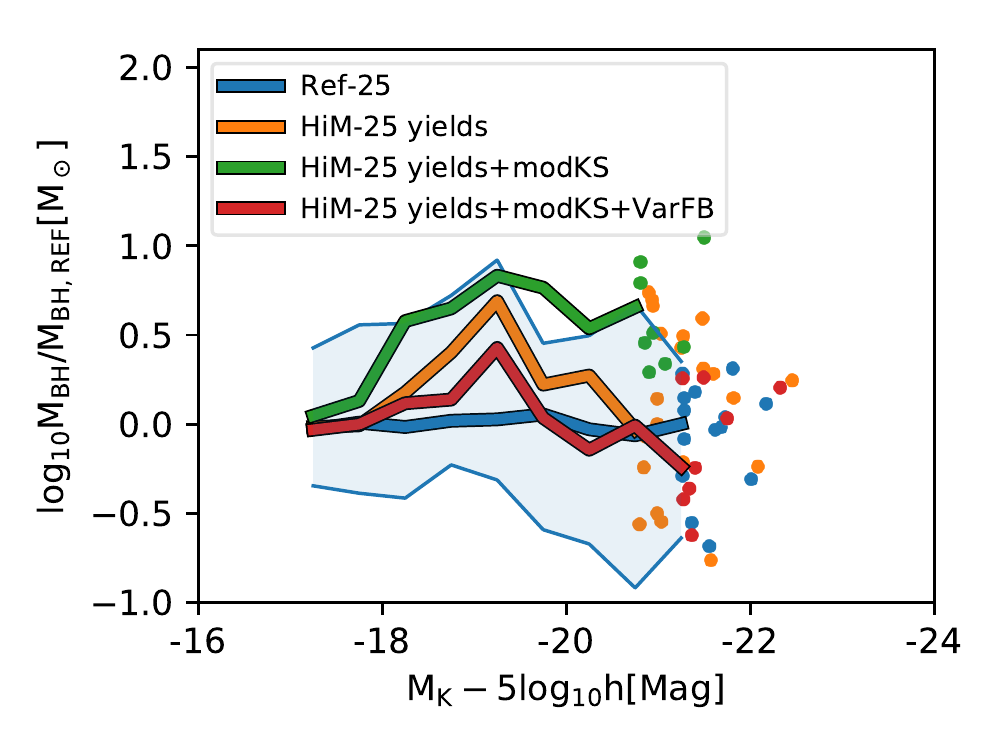}
\caption{As in \Fig{calibration_25_LoM} but for the HiM prescription. Out of these three models, our fiducial, self-consistent model does the best job at matching the reference model and hence the calibration diagnostics, especially for BH masses.}
\label{fig:calibration_25_HiM}
\end{figure}

\Fig{calibration_25_HiM} shows the same results but for the HiM IMF and corresponding HiM-25 simulations. Modifying the yields (orange curve) increases the sizes slightly and BH masses more strongly. Since the HiM IMF affects metallicities much more strongly than the LoM IMF, this can be seen as the effect of increasing the cooling rate due to higher metal yields, increasing the ability of gas to accrete onto BH particles and the importance of AGN feedback. Reducing the physical star-formation law at high pressure so as to maintain the same observed law makes the situation worse, further increasing BH masses and sizes, while also strongly suppressing the bright end of the $M_K$ function. The situation is essentially reversed relative to the LoM-25 case: the gas density is increased in order to obtain strong enough stellar feedback for self-regulation, which increases the BH masses and AGN feedback, which lowers stellar masses at the high-mass end of the GSMF. Again, including self-consistent stellar feedback (red curve) removes the need to change the gas densities, correcting the stellar and BH masses. The sizes are still slightly larger, but this is not a large difference from Ref-25.

\section{ The dwarf-to-giant ratio}
\label{sec:appendix_F05}
\setcounter{figure}{0}  

Spectroscopic IMF studies are sensitive to the ratio of dwarf-to-giant stars in the present-day stellar populations of the galaxies for which they constrain the IMF. \citet{LaBarbera2013} have concluded that, as long as models that differ only in IMF parameterization fit the IMF-sensitive stellar absorption features equally well, they yield the same dwarf-to-giant ratio. In that study, this ratio is defined as (their eq. 4)
\be
F_{0.5} = \frac{ \int_{0.1}^{0.5}M\Phi(M)dM}{\int_{0.1}^{100}M\Phi(M)dM}
\label{eqn:F05}
\ee
where $\Phi(M)$ is the IMF. However, as noted by \citet{Clauwens2016}, spectroscopic IMF studies are not able to constrain the denominator since in most ETGs, the stellar populations are so old that the highest stellar mass remaining is $\approx 1 \Msun$. Thus, while this definition of $F_{0.5}$ is unique for a given IMF, it is not clear that it is unique for a given mass fraction of dwarf stars at the present day. To investigate this, we employ the definition used by \citet{Clauwens2016}, where the ratio is instead defined with respect to stars with $m<1 \Msun$, $F_{0.5,1}$ (Equation \ref{eqn:F051}). $F_{0.5,1}$ is a more accurate representation of the present-day dwarf-to-giant mass ratio, to which spectroscopic IMF studies are sensitive. If it is true that as long as the present-day IMF-sensitive spectral features are well-fit, the choice of IMF parameterization does not affect $F_{0.5}$, it must also be true that it does not affect $F_{0.5,1}$.

\begin{figure}
  \centering 
\includegraphics[width=0.45\textwidth]{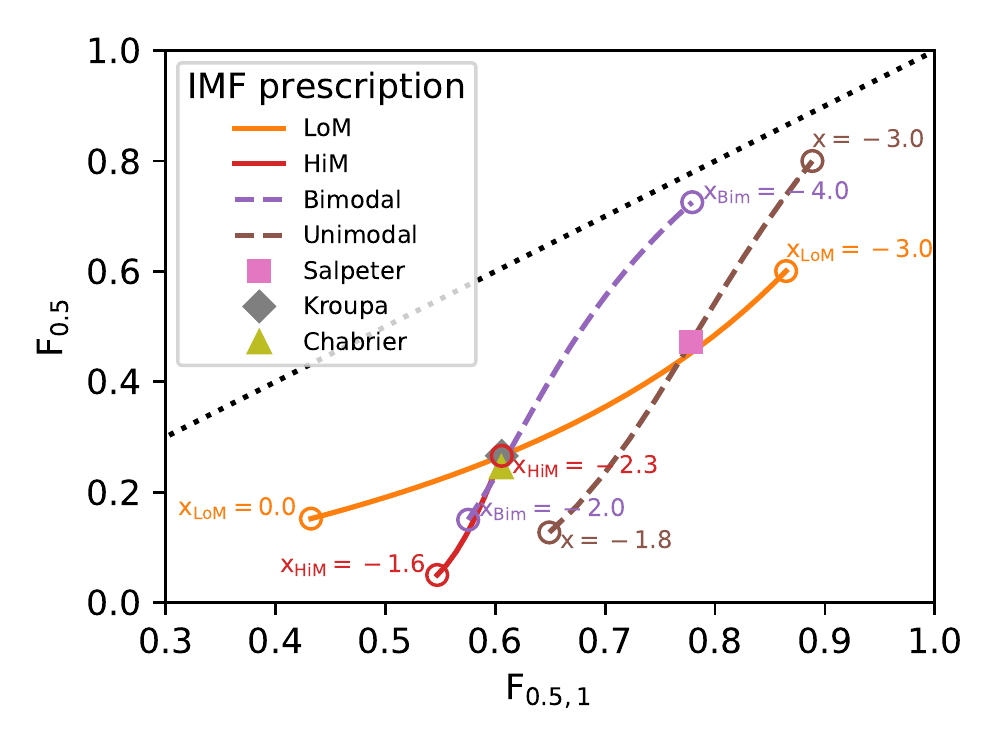}

\caption{Effect of IMF parameterization on the inferred dwarf-to-giant ratio. We show the mass fraction of stars with $m<0.5\Msun$ relative to the total mass of the IMF, $F_{0.5}$ (Equation \ref{eqn:F05}), as a function of the same fraction but relative to the mass $< 1 \Msun$, $F_{0.5,1}$ (Equation \ref{eqn:F051}). The former is the dwarf-to-giant ratio defined by \citet[][their Equation 4]{LaBarbera2013}, while the latter is that defined in \citet[][]{Clauwens2016}. A dotted black line shows the 1:1 relation. As orange and red solid lines we show the two variable IMF prescriptions used in this work, LoM and HiM, respectively. The dashed purple line shows the bimodal IMF prescription of \citet{Vazdekis1996}, while the dashed brown line shows the relation for a single power law IMF slope. The values for Salpeter, Kroupa, and Chabrier IMFs are indicated with filled symbols (see legend). Open circles mark the values of the IMF slope for the range over which the IMF is varied in each parameterization. At fixed $F_{0.5,1}$, the $F_{0.5}$ depends sensitively on the parameterization assumed.  }
\label{fig:F05}
\end{figure}

In \Fig{F05}, we plot $F_{0.5}$ as a function of $F_{0.5,1}$ for a variety of IMF parameterizations. Note that both of these quantities are unique for a given IMF. As orange and red solid lines we show results for the LoM and HiM IMF parameterizations respectively, where the range of values plotted corresponds to the ranges over which the IMFs are allowed to vary in our model (see \Sec{calibration}). As expected, for LoM the dwarf-to-giant ratio spans a wide range under either definition, with both increasing with the steepening of the low-mass slope. HiM also decreases for both dwarf-to-giant ratio definitions as the high-mass slope becomes shallower, but follows a much steeper trend than in the LoM case. This shows that $F_{0.5}$ is much more sensitive to changes in the high-mass slope than is $F_{0.5,1}$.  This result indicates that, at a fixed present-day dwarf-to-giant ratio, $F_{0.5}$ is sensitive to the parameterization of the IMF.

To further illustrate this point, we also include in \Fig{F05} results for the Bimodal IMF parameterization of \citet{Vazdekis1996}, where we plot results for the range of high-mass slopes recovered for high-mass early-type galaxies found by \citet{LaBarbera2013}, corresponding to $\Gamma_b = 1$ to 3 (or $x_{\rm Bimodal} = -2$ to $-4$ according to the IMF slope convention used in our paper). Unsurprisingly, the Bimodal IMF line follows the HiM trend and extends it to higher dwarf-to-giant ratios as the high-mass slope steepens. We also include the same trend for a single power-law (or ``unimodal'') IMF in brown. Here we also show results for slopes recovered for high-mass early-type galaxies by \citet{LaBarbera2013}, with $\Gamma=0.8$ to $2$ (or $x = -1.8$ to $-3$). The trend between $F_{0.5}$ and $F_{0.5,1}$ is also monotonic for this prescription, but again follows a separate track.

It is clear that, for a given IMF parameterization, $F_{0.5}$ is a good tracer of the present-day dwarf-to-giant ratio. However, for a fixed present-day dwarf-to-giant ratio, $F_{0.5,1}$, the corresponding zero-age value, $F_{0.5}$, is extremely sensitive to the choice of IMF parameterization. This is particularly true at $F_{0.5,1} \approx 0.7$, where the range of $F_{0.5}$ values ranges from sub-Chabrier to super-Salpeter, depending on the IMF parameterization employed. 

It is thus interesting that \citet{LaBarbera2013} find that models with different IMF parameterization are consistent in $F_{0.5}$. However, at fixed $F_{0.5}$, the difference in $F_{0.5,1}$ values between bimodal and unimodal IMFs is not large, and is comparable to the typical differences in $F_{0.5}$ seen between these two IMF parameterizations found by \citet{LaBarbera2013}. This is because IMF prescriptions that vary the high-mass slope are more sensitive to $F_{0.5}$ than to $F_{0.5,1}$, as can be seen by the steep slopes of the Bimodal and Unimodal lines in \Fig{F05}. It would thus be interesting to see if spectroscopic analyses would still yield consistent $F_{0.5}$ values under the assumption of a LoM-like IMF parameterization, as the differences could be greater.

\bsp	
\label{lastpage}
\end{document}